\documentclass{rQUF2e}

\usepackage{blindtext}
\usepackage[T1]{fontenc}
\usepackage{lmodern}
\usepackage[utf8]{inputenc}
\usepackage{mathrsfs}
\usepackage{epsfig}
\usepackage{epstopdf}
\usepackage{bigints}
\usepackage{lastpage} 
\usepackage{comment} 
\usepackage{setspace}
\usepackage{enumerate}
\usepackage{amsmath}
\usepackage{float,lscape}
\usepackage{array}
\usepackage{dcolumn}
\usepackage{multirow}
\usepackage{xcolor}
\usepackage{verbatim} 
\usepackage{lipsum}
\usepackage{setspace}
\usepackage{nth}
\usepackage{framed}
\usepackage{amsthm}
\usepackage{bm}
\usepackage{mathtools}
\usepackage{amssymb}
\usepackage{relsize}
\usepackage{graphicx}
\usepackage{algorithmicx}
\usepackage{algorithm}
\usepackage{algpseudocode}
\usepackage{hyperref}
\usepackage{lscape}
\usepackage{pdflscape}
\usepackage{booktabs}
\usepackage{threeparttable}

\newcommand{\E}{\mathop{\mbox{\sf E}}} 
     
\newcommand{\Cov}{\mathop{\mbox{\sf Cov}}}     
\newcommand{\QQ}{\mathop{\mbox{\sf Q}}}

\theoremstyle{plain}

\theoremstyle{definition}

\newcolumntype{.}{D{.}{.}{-1}}

\definecolor{dblue}{HTML}{000099}
\definecolor{spycol}{HTML}{000099}
\definecolor{ssocol}{HTML}{FF0000}
\definecolor{uprocol}{HTML}{FF33FF}
\definecolor{sdscol}{HTML}{003300}
\definecolor{vixcol}{HTML}{FF1493}
\definecolor{darkgreen}{HTML}{007F00}

\begin{document}

	\title{Model-driven statistical arbitrage on LETF option markets}
	
	\author{S. NASEKIN$^{\ast}$ ${\dag}$\thanks{$^\ast$Corresponding author.
			Email: sergey.nasekin@gmail.com} and W. K. H\"{A}RDLE$\ddag$\\
		\affil{$\dag$Lancaster University Management School, Lancaster University, Lancaster, United Kingdom, LA14YX\\
			$\ddag$ C.A.S.E.- Center for Applied Statistics \& Economics,
			Humboldt-Universit\"{a}t zu Berlin, Spandauer Str. 1, 10178 Berlin, Germany} }
	
	\maketitle

	
	\begin{abstract}
		In this paper, we study the statistical properties of the moneyness scaling transformation by \citet{leung_sircar}. This transformation adjusts the moneyness coordinate of the implied volatility smile in an attempt to remove the discrepancy between the IV smiles for levered and unlevered ETF options. We construct bootstrap uniform confidence bands which indicate that the implied volatility smiles are statistically different after moneyness scaling has been performed. 
		An empirical application shows that there are trading opportunities possible on the LETF market. A statistical arbitrage type strategy based on a dynamic semiparametric factor model is presented. This strategy presents a statistical decision algorithm which generates trade recommendations based on comparison of model and observed LETF implied volatility surface. It is shown to generate positive returns with a high probability.
		Extensive econometric analysis of LETF implied volatility process is performed including out-of-sample forecasting based on a semiparametric factor model and uniform confidence bands' study. It provides new insights into the latent dynamics of the implied volatility surface. We also incorporate Heston stochastic volatility into the moneyness scaling method for better tractability of the model.
	\end{abstract}
	
	
	\begin{classcode}C00, C14, C50\end{classcode}

	\section{Introduction}
	Exchange-traded funds (ETFs) are financial products that track indices, commodities, bonds, baskets of assets. They have become increasingly popular due to diversification benefits as well as the investor's ability to perform short-selling, buying on margin and lower expense ratios than, for instance, those of mutual funds.
	
	Leveraged ETFs (LETFs) are used to generate multiples or inverse multiples of returns on the underlying asset. For instance, the LETF ProShares Ultra S\&P500 (SSO) with a leverage ratio $\beta = +2$ is supposed to grow 2\% for every 1\% daily gain in the price of the S\&P500 index, minus an expense fee. An inverse leveraged ETF would invert the gain/loss of the underlying index and amplify it proportionally to the ratio: the ProShares UltraShort S\&P500 (SDS) with leverage ratio $\beta = -2$ would generate a 2\% gain for every 1\% daily loss in the price of the underlying S\&P500 index. Selected financial information on LETFs relevant for this study, is summarized in Table \ref{letfs_desc_info}.
	
	Due to their growing popularity and the nature of ETF and LETF similar dynamics, recently there has been growing research on leveraged ETFs and their consistent pricing. \citet{avellaneda_zhang} found that the terminal value of an option on a LETF depends on the integrated variance of the underlying LETF. \citet{ahn_et_al} show that given Heston dynamics for the underlying ETF, the corresponding LETF also has Heston dynamics with different parameters. \citet{leung_santoli} give a broad overview of LETFs, LETF options, their pricing and implied volatility.
	
	Studies on LETF options' implied volatility should be mentioned as most relevant for this work. \citet{figueroa_lopez_et_al} derived asymptotic equivalence of options on ETFs and LETFs under restricted conditions. \citet{lee_wang} studied a direct relationship between volatility skews of leveraged and unlevered products, providing asymptotic error estimates. \citet{leung_et_al} studied the relationship between the ETF and LETF implied volatility surfaces when the underlying ETF is modeled by a general class of local-stochastic volatility models.
	
	Various studies including \citet{ait_sahalia_et_al}, \citet{cont_da_fonseca}, \citet{fengler_haerdle_villa}, \citet{fengler_haerdle_mammen}  apply non- and semiparametric approaches to model implied volatility surfaces (IVS). Some important issues of IVS estimation and modelling include choice between parametric and non-parametric methods, model selection, out-of-sample forecasting.
	
	\citet{leung_sircar} introduced the so-called "moneyness scaling" technique which links implied volatilities (IV) between ETF and LETF in the way that the discrepancy between the implied volatility "smile" pattern is removed. The question arises whether the moneyness scaling method indeed removes discrepancies consistently in time. To answer this question, we need to verify whether IV deviations are significant from the statistical point of view. 
	
	In this study, we use an econometric approach to study the issue of errors describing the difference between leveraged and unleveraged volatility smiles. Unlike \citet{lee_wang}, who derive an asymptotic error estimate directly for implied volatilities, we investigate the indirect approach of \citet{leung_sircar} further, invoking stochastic volatility framework. This approach allows to apply the moneyness scaling technique under more general assumptions. 
	
	We consider the statistical properties of the moneyness scaling transformation which adjusts the moneyness coordinate of the implied volatility smile in an attempt to remove the discrepancy between the IV smiles for levered and unlevered ETF options. We construct bootstrap uniform confidence bands which allow for more flexible error analysis. The results indicate that the implied volatility smiles are statistically different, even after moneyness scaling has been performed.
	
	Furthermore, we develop a trading strategy based on a dynamic semiparametric factor model. This strategy utilizes the dynamic structure of implied volatility surface allowing out-of-sample forecasting and information on unleveraged ETF options to construct theoretical one-step-ahead implied volatility surfaces. This strategy exploits statistical discrepancies on (L)ETF markets and falls within the class of model-driven statistical arbitrage described in \citet{avellaneda_lee}.

	\section{Confidence analysis of moneyness scaling}
	
	\subsection{Moneyness scaling}
	We begin by introducing basic results on (L)ETF options and moneyness scaling. The dynamics of the underlying asset is assumed to follow a stochastic process under a risk-neutral measure $\QQ$:
	\begin{equation}
	\frac{dS_t}{S_t} = (r-\delta)dt + \sigma_t dW^{\QQ}_t,
	\label{underl}
	\end{equation}
	where $r$ is the risk-free interest rate, $\delta$ the dividend yield, $(\sigma_t)_{t \geq 0}$ is some stochastic volatility process.
	
	The moneyness scaling technique proposed by \citet{leung_sircar} proposes a coordinate transformation for the LETF option implied volatility and potentially reflects the increase of risk in the underlying index. 
	
\begin{figure}[H]
		\begin{center}
			\minipage{0.4\textwidth}
			\includegraphics[width=\linewidth]{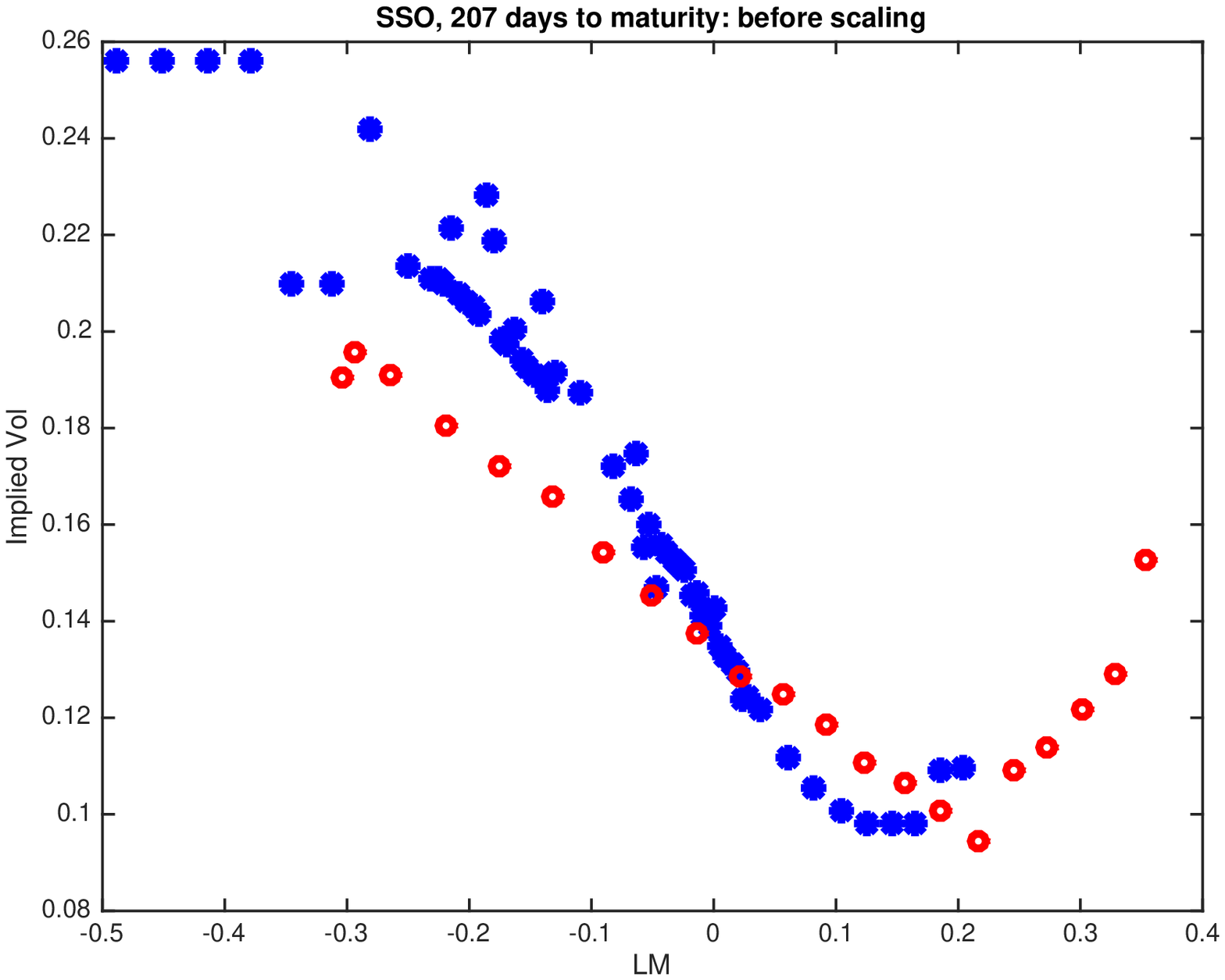}
			\endminipage
			\minipage{0.4\textwidth}
			\includegraphics[width=\linewidth]{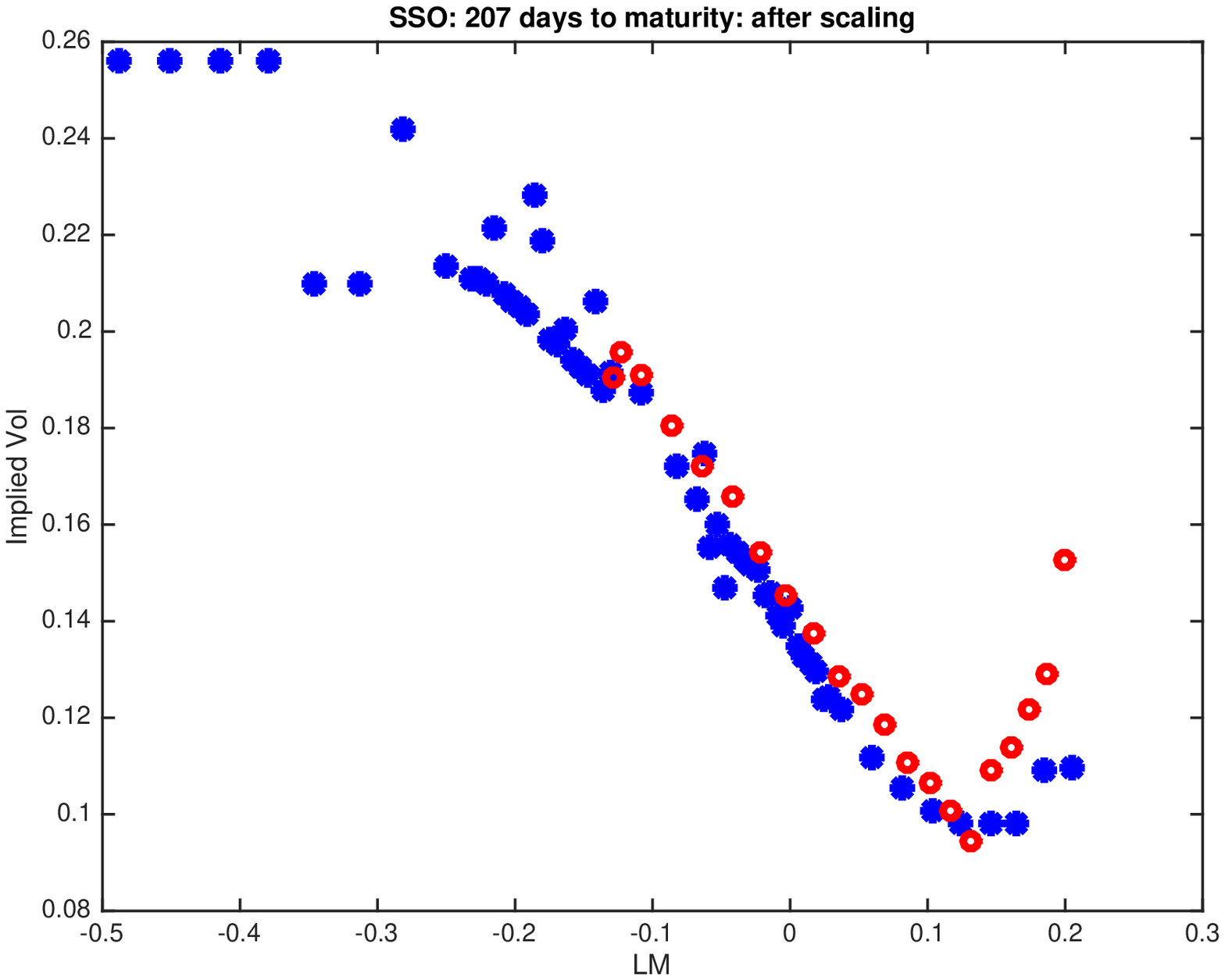}
			\endminipage\\
			\minipage{0.4\textwidth}
			\includegraphics[width=\linewidth]{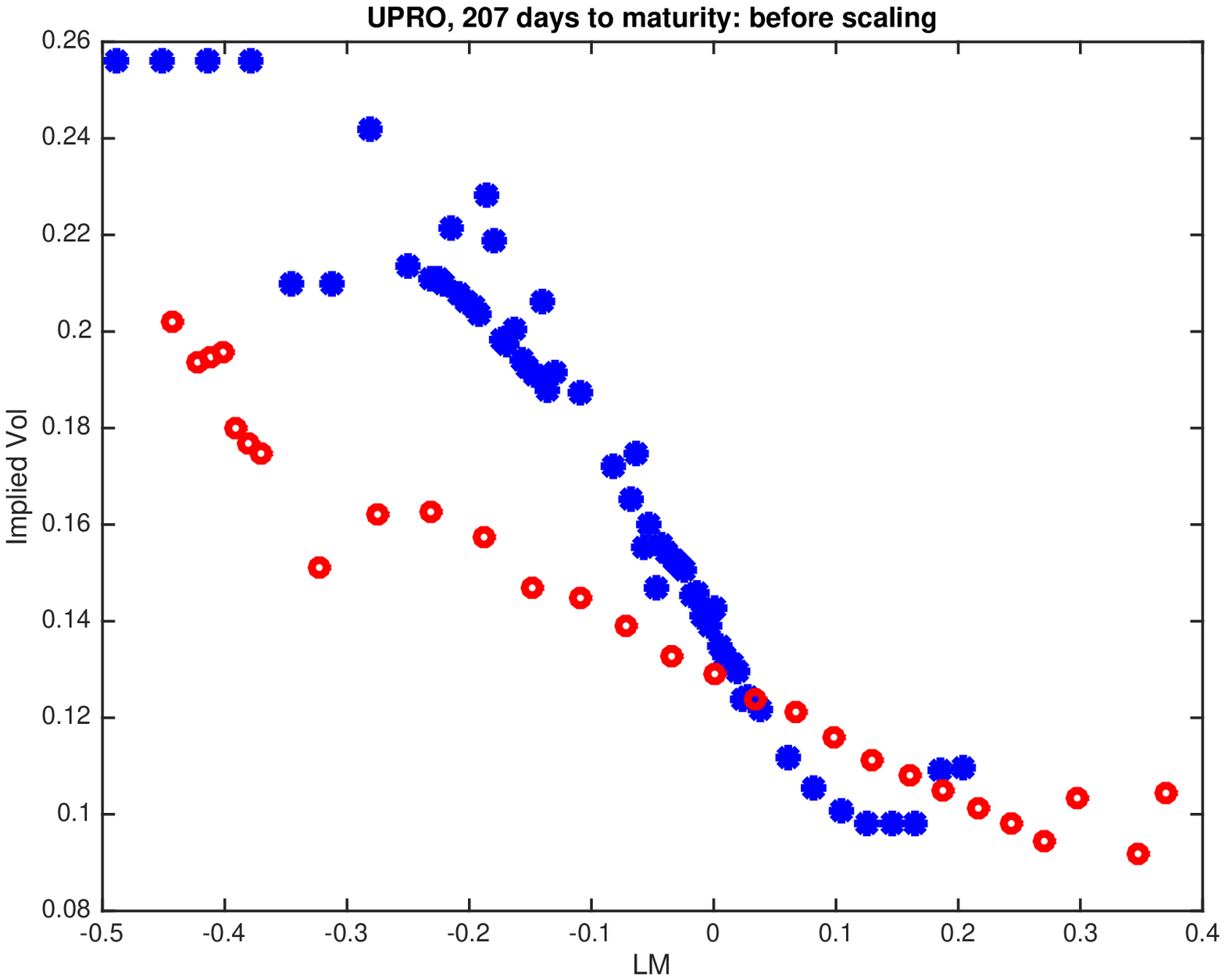}
			\endminipage
			\minipage{0.4\textwidth}
			\includegraphics[width=\linewidth]{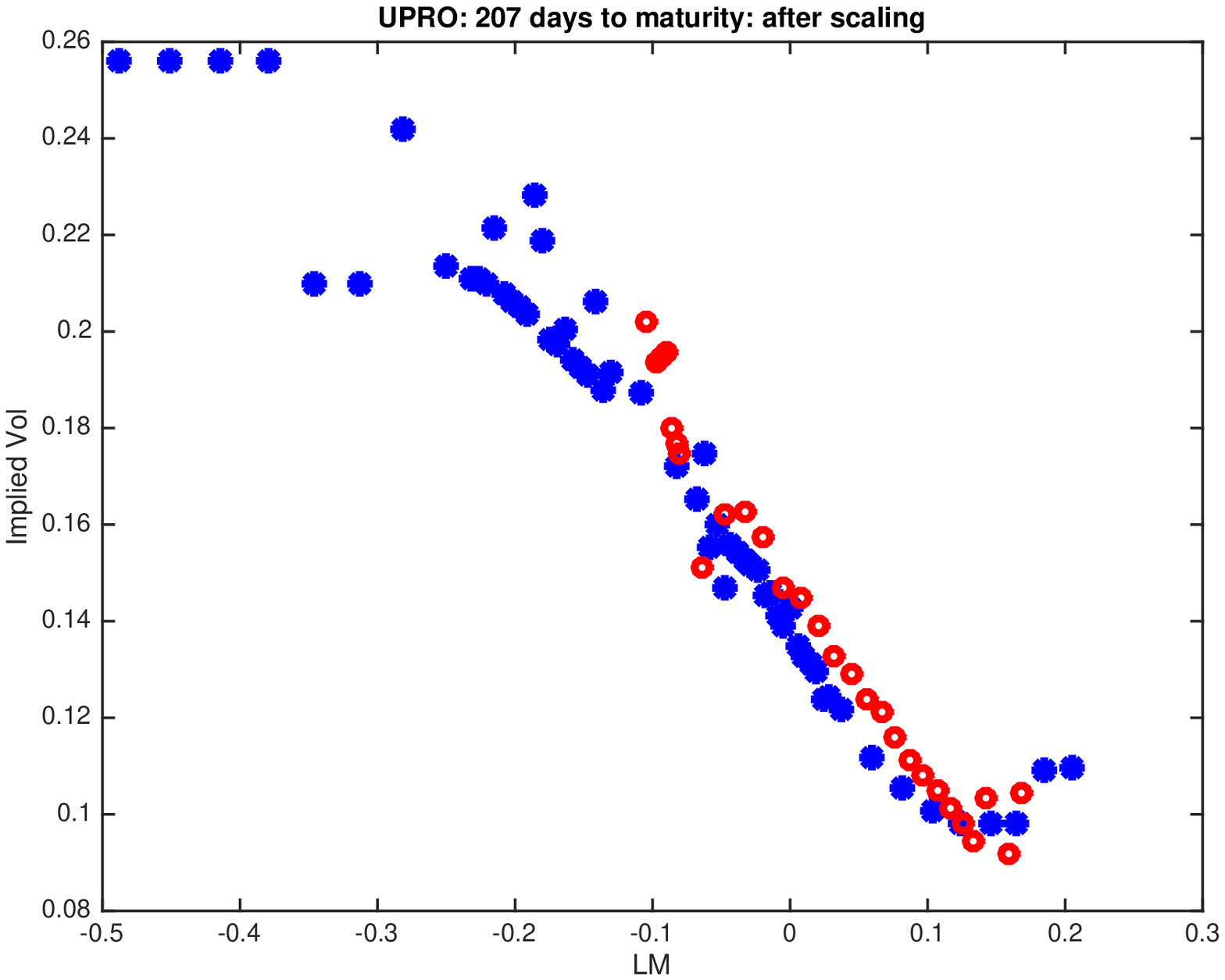}
			\endminipage\\
			\minipage{0.4\textwidth}
			\includegraphics[width=\linewidth]{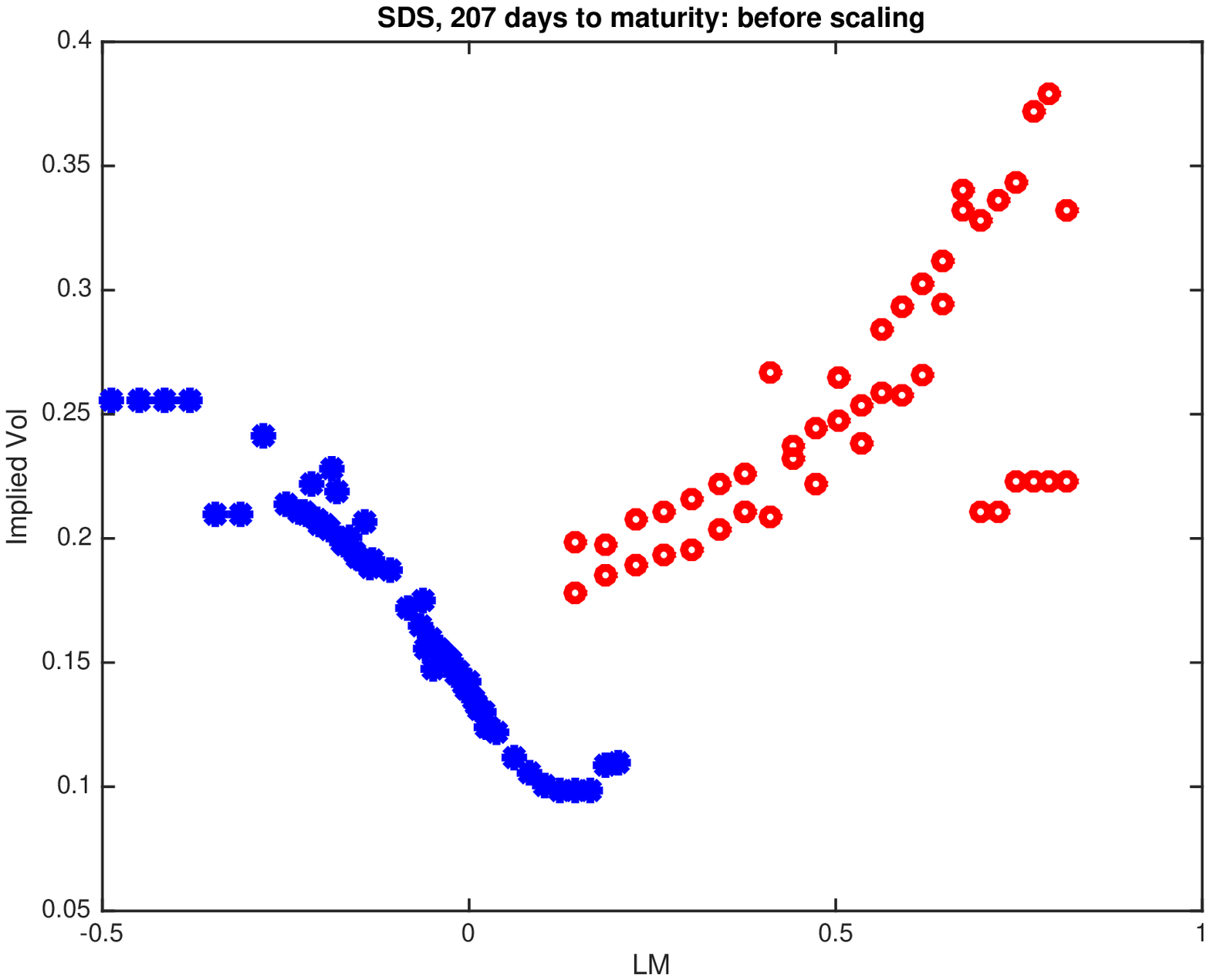}
			\endminipage
			\minipage{0.4\textwidth}
			\includegraphics[width=\linewidth]{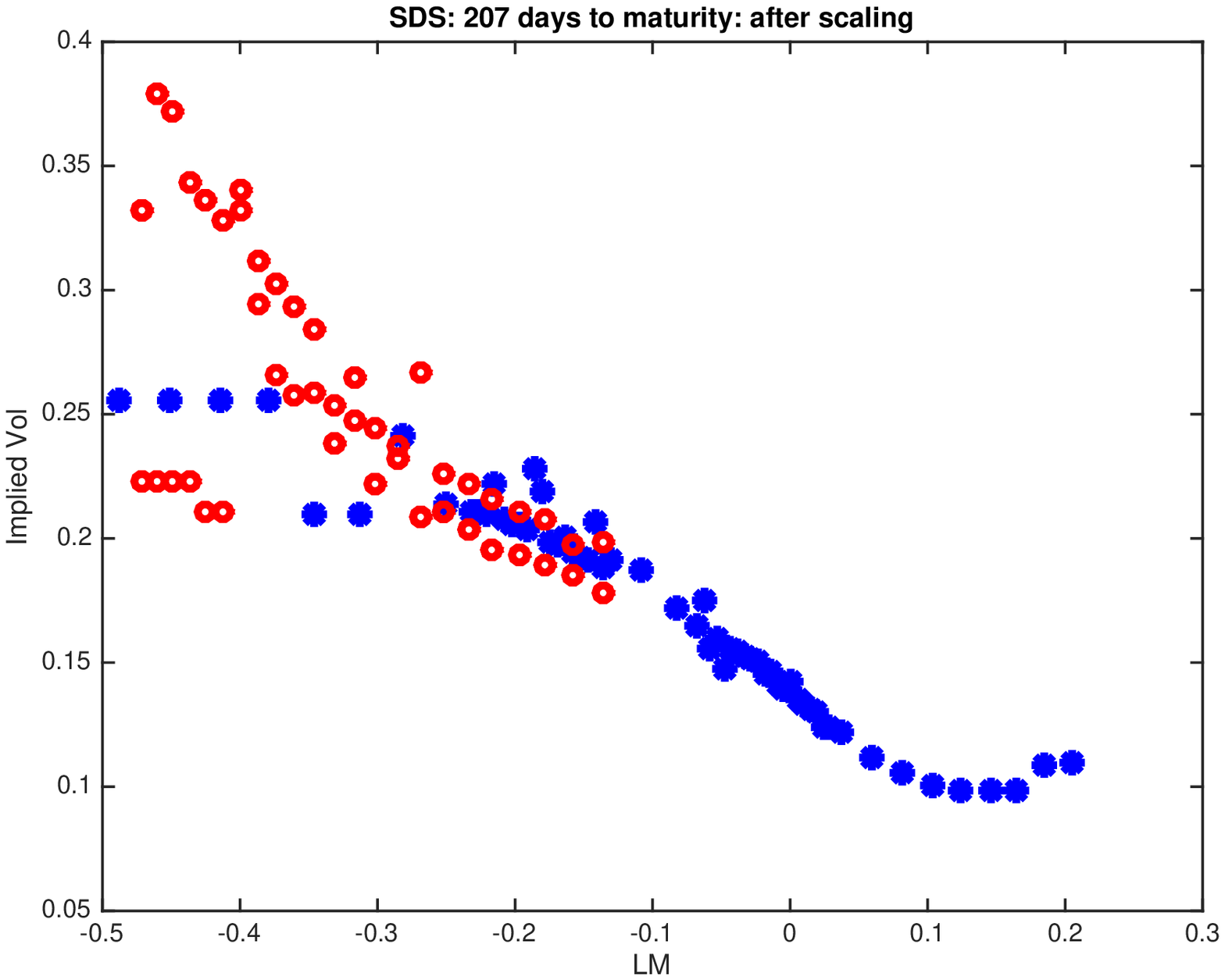}
			\endminipage\\
			\minipage{0.4\textwidth}
			\includegraphics[width=\linewidth]{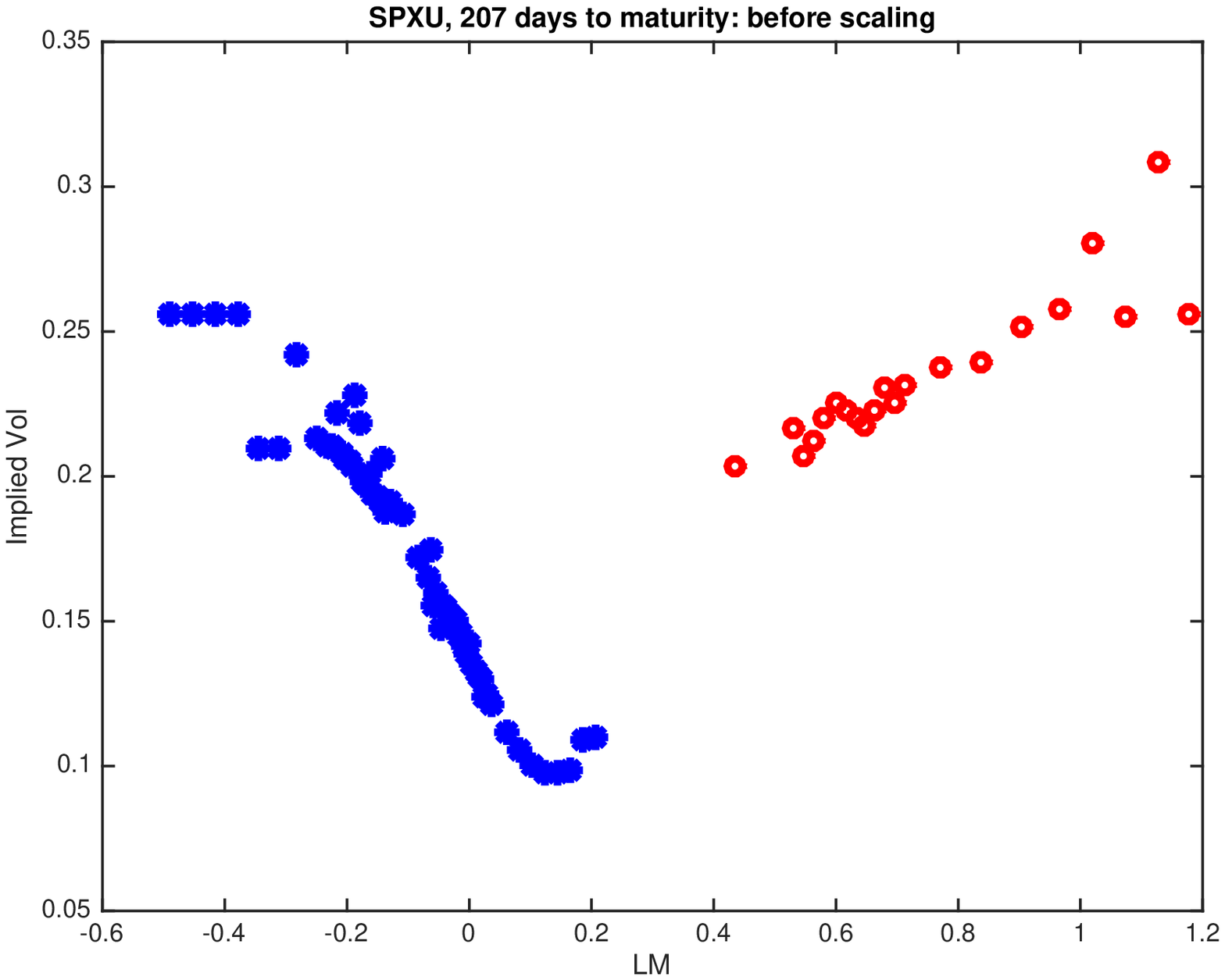}
			\endminipage
			\minipage{0.4\textwidth}
			\includegraphics[width=\linewidth]{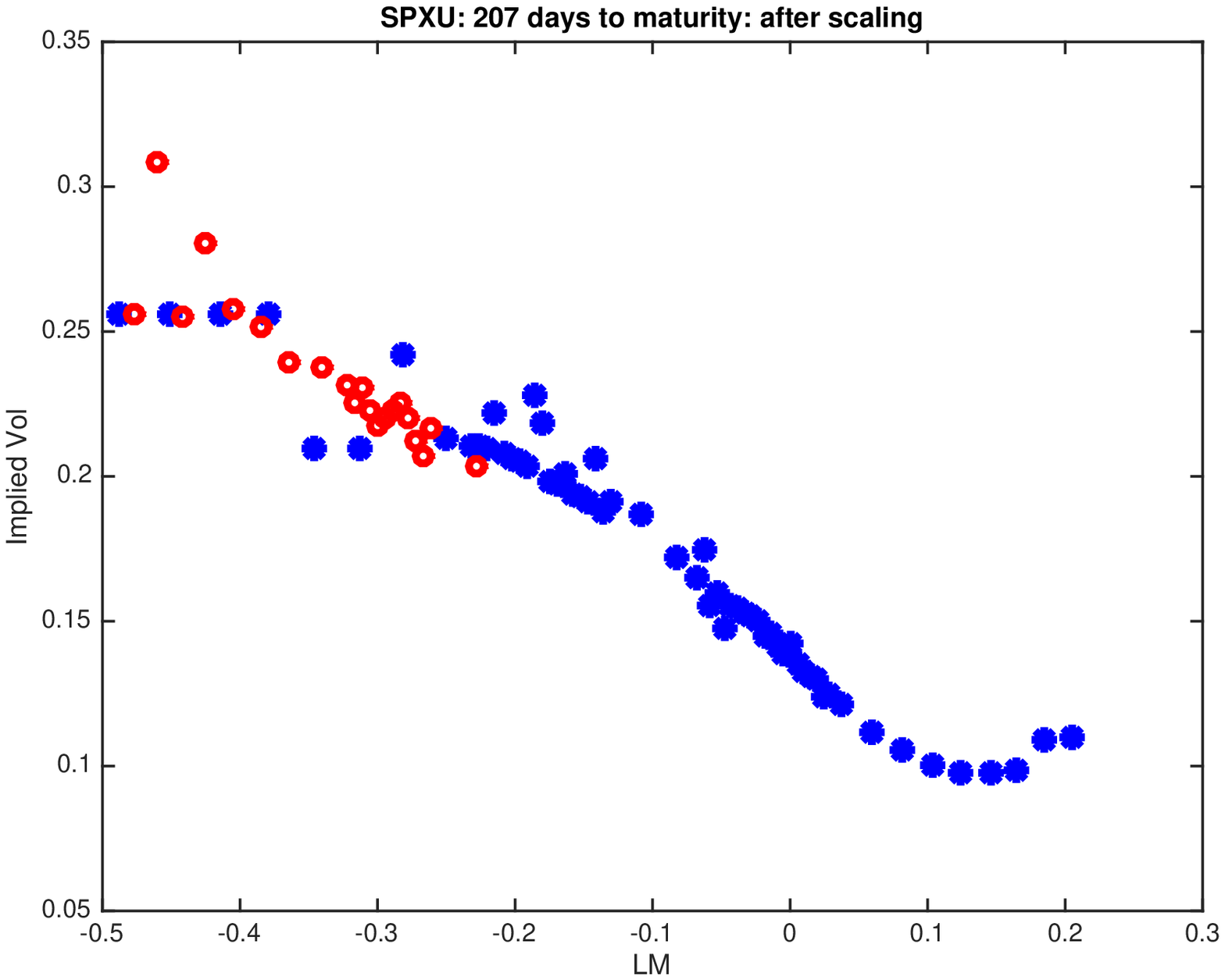}
			\endminipage
		\end{center}
		\caption{SPY (blue) and LETFs (red) implied volatilities before (left column) and after scaling (right column) on June 23, 2015 with 207 days to maturity, plotted against log-moneyness } 
		\label{letf_sc_unsc}
	\end{figure}

	Figure \ref{letf_sc_unsc} compares empirical implied volatilities for SSO, UPRO, SDS, SPXU before moneyness scaling has been applied and afterwards. In this example, the log-moneyness $LM \stackrel{\operatorname{def}}{=} \log (K/L_t)$ is used, where $K$ is the strike of the LETF option and $L_t$ the LETF price at time $t$. After re-scaling, there are still visible discrepancies between the implied volatilities for the  SPY ETF and its leveraged counterparts. The moneyness scaling procedure yields a more coherent picture when the LETF and ETF implied volatilities overlap visually better.
	
	Based on the assumption that the distribution of the terminal price of the $\beta$-LETF depends on the leverage ratio $\beta$, the moneyness scaling formula includes an expectation of the $\beta$-LETF log-moneyness conditional on the terminal value of the unleveraged counterpart. For the LETF log-moneyness $LM^{(\beta)}$ (consider ETFs as LETFs with $\beta = 1$) the result linking the log-moneyness coordinates $LM^{(\beta)}$ and $LM^{(1)}$ of the leveraged and unleveraged ETF is written as follows:
	\begin{equation}
	LM^{(\beta)} = \beta LM^{(1)} - \{r(\beta - 1) + c^{*}\} T - \frac{\beta(\beta - 1)}{2} {\E}^{\QQ}\left\{ \int_{0}^{T}\sigma_t^2 \text{d} t \left| \log \left(\frac{S_T}{S_0}\right) = LM^{(1)} \right.\right\}, 
	\label{monsc_unlev}
	\end{equation}
	where $T$ is the time to maturity/expiration (TTM), $c^{*} = c + \delta$ is the LETF expense ratio $c$ corrected for dividend yield $\delta$. The expense ratio $c$ is expressed in percent and approximates an annual fee charged by the ETF from the shareholders to cover the fund's operating expenses.
	
	More generally, for two LETFs with different leverage ratios $\beta_1$, $\beta_2$ the expression (\ref{monsc_unlev}) takes the form:
	\begin{multline}
	LM^{(\beta_1)} = \frac{\beta_1}{\beta_2} LM^{(\beta_2)} + \left[ \left\{\frac{\beta_1}{\beta_2} (\beta_2-1) - (\beta_1-1) \right\} r + \frac{\beta_1}{\beta_2}c_2^{*} - c_1^{*} \right]T + \\ \frac{\beta_1(\beta_2-1) - \beta_1(\beta_1-1)}{2}  {\E}^{\QQ}\left\{ \int_{0}^{T}\sigma_t^2 \text{d} t \left| \log \left(\frac{S_T}{S_0}\right) = LM^{(1)} \right.\right\},
	\label{monsc}
	\end{multline}
	with $c^{*}_k = c_k + \delta_k$, $k = 1,2$.
	
	It is worth mentioning that declared leverage ratios are not always the same as the empirical ones. \citet{leung_santoli} introduce a novel method to estimate the empirical leverage ratio realized by ETFs. Particularly for longer holding periods, return discrepancies between ETFs and LETFs increase. This can cause, e.g. for both long and short ETFs to have negative cumulative returns over a longer horizon. In this study we use declared leverage ratios. This is motivated by the use of short time horizons in the empirical study of LETF option portfolios' returns in Section \ref{empirical_section}.

	\subsection{Confidence bands} 
	\label{confb}
	
	\citet{cont_da_fonseca}, \citet{fengler_haerdle_mammen}, \citet{park_et_al} studied the implied volatility as a random process in time, so that the data generating process includes some non-parametric function $m$:
	\begin{equation}
	Y_t = m(X_t) + \varepsilon_t, \hspace{3mm} t = 1,\ldots,T,
	\label{ivnonpar}
	\end{equation}
	or can be driven by a latent factor process $\mathcal{Z}_t$:
	\begin{equation}
	Y_t = \mathcal{Z}_t^{\top} m(X_t) + \varepsilon_t, \hspace{3mm} t = 1,\ldots,T,
	\label{ivnonpardsfm}
	\end{equation}
	where $Y_t$ stands for an implied volatility process, the covariates $X_t$ can be one- or multi-dimensional, including, for instance, moneyness and time-to-maturity.
	
	The statistical properties of the estimators $\widehat{m}(X_t)$ and $\widehat{\mathcal{Z}}^{\top}_t \widehat{m}(X_t)$ for the models (\ref{ivnonpar}) and (\ref{ivnonpardsfm}) have been outlined, respectively, in, e.g., \citet{haerdle_book_reg}, \citet{ruppert_wand} and \citet{park_et_al}. To study the consistency of the implied volatility difference between the ETF and the moneyness-scaled LETF case, one needs to consider statistical differences of the corresponding estimators. Confidence band analysis may provide an insight into the matter. An important issue for smooth confidence bands for functions is the correct probability of covering the "true" curve.
	
	The approach of \citet{haerdle_ritov_wang} proposes a uniform bootstrap bands construction for a wide class of non-parametric $M$ and $L$-estimates. It is logical to use a robust $M$-type smoother for the estimation of (\ref{ivnonpar}) for implied volatility, as IV data often suffer from outliers. The procedure runs as follows: considering the sample $\{X_t,Y_t\}_{t=1}^T$, where $Y_t$ denotes the IV process, $X_t$ is taken to be one-dimensional and includes the log-moneyness covariate $LM^{(\beta)}$, do the following:
	
	\begin{enumerate}
		\item
		compute the estimate $\widehat{m}_h(X_t)$ by a local linear $M$-smoothing procedure (see Appendix \ref{app2}) with some kernel function and bandwidth $h$ chosen by, e.g., cross-validation, and obtain residuals $\widehat{\varepsilon}_t \stackrel{\operatorname{def}}{=} Y_t - \widehat{m}_h(X_t)$, 
		\item
		do bootstrap resampling from $\widehat{\varepsilon}_t$: for each $t = 1,\ldots,T$, generate random variables $\varepsilon_{t,b}^{*} \sim \widehat{F}_{\varepsilon|X_t}(z)$ for $b = 1, \ldots, B$ according to the conditional edf
		\begin{equation}
		\widehat{F}_{\varepsilon|x}(z) \stackrel{\operatorname{def}}{=} \frac{\sum_{t=1}^{T} K_h (x - X_t) \mathbf{1}\{\widehat{\varepsilon}_t \leq z\}}{\sum_{t=1}^{T} K_h (x - X_t)},
		\label{bootstrap_edf}
		\end{equation}
		which is further centered as shown in \citet{haerdle_ritov_wang}. Then construct the bootstrap sample $Y_{t,b}^{*}$ as follows:
		\begin{equation}
		Y_{t,b}^{*} = \widehat{m}_g(X_t) + \varepsilon_{t,b}^{*},
		\label{ybootresamp}
		\end{equation}
		with an "oversmoothing" bandwidth $g \gg h$ such as $g = \mathcal{O}(T^{-1/9})$ to allow for bias correction,
		\item
		for each bootstrap sample $\{X_t,Y_{t,b}^{*}\}_{t=1}^T$ compute $\widehat{m}_{h,g}^{*}$ using the bandwidth $h$ and construct the random variable
		\begin{equation}
		d_{b} \stackrel{\operatorname{def}}{=} \underset{x \in J}{\sup} \left[ \frac{|\widehat{m}_{h,g}^{*}(x) - \widehat{m}_{g}(x)| \sqrt{\widehat{f}_X (x)} \widehat{f}_{\varepsilon|X_t}(\varepsilon_t^{*})}{\sqrt{\widehat{\E}_{Y|x}\{\psi^2(\varepsilon_t^{*})\}}} \right],
		\end{equation}
		where $J$ is a finite compact support set of $\widehat{f}_X$ and $\psi(u) = \rho '(\cdot)$ as described in Appendix \ref{app2}; the conditional expectation $\widehat{\E}_{Y|x}(\cdot)$ is defined with respect to the edf
		\begin{equation}
		\widehat{F}_{Y|x}(z) \stackrel{\operatorname{def}}{=} \frac{\sum_{t=1}^{T} K_h (x - X_t) \mathbf{1}\{Y_t \leq z\}}{\sum_{t=1}^{T} K_h (x - X_t)},
		\label{expec_edf}
		\end{equation}
		where $\widehat{f}_{\varepsilon|X_t}(\cdot)$ and $\widehat{f}_X (x)$ are consistent estimators of conditional density corresponding to (\ref{bootstrap_edf}) and the density $f_X (x)$, respectively; for more details, see \citet{haerdle_ritov_wang},
		\item
		calculate the $1-\alpha$ quantile $d^*_{\alpha}$ of $d_1,\ldots,d_B$,
		\item
		construct the bootstrap uniform confidence band centered around $\widehat{m}_{h}(x)$:
		\begin{equation}
		\widehat{m}_{h}(x) \pm \left[ \frac{ \sqrt{\widehat{\E}_{Y|x}\{\psi^2(\varepsilon_t^{*})\}} d^*_{\alpha}}{\sqrt{\widehat{f}_X (x)} \widehat{f}_{\varepsilon|X_t}(\varepsilon_t^{*})} \right]. 
		\end{equation}
	\end{enumerate}
	
	Such an approach utilizes bootstrap confidence bands while the distribution of the original data is "mimicked" via a pre-specified random mechanism achieving both uniformity and better coverage. Additionally, it performs better than asymptotic confidence bands which generally tend to underestimate the true coverage probability, see \citet{hall_horowitz}. Compared to a Bonferroni approach, bootstrap uniform confidence bands would be less conservative and make use of the substantial positive correlation of the curve estimates at nearby points, see \citet{haerdle_book_reg}.

	\section{Moneyness scaling under Heston stochastic volatility} 
	
	\subsection{An analytical approach}
	
	In the study of moneyness scaling, one needs to estimate the following conditional expectation:
	\begin{equation}
	{\E}^{\QQ}\left\{ \int_{0}^{T}\sigma_t^2 dt \left| \log \left(\frac{S_T}{S_0}\right) = LM^{(1)} \right.\right\}.
	\label{monscex}
	\end{equation}
	Taking $\sigma_t = \sigma$ constant, one obtains $\sigma^2 T$. As empirical evidence shows, constant volatility is not a plausible assumption, therefore one needs to determine the measure $\QQ$ for the case of random volatility under a model which allows random dynamics of $\sigma_t$. Second, one needs to estimate the integrated variance
	\begin{equation}
	\int_{0}^{T}\sigma_t^2 dt,
	\label{intvar}
	\end{equation}
	
	Stochastic volatility presents a viable alternative to the constant case. One could choose among different specifications of stochastic volatility models. Popular special cases include specifications of \citet{heston}, \citet{hull_white}, \citet{schoebel_zhu}. An example of a more general stochastic volatility system is given in \citet{leung_sircar}. Simpler models tend to generate semi-closed-form solutions for return distributions. For instance, a solution for the Heston model by \citet{heston} was proposed by \citet{dragulescu_yakovenko}.
	
	We use the Heston model to compute the quantity in (\ref{monscex}). As noticed in \citet{leung_santoli}, this approach allows for tractability and efficient numerical pricing of options on LETFs. Stochastic volatility framework also allows to better assess volatility decay, i.e. value erosion due to the increase of the realized variance with the holding horizon.
	
	The Heston model with risk-neutral dynamics under a risk-neutral measure $\QQ$ and zero volatility risk premium is described by a two-dimensional system of stochastic differential equations
	\begin{align}
	dS_t &= (r-c-0.5)S_t dt + \sqrt{V_t} S_t dW_{S,t}^{\QQ},\label{sz1}\\
	dV_t &= \kappa(\theta - V_t)dt + \sigma dW_{V,t}^{\QQ},\label{sz2}
	\end{align}
	where we have put $V_t = \sigma_t^2$; $r-c$ are costs of carry on $S_t$, $\theta$ is the long-run variance level, $\kappa$ is the rate of reversion to $\theta$, $\sigma$ is the "volatility of the volatility" parameter which determines the variance of $V_t$; $W_{S,t}$, $W_{V,t}$ are correlated with parameter $\rho$. The tails of the Heston-implied densities for log-returns $x_t = \log(S_t/S_{t-1})$ are exponential and heavier than those of the normal distribution with the dispersion parameter equal to the long-term variance $\theta$, see, i.e. \citet{cizek_et_al}.
	
	An analytical solution to (\ref{monscex}) requires knowledge of the conditional distribution of the integrated variance (\ref{intvar}) given the logarithm of terminal stock price $\log (S_T)$. If we define:
	\begin{align*}
	\widetilde{V} &\stackrel{\operatorname{def}}{=} \int_{0}^{T}V_t dt,\\
	X_T &\stackrel{\operatorname{def}}{=} \log(S_T),
	\end{align*}
	then we can write
	\begin{align}
	{\E}^{\QQ}\left\{ \int_{0}^{T}V_t dt \left| \log \left(\frac{S_T}{S_0}\right) = LM^{(1)} \right.\right\} &= {\E}^{\QQ}\left\{ \widetilde{V} \left| X_T = \widetilde{LM}^{(1)} \right.\right\} \nonumber \\
	&= \int_{0}^{\infty} f_{\widetilde{V}|X_T}\left(\widetilde{v}\left|x_T=\widetilde{LM}^{(1)}\right.\right) d\widetilde{v}, \label{expconddens}
	\end{align}
	where $f_{\widetilde{V}|X_T}(\widetilde{v}|x_T)$ is the conditional density of $\widetilde{V}$ given $X_T$ under the measure $\QQ$ and $\widetilde{LM}^{(1)} = \log(S_0) + LM^{(1)}$.
	
	Unfortunately, $f_{\widetilde{V}|X_T}(\widetilde{v}|x_T)$ does not assume a simple form and is ultimately expressed in terms of Fourier transforms of characteristic functions of these quantities. Technical details are given in Appendix \ref{hestproof}. As follows from the details, four improper integrals have to be estimated. Numeric integration methods can be used to approximate (\ref{monscex}).
	
	Additional complexity arises from the necessity to evaluate a modified Bessel function of the first kind which takes a complex argument. Numerical approximation methods for such evaluations such as the trapezoidal rule are outlined in \citet{broadie_kaya}.

	Considering the complexity of the density estimation, we consider a Monte-Carlo approach to evaluate (\ref{monscex}). This method is feasible and straightforward from a practical point of view.
	
	\subsection{A Monte-Carlo approach}
	Alternatively, the conditional expectation in (\ref{monscex}) can be computed using Monte-Carlo simulations. The simulations are performed using the Heston model and the calibrated parameters obtained minimizing the squared difference between theoretical Heston prices $C^{\Theta}(K,\tau)$ obtained from the model and observed market prices $C^{M}(K,\tau)$, 
	\begin{equation}
	\underset{\Theta \in \mathbb{R}^5}{\operatorname{min}} \sum_{i=1}^N \left(  C_i^{\Theta}(K_i,\tau_i) - C_i^{M}(K_i,\tau_i) \right) 
	\end{equation}
	where $\Theta \stackrel{\operatorname{def}}{=} (\kappa, \theta, \sigma, v_0, \rho)$ Heston parameters, $N$ number of options used for calibration, $K$ strikes and $\tau$ times-to-maturity. Theoretical prices $C^{\Theta}(K,\tau)$ are obtained via numeric integration of the Heston characteristic function.
	
	The Monte-Carlo algorithm is motivated by \citet{van_der_stoep_et_al} and can be formulated as follows:
	\begin{enumerate}
		\item
		Generate $N$ pairs of observations $(s_i,v_i)$, $i = 1,\ldots,N$.
		\item
		Order the realizations $s_i$: $s_1 \leq x_2 \leq \ldots \leq s_N$.
		\item
		Determine the boundaries of $M$ bins $(l_k,l_{k+1}]$, $k = 1,\ldots,M$ on an equidistant grid of values $S^* \stackrel{\operatorname{def}}{=} S_0 e^{LM^{(1)}}$
		\item
		For the $k$th bin approximate the conditional expectation (\ref{monscex}) by 
		\begin{equation}
		{\E}^{\QQ} \left(  \int_{0}^{T}\sigma_t^2 \text{d} t \left| S_T \in (l_k,l_{k+1}] \right. \right) \approx \frac{h}{N Q(k)} \sum_{i = 1}^H \sum_{j \in \mathcal{J}_k} V_{ij},
		\end{equation}
		where $h$ is the discretization step for $V_t$, $\mathcal{J}_k$ the set of numbers $j$, for which the observations $S_T$ are in the $k$th bin and $Q(k)$ is the probability of $S_T$ being in the $k$th bin. 
	\end{enumerate}
	
\begin{figure}[H]
		\begin{center}
			\minipage{0.7\textwidth}
			\includegraphics[width=\linewidth]{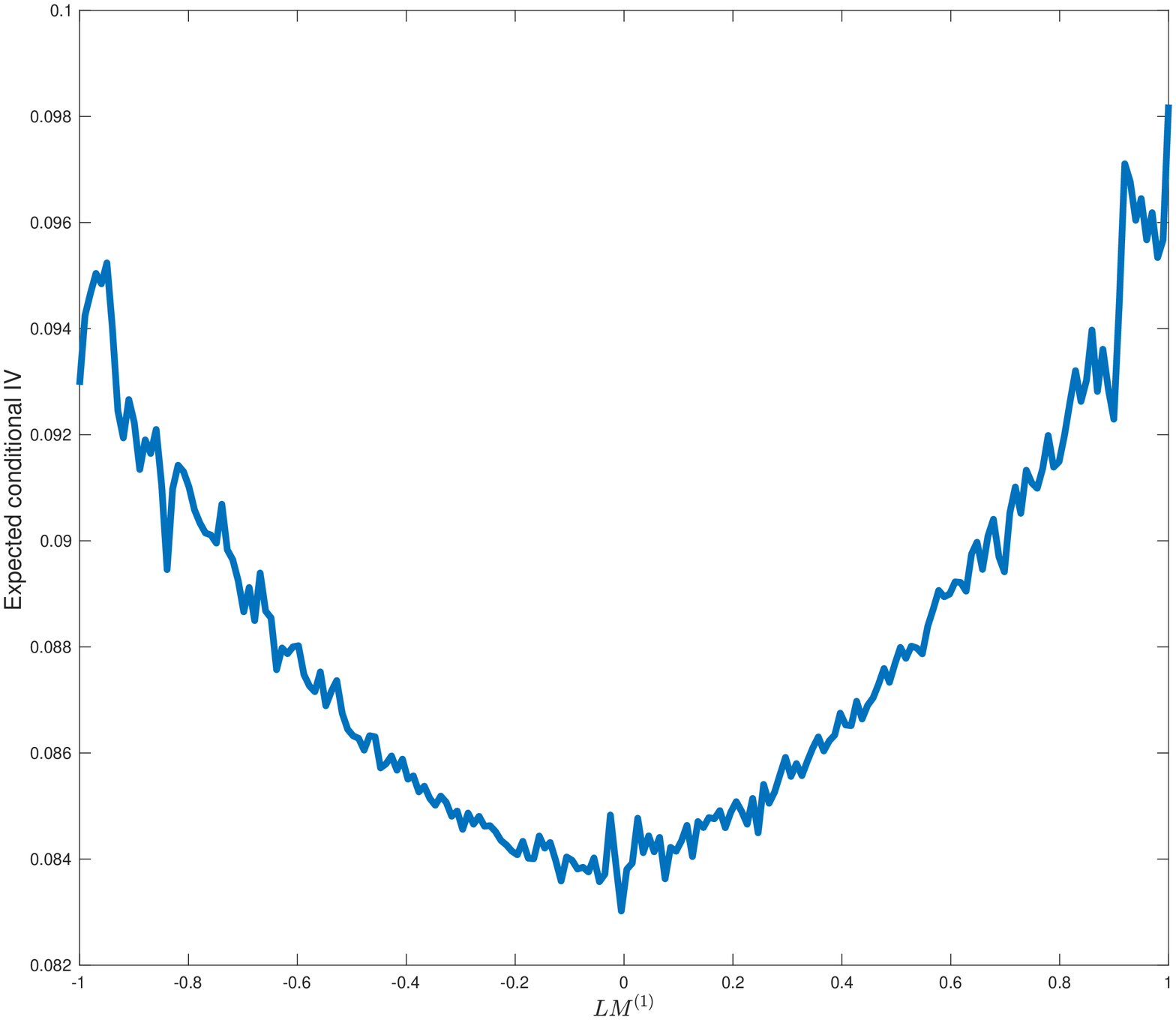}
			\endminipage\\
			\minipage{0.7\textwidth}
			\includegraphics[width=\linewidth]{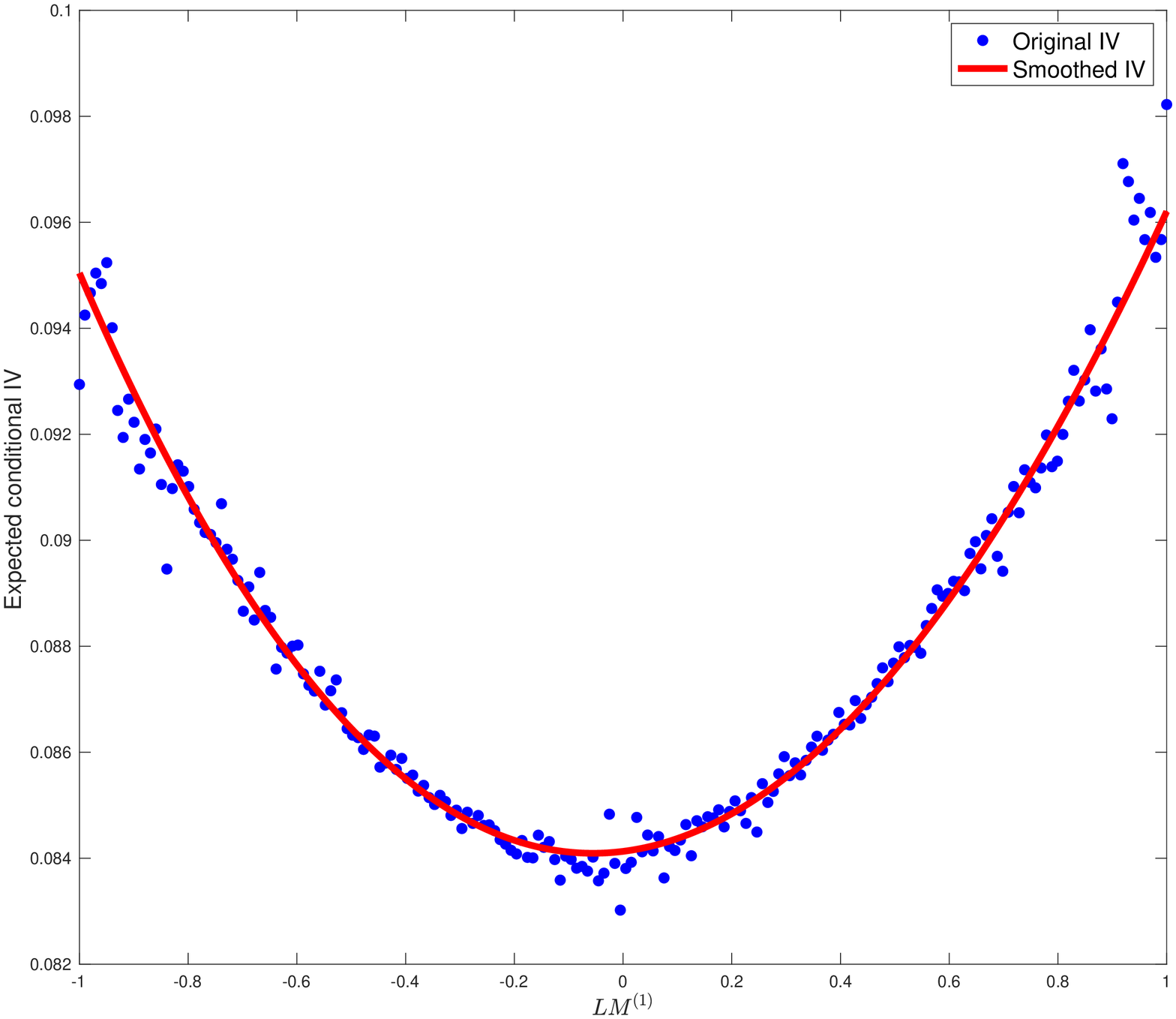}
			\endminipage
		\end{center}
		\caption{Upper panel: estimated value of ${\E}^{\QQ}( \int_{0}^{T}\sigma_t^2 \text{d} t | \log (S_T/S_0) = LM^{(1)} )$; lower panel: smoothed estimate} 
		\label{expFun}
	\end{figure}

	The results of the simulation are presented in Figure \ref{expFun}. Polynomial smoothing is applied to produce the smoothed version of SCO LETF realized variance.  The generated expected realized variance has the form of a "smile" which confirms the intuition behind using average square implied volatility in the case of constant-volatility moneyness scaling approach.
	
	We use the Monte-Carlo approach given Euler discretization scheme for the empirical application in later sections given its tractability and theoretical justification. Both methods, analytical and Monte-Carlo, introduce errors into the calculation of (\ref{monscex}). For the analytical method, discretization and truncation errors appear when the integral is estimated at discrete points and is truncated to be approximated as a finite sum. If the trapezoidal rule is used to approximate the integrals in the analytical method, then the discretization error is of order $\mathcal{O}(M^{-2})$ where $M-1$ is the number of discrete intervals. However, as the integral dimensionality $d$ increases, the discretization error order increases to $\mathcal{O}(M^{-2/d})$ ("curse of dimensionality"). 
	
	For the current example, the discretization error order for (\ref{monscex}) becomes $\mathcal{O}(M^{-1/2})$, which matches the convergence order of Monte-Carlo discretization bias. Additionally, analytical approximation of (\ref{monscex}) effects a truncation error which is potentially significant due to the oscillatory nature of the integrand. Monte-Carlo approach inherently induces a statistical error of order $\mathcal{O}(N^{-1/2})$ which can be made sufficiently small by taking a large number $N$ of samples. As noted in \citet{higham_mao}, \citet{lord_et_al}, a simple discretization scheme such as the Euler scheme converges to the true process under certain conditions on the discretization size. This was shown to be true for the Heston in particular by \cite{higham_mao}.

	\section{Dynamic semiparametric factor model}
	
	\subsection{Model description}
	A generalized version of the model in (\ref{ivnonpar}) represented by (\ref{ivnonpardsfm}) assumes the implied volatility $Y_t$ to be a stochastic process driven by a latent stochastic factor process $\mathcal{Z}_t$ contaminated by noise $\varepsilon_t$. To be more specific, define $\mathcal{J} \stackrel{\operatorname{def}}{=} [{\kappa}_{min},{\kappa}_{max}] \times [{\tau}_{min},{\tau}_{max}]$, $Y_{t,j}$ implied volatility, $t = 1,\ldots,T$ time index, $j = 1,\ldots,J_t$ option intraday numbering on day $t$, $X_{t,j} \stackrel{\operatorname{def}}{=} ({\kappa}_{t,j}, \tau_{t,j})^{\top}$, ${\kappa}_{t,j}$, ${\tau}_{t,j}$ are, respectively, a moneyness measure (log-, forward, etc.) and time-to-maturity at time point $t$ for option $j$. Then the \textit{dynamic semiparametric factor model (DSFM)} is defined as follows: assume
	\begin{equation}
	Y_{t,j} = \mathcal{Z}^{\top}_t m(X_{t,j}) + \varepsilon_{t,j},
	\label{dsfm}
	\end{equation}
	where $\mathcal{Z}_t = (1,Z_t^{\top})$, $Z_t = (Z_{t,1},\ldots,Z_{t,L})^{\top}$ unobservable $L$-dimensional stochastic process, $m = (m_0,\ldots,m_L)^{\top}$, real-valued functions; $m_l$, $l = 1,\ldots,L+1$ are defined on a subset of $\mathbb{R}^d$. One can estimate:
	\begin{align}
	\widehat{Y}_t &= \widehat{\mathcal{Z}}^{\top}_t \widehat{m}(X_t)\\
	&= \widehat{\mathcal{Z}}^{\top}_t \widehat{\mathcal{A}} \psi(X_{t}),
	\end{align}
	with $\psi (X_t) \stackrel{\operatorname{def}}{=} \left\{\psi_1(X_t),\ldots,\psi_K(X_t)\right\}^{\top}$ being a space basis such as a tensor B-spline basis, $\mathcal{A}$ is the $(L+1) \times K$ coefficient matrix. In this case $K$ denotes the number of tensor B-spline sites: let $(s_u)_{u=1}^{U}$, $(s_v)_{v=1}^{V}$ be the B-spline sites for moneyness and time-to-maturity coordinates, respectively, then $K = U \cdot V$. Given some spline orders $n_{\kappa}$ and $n_{\tau}$ for both coordinates and sets of knots $(t^{\kappa}_i)_{i=1}^{M}$, $(t^{\tau}_j)_{j=1}^{N}$, one of the Schoenberg-Whitney conditions requires that $U = M - n_{\kappa}$, $V = N - n_{\tau}$, see \citet{de_boor}. The usage of the parameter $K$ is roughly analogous to the bandwidth choice in \citet{fengler_haerdle_villa} and \citet{fengler_haerdle_mammen}; however the results of \citet{park_et_al} demonstrate insensitivity of DSFM estimation results to the choice of $K$, $n$.   
	
	The estimates for the IV surfaces $\widehat{m}_l$ are re-calculated on a fine 2-dimensional grid of tensor B-spline sites: the estimated coefficient matrix $\widehat{\mathcal{A}}$ is reshaped into a $U \times V \times L+1$ array of $L+1$ matrices $\widehat{A}$ of dimension $U \times V$. Factor functions $m_l$ can then be estimated as follows:
	
	\begin{equation}
	\widehat{m}_{l;i,j} = \sum_{i}^{U} \sum_{j}^{V} \widehat{A}_{l;i,j} \psi_{i,k_{\kappa}} (\kappa_i) \psi_{j,k_{\tau}} (\tau_j),
	\label{facfun}
	\end{equation}
	where $k_{\kappa}$, $k_{\tau}$ are knot sequences for the moneyness and time-to-maturity coordinates, respectively. 
	
	The estimated factor functions $\widehat{m}_{l}$ together with stochastic factor loadings $\widehat{\mathcal{Z}}_t$ are combined into the dynamic estimator of the implied volatility surface:
	
	\begin{equation}
	\widehat{IV}_{t;i,j} = \widehat{m}_{0;i,j} + \sum_{l=1}^{L} \widehat{\mathcal{Z}}_{l,t} \widehat{m}_{l;i,j},
	\label{dynivest}
	\end{equation}
	where $\widehat{\mathcal{Z}}_{l,t}$ can be modeled as a vector autoregressive process. It should be noted that $\widehat{m}_{l}$ and $\widehat{\mathcal{Z}}_{l,t}$ are not uniquely defined, so an orthonormalization procedure must be applied.
	
	An indication of possible mispricing of LETF options allows to test a trading strategy based on the comparison of the theoretical price obtained from the moneyness scaling correction as well as the application of the DSFM model and the market price. Such a strategy would mainly exploit the two essential elements of information from these two approaches. The first element is obtaining evidence of statistical discrepancies resulting from the mismatch between ETF and LETF IVs. The moneyness scaling approach allows to estimate LETF IV using richer unleveraged ETF data which also would make the DSFM IV estimator more consistent. The second element is implied volatility forecasting. The DSFM model allows to forecast a whole IV surface via the dynamics of stochastic factor loadings $\mathcal{Z}_t$.

	\subsection{Model estimation}
	
	The DSFM model is estimated numerically. The number of factors has to be chosen in advance. One should also notice that for $m_l$ to be chosen as eigenfunctions of the covariance operator $K(u,v) \stackrel{\operatorname{def}}{=} \Cov \{Y(u),Y(v)\}$ in an $L$-dimensional approximating linear space, where $Y$ is understood to be the random IV surface, they should be properly normalized, such that $\|m_l(\cdot)\| = 1$ and $\langle m_l,m_k \rangle = 0$ for $l \neq k$.
	
	The choice of $L$ can be based on the explained variance by factors:
	\begin{equation}
	EV(L) \stackrel{\operatorname{def}}{=} 1 - \frac{\sum_{t=1}^{T} \sum_{j=1}^{J_t} \left\{Y_{t,j} - \sum_{l=0}^{L} \widehat{Z}_{t,l}\widehat{m}_l(X_{t,j})\right\}^2 }{\sum_{t=1}^{T} \sum_{j=1}^{J_t} (Y_{t,j} - \overline{Y})^2}.
	\label{explvar}
	\end{equation}  
	
	The model's goodness-of-fit is evaluated by the root mean squared error (RMSE) criterion: 
	\begin{equation}
	RMSE \stackrel{\operatorname{def}}{=} \sqrt{\frac{1}{\sum_t J_t}\sum_{t=1}^{T} \sum_{j=1}^{J_t} \left\{Y_{t,j} - \sum_{l=0}^{L} \widehat{Z}_{t,l}\widehat{m}_l(X_{t,j})\right\}^2 }.
	\label{rmsecrit}
	\end{equation} 
	
	The prediction quality at time point $t+1$ is measured by the root mean squared prediction error (RMSPE) given by
	\begin{equation}
	RMSPE \stackrel{\operatorname{def}}{=} \sqrt{\frac{1}{J_{t+1}} \sum_{j=1}^{J_{t+1}} \left\{Y_{t+1,j} - \sum_{l=0}^{L} \widehat{Z}_{t+1,l}\widehat{m}_l(X_{t+1,j})\right\}^2 }.
	\label{rmspecrit}
	\end{equation}

	\section{Empirical application}
	\label{empirical_section}
	
	\subsection{Data description}
	
	For the purpose of an empirical application, we use data on SPY, SSO, UPRO and SDS (L)ETF call options in the period Nov 2014 - June 2015. The data summary statistics are outlined in Table \ref{sumstat} below. The data were taken from the Datastream database by Thomson Reuters. 
	
\begin{figure}[H]
		\begin{center}
			\minipage{0.80\textwidth}
			\includegraphics[width=\linewidth]{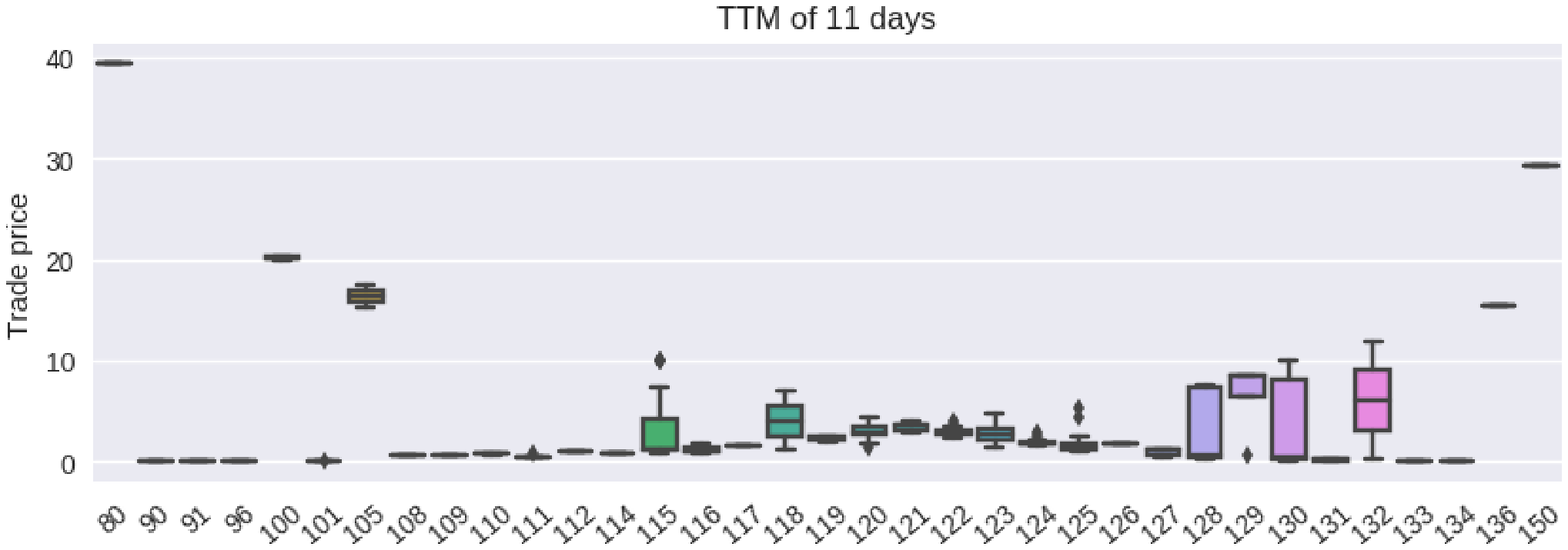}
			\endminipage\\
			\minipage{0.80\textwidth}
			\includegraphics[width=\linewidth]{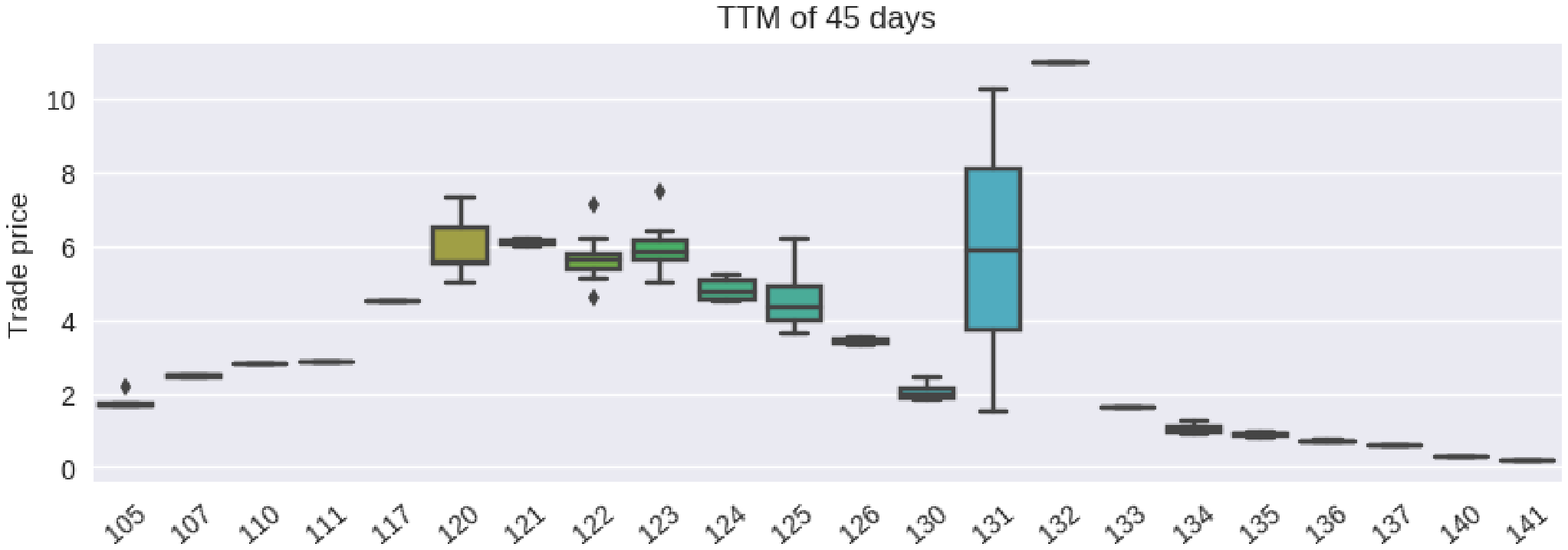}
			\endminipage\\
			\minipage{0.80\textwidth}
			\includegraphics[width=\linewidth]{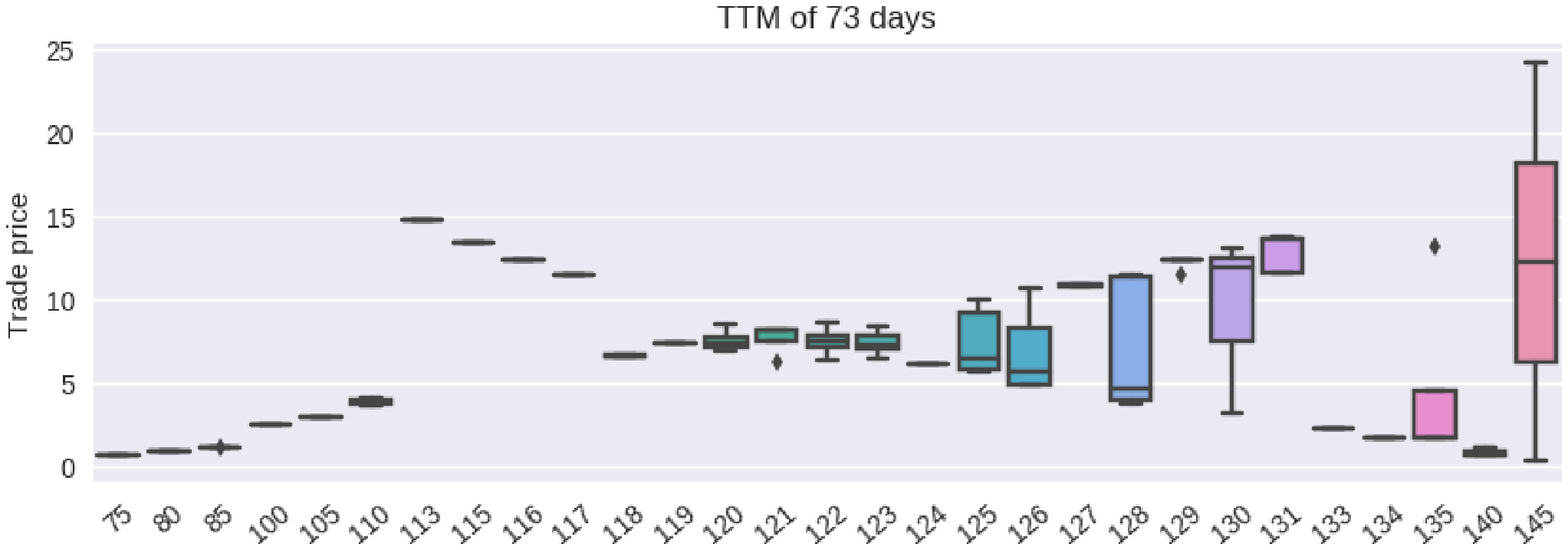}
			\endminipage\\
			\minipage{0.80\textwidth}
			\includegraphics[width=\linewidth]{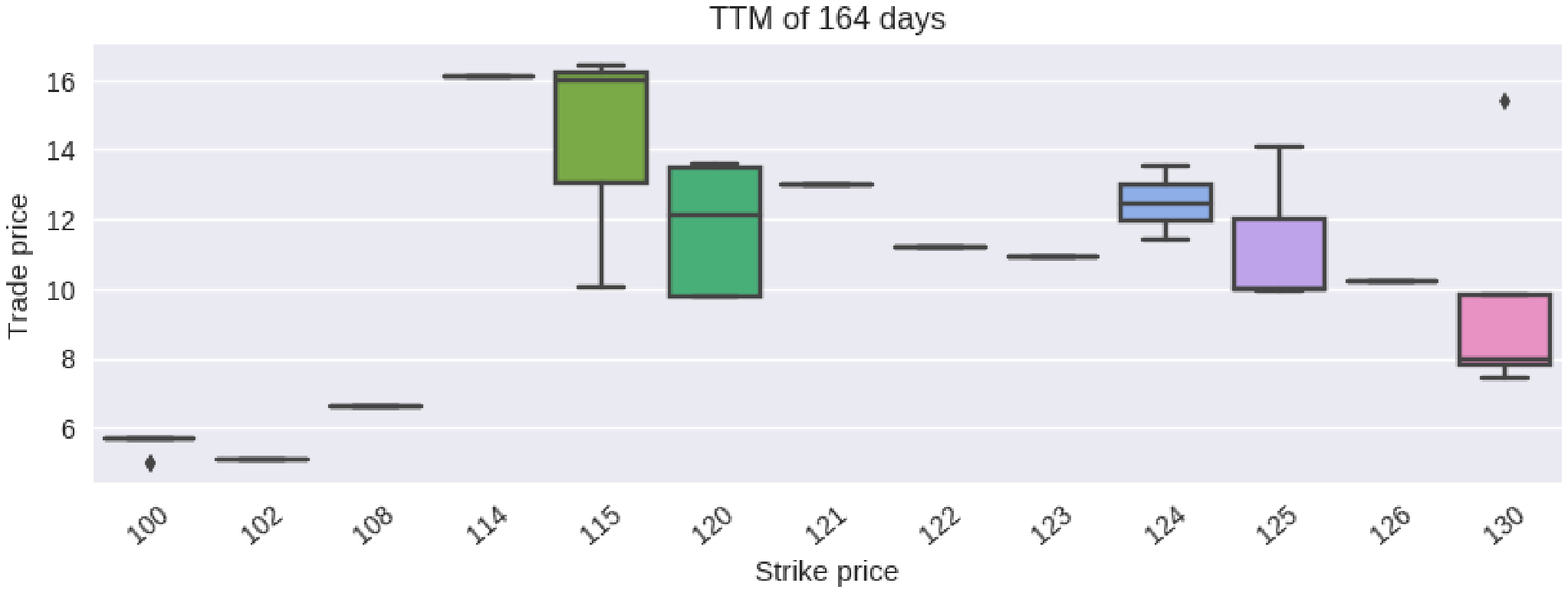}
			\endminipage
		\end{center}
		\caption{Intraday trade prices for various option contracts on SSO on 6 Jan, 2015; data source: Option Price Reporting Authority } 
		\label{trades_opra}
	\end{figure}

\begin{figure}[H]
		\begin{center}
			\minipage{0.80\textwidth}
			\includegraphics[width=\linewidth]{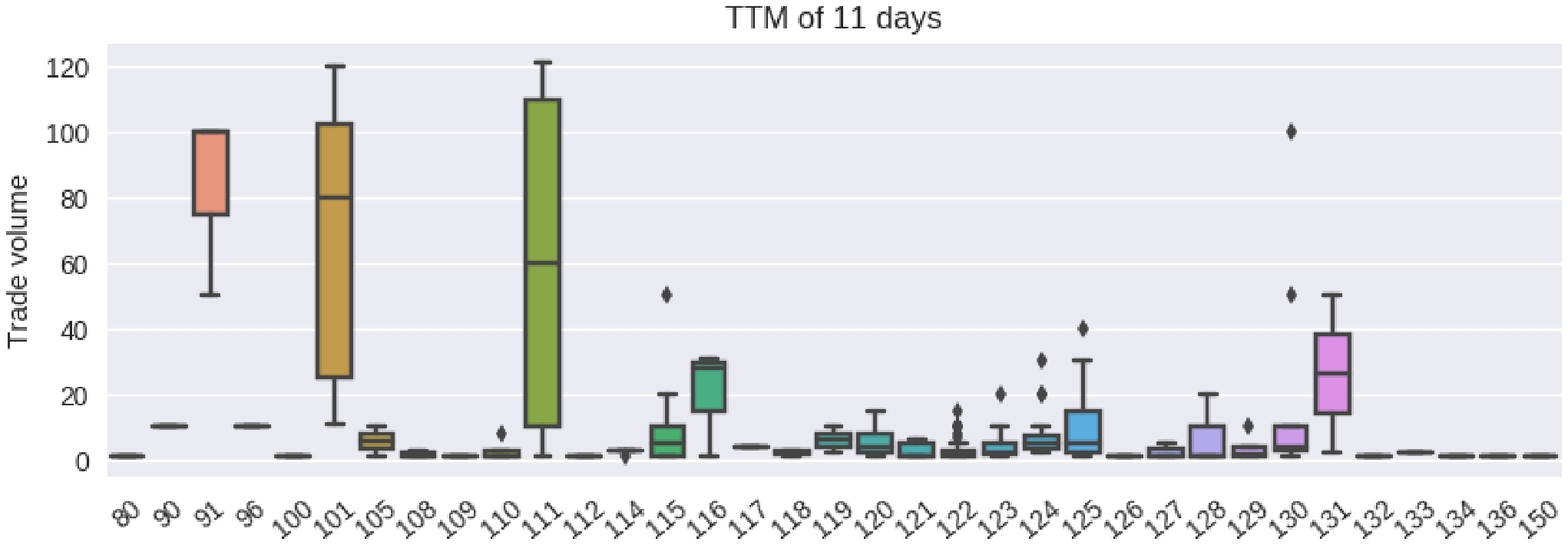}
			\endminipage\\
			\minipage{0.80\textwidth}
			\includegraphics[width=\linewidth]{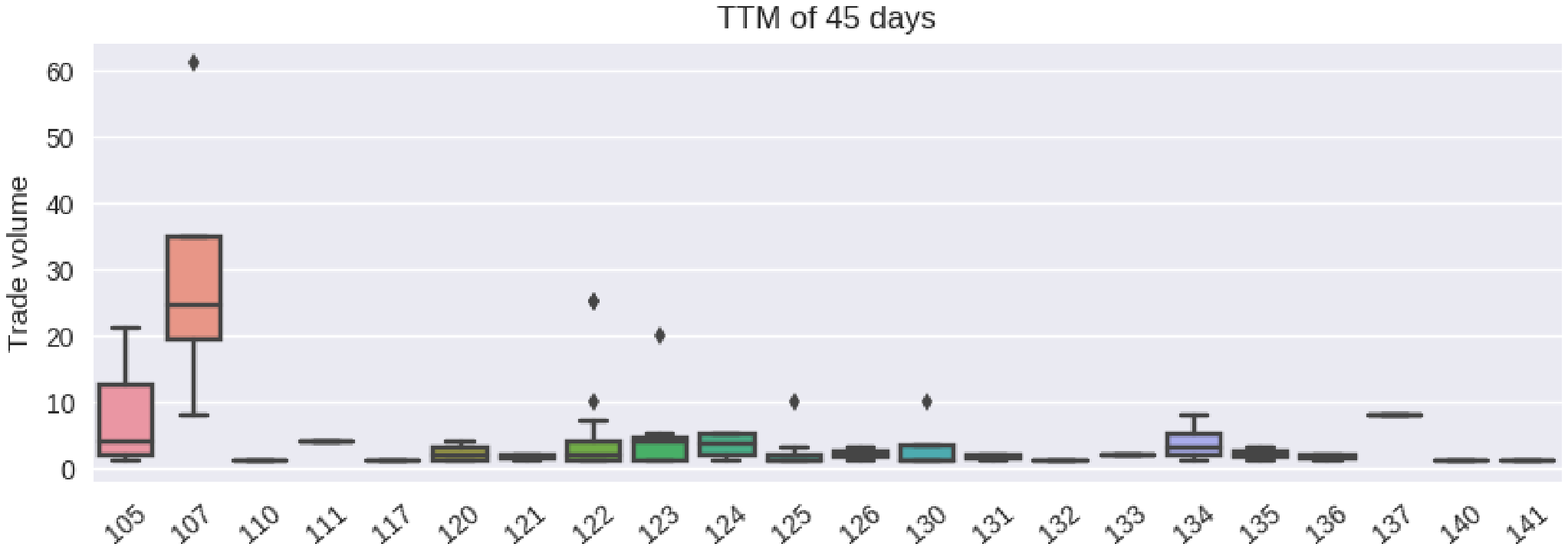}
			\endminipage\\
			\minipage{0.80\textwidth}
			\includegraphics[width=\linewidth]{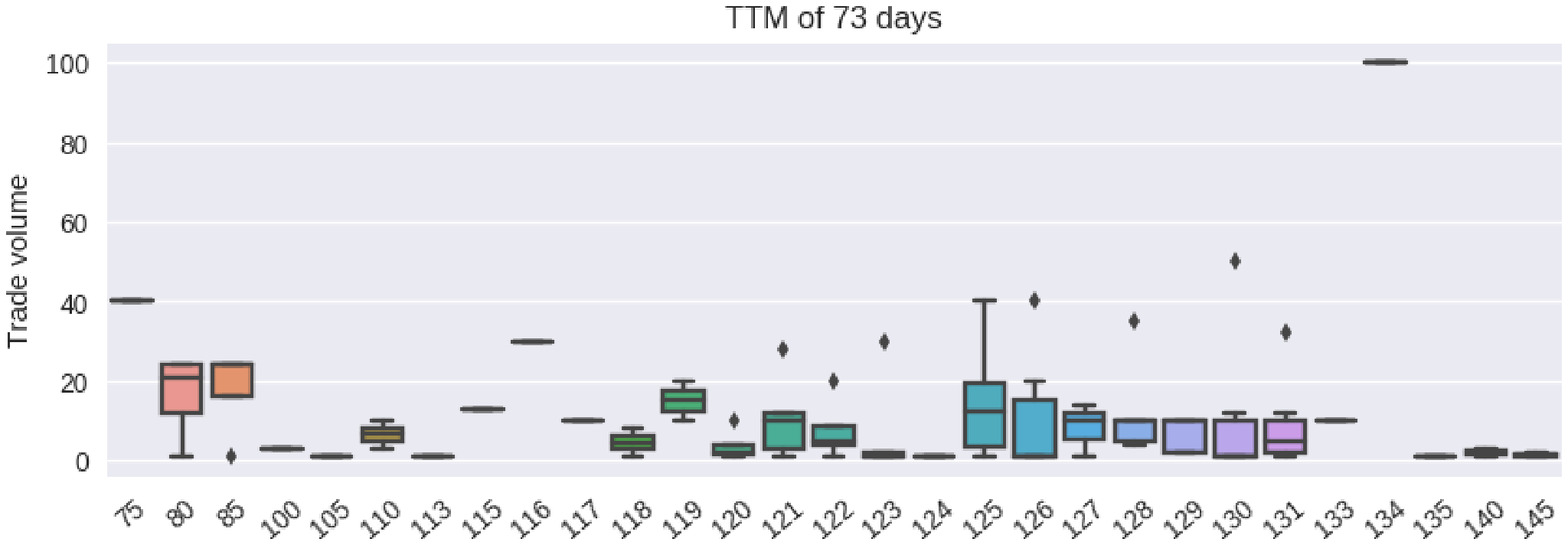}
			\endminipage\\
			\minipage{0.80\textwidth}
			\includegraphics[width=\linewidth]{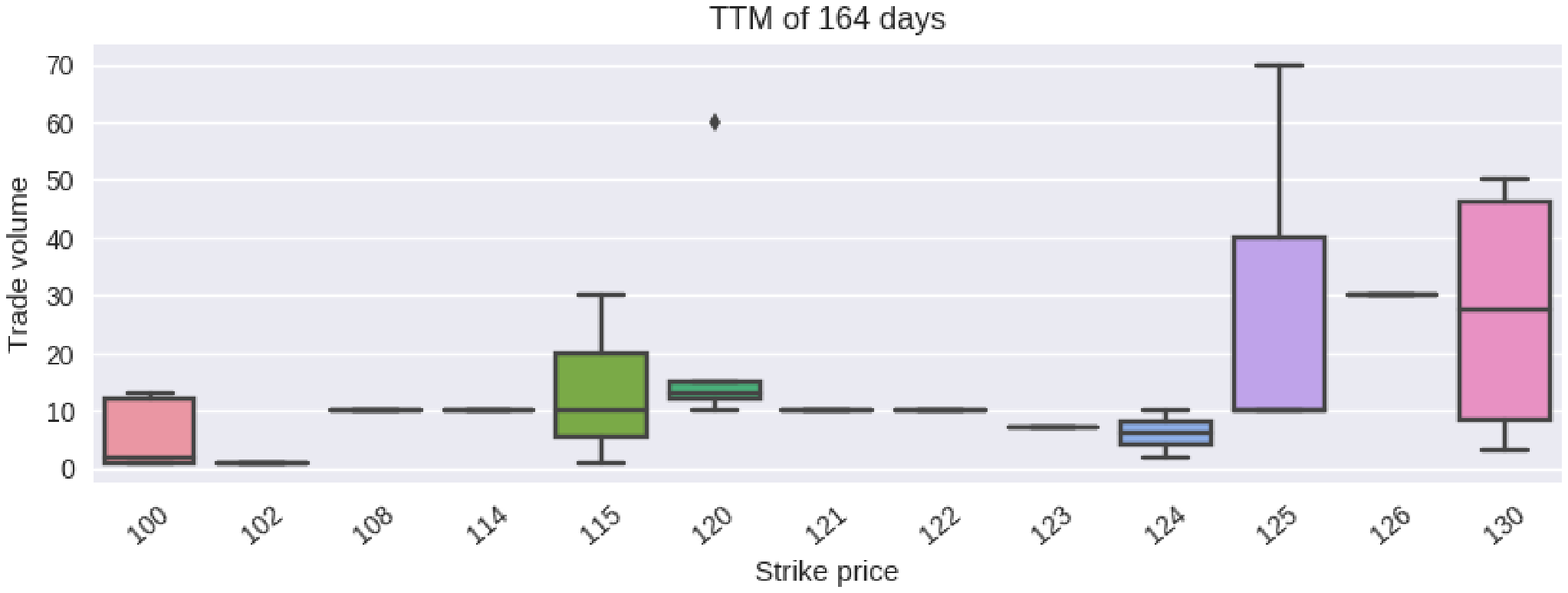}
			\endminipage
		\end{center}
		\caption{Intraday trade volumes for various option contracts on SSO on 6 Jan, 2015; data source: Option Price Reporting Authority } 
		\label{volumes_opra}
	\end{figure}

\begin{figure}[H]
		\begin{center}
			\minipage{0.80\textwidth}
			\includegraphics[width=\linewidth]{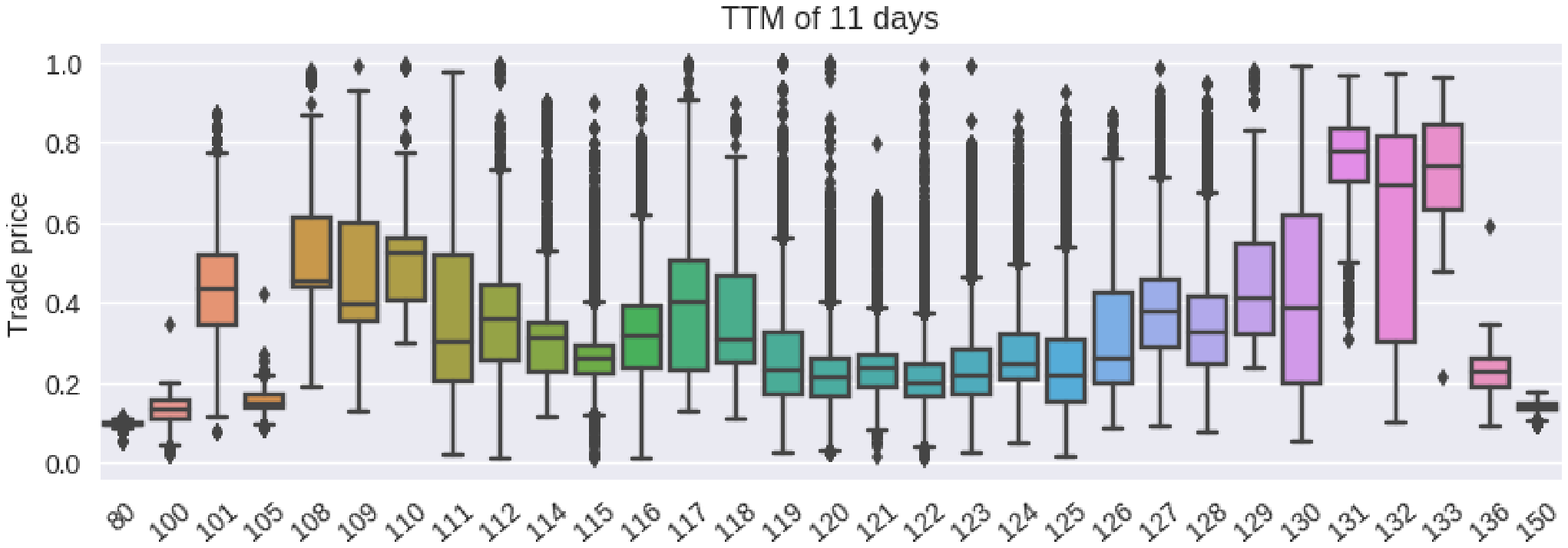}
			\endminipage\\
			\minipage{0.80\textwidth}
			\includegraphics[width=\linewidth]{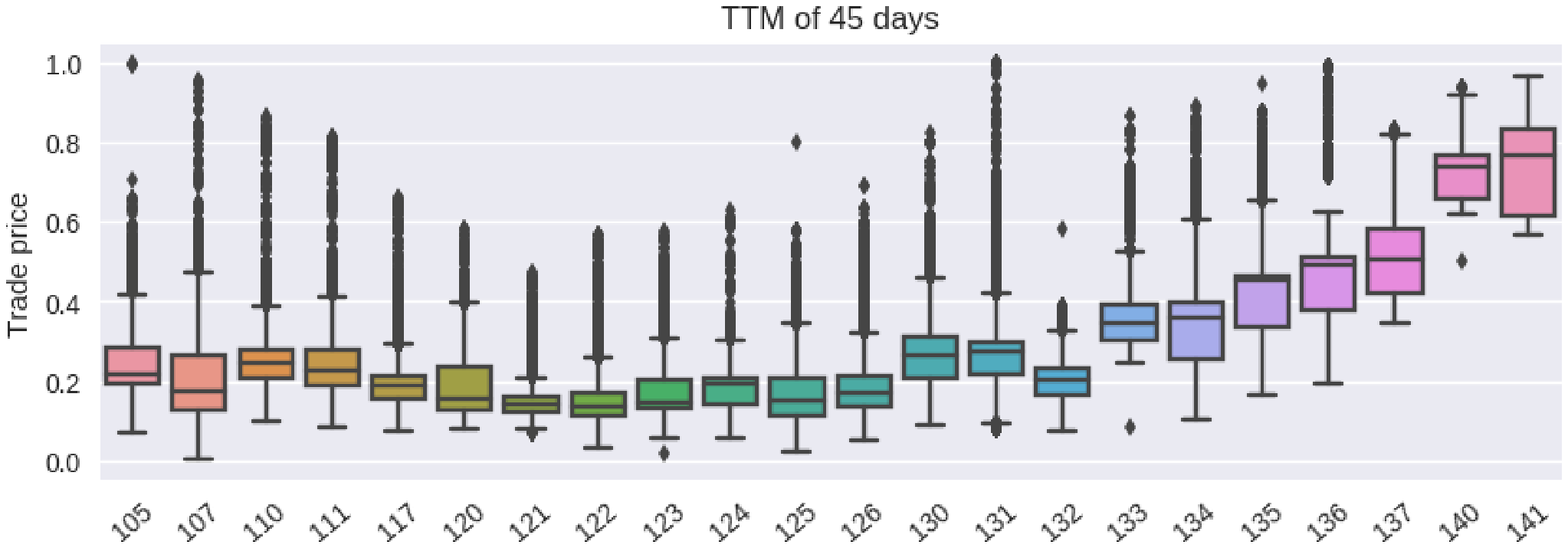}
			\endminipage\\
			\minipage{0.80\textwidth}
			\includegraphics[width=\linewidth]{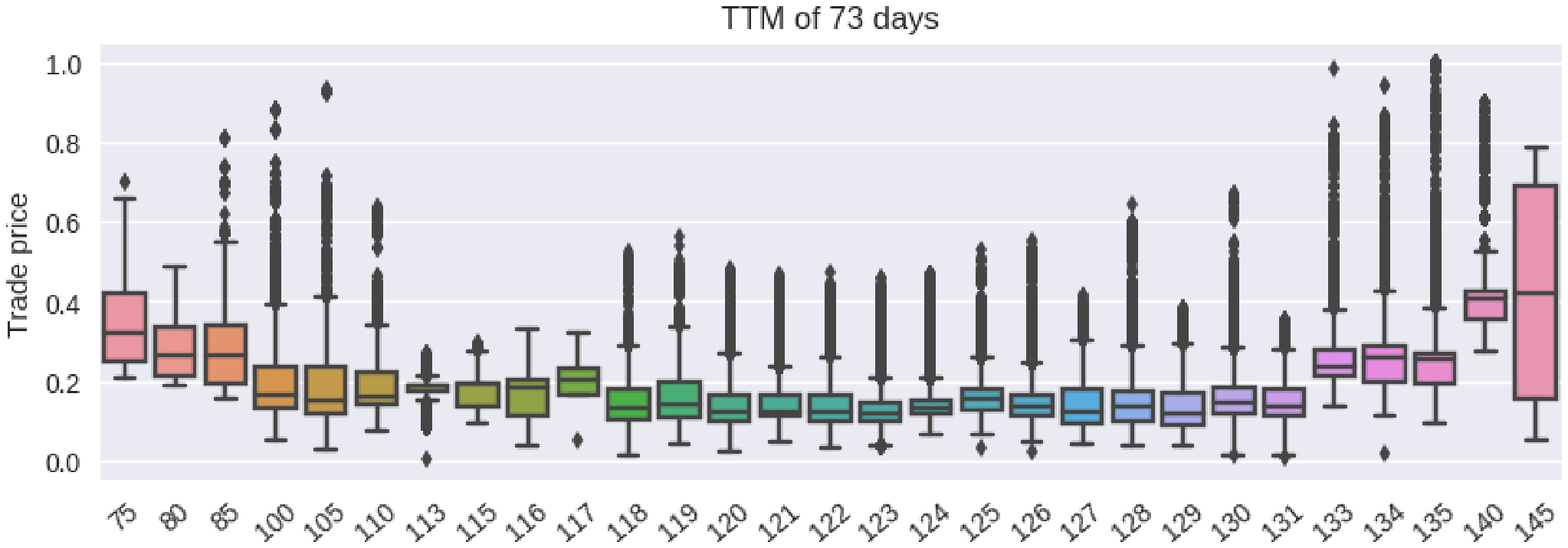}
			\endminipage\\
			\minipage{0.80\textwidth}
			\includegraphics[width=\linewidth]{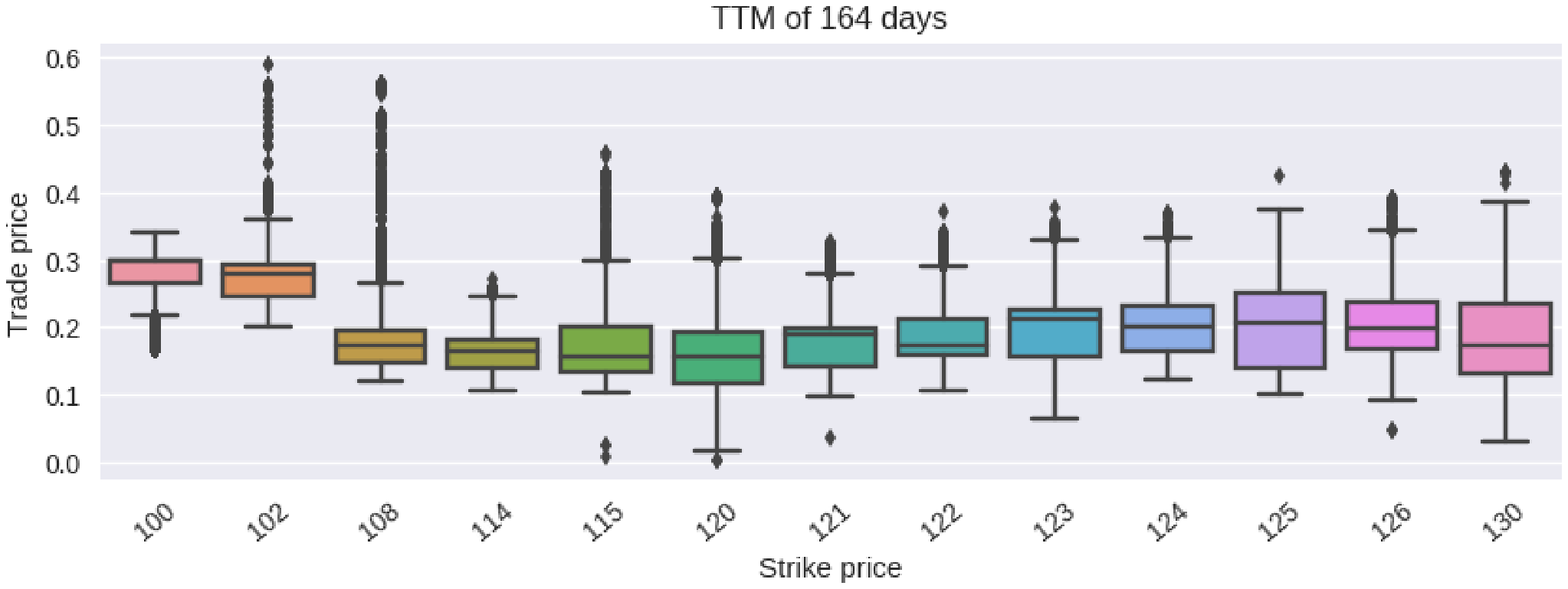}
			\endminipage
		\end{center}
		\caption{Intraday bid-ask spreads (percentage of the ask price) for various option contracts on SSO on 6 Jan, 2015; data source: Option Price Reporting Authority } 
		\label{spreads_opra}
	\end{figure}

	To give an impression of leveraged ETF option tradability, we give an illustration of the existing bid-ask spreads and actual trades of the SSO LETF, as these data will be used for the trading strategy example below. Figures \ref{trades_opra} and \ref{volumes_opra} show variation of existing trade prices and volumes for various option contacts based on exercise price and time to expiration. We can see that shorter-term contracts are traded more broadly. It has been also found that trades predominantly occur at or near mid-quotes. In Figure \ref{spreads_opra} we show bid-ask spreads for the same range of option contracts, which tend to be quite high, but somewhat lower for longer-term contracts. 
	
	The option data we use for the empirical application are trade-based data, i.e. each observation corresponds to an actual trade, not price quotes or settlement data. Implied volatility and option prices are taken from the database and computed in accordance with standard conventions used by market participants using the midpoint of the best closing bid price and best closing offer price for the options, taking account of liquidity and dividends.
	
	Additionally, we remove data which may contain noise, potential misprints and other errors. Such data include anomalous and outlier data resulting, e.g., from artificial extrapolation of implied volatilities for non-traded options or feature lower liquidity for the out-of-the-money or options which are deeply in-the-money.

	\subsection{Confidence bands}
	\label{confb_appl}
	
\begin{landscape}
		\begin{figure}[H]
			\begin{center}
				\minipage{0.8\textwidth}
				\includegraphics[width=\linewidth]{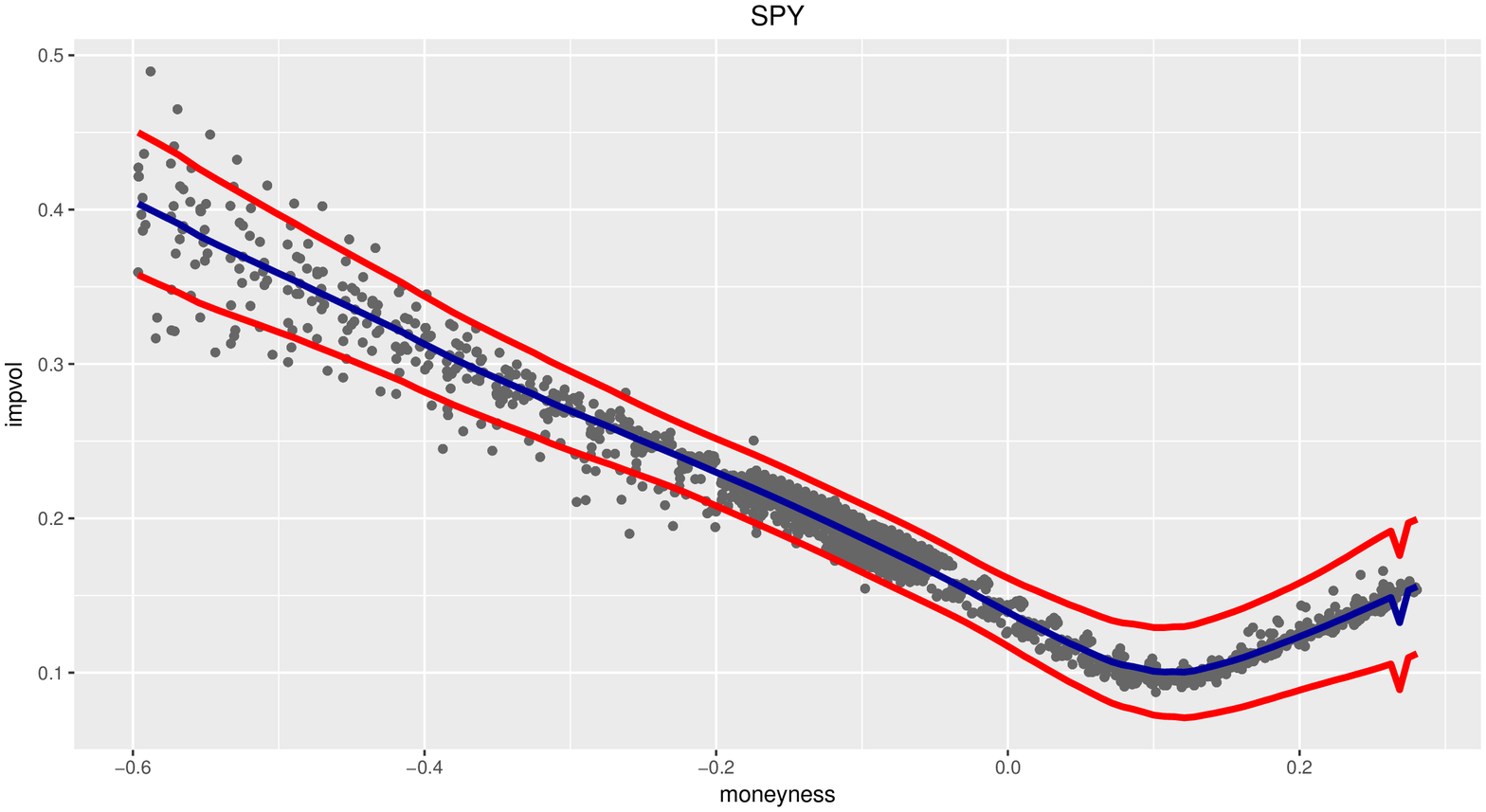}
				\endminipage
				\minipage{0.8\textwidth}
				\includegraphics[width=\linewidth]{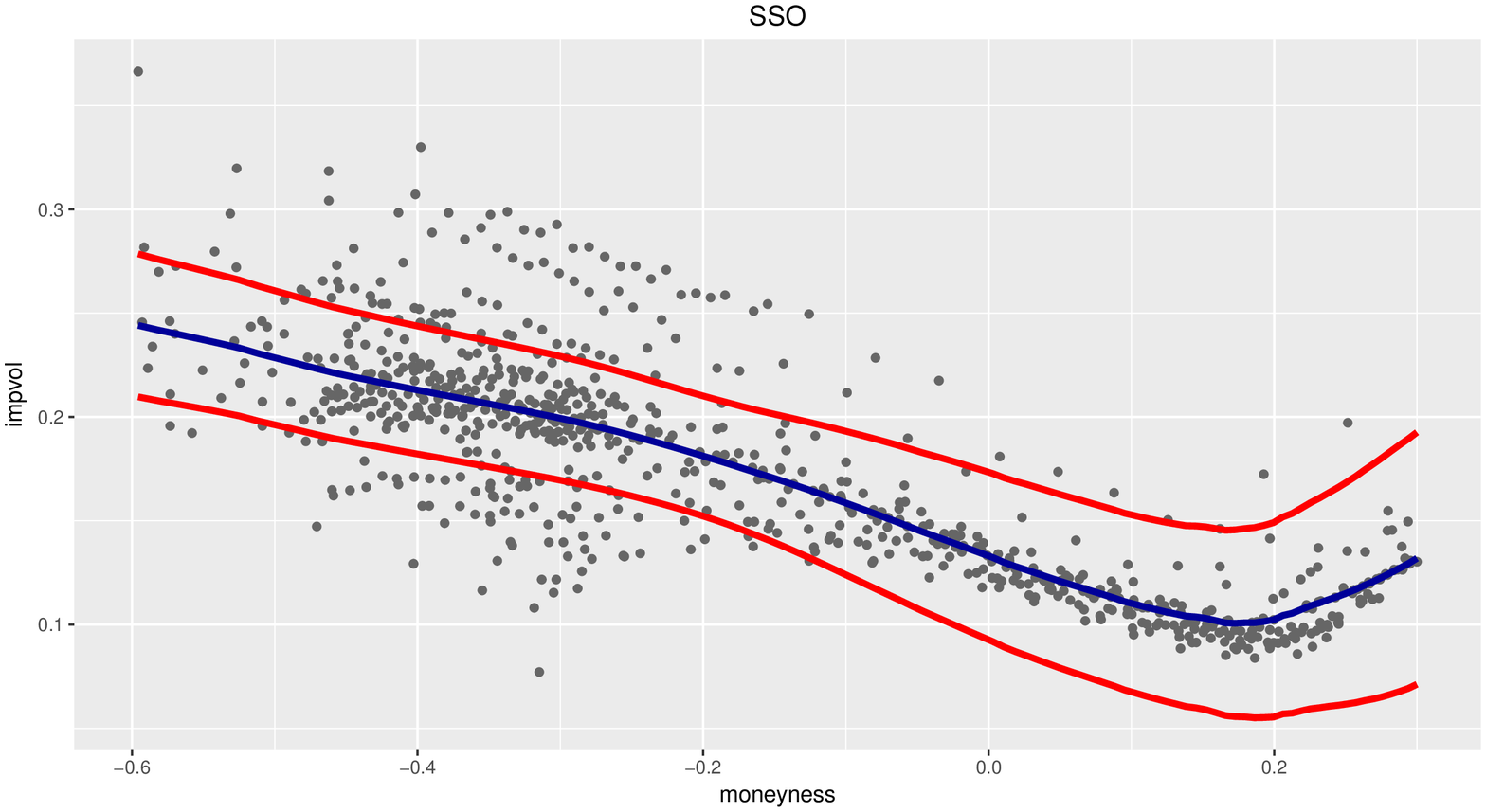}
				\endminipage\\
				\minipage{0.8\textwidth}
				\includegraphics[width=\linewidth]{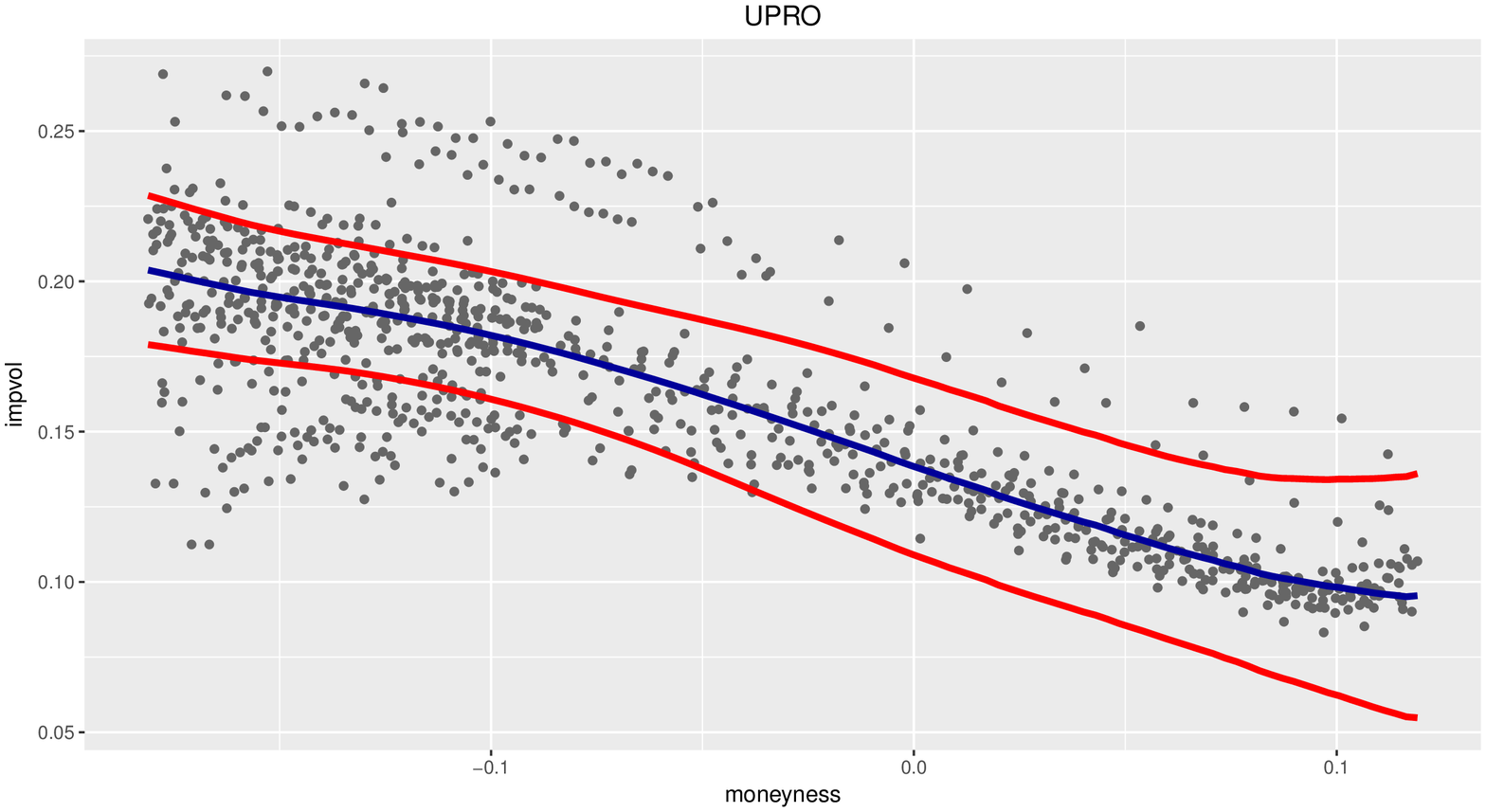}
				\endminipage
				\minipage{0.8\textwidth}
				\includegraphics[width=\linewidth]{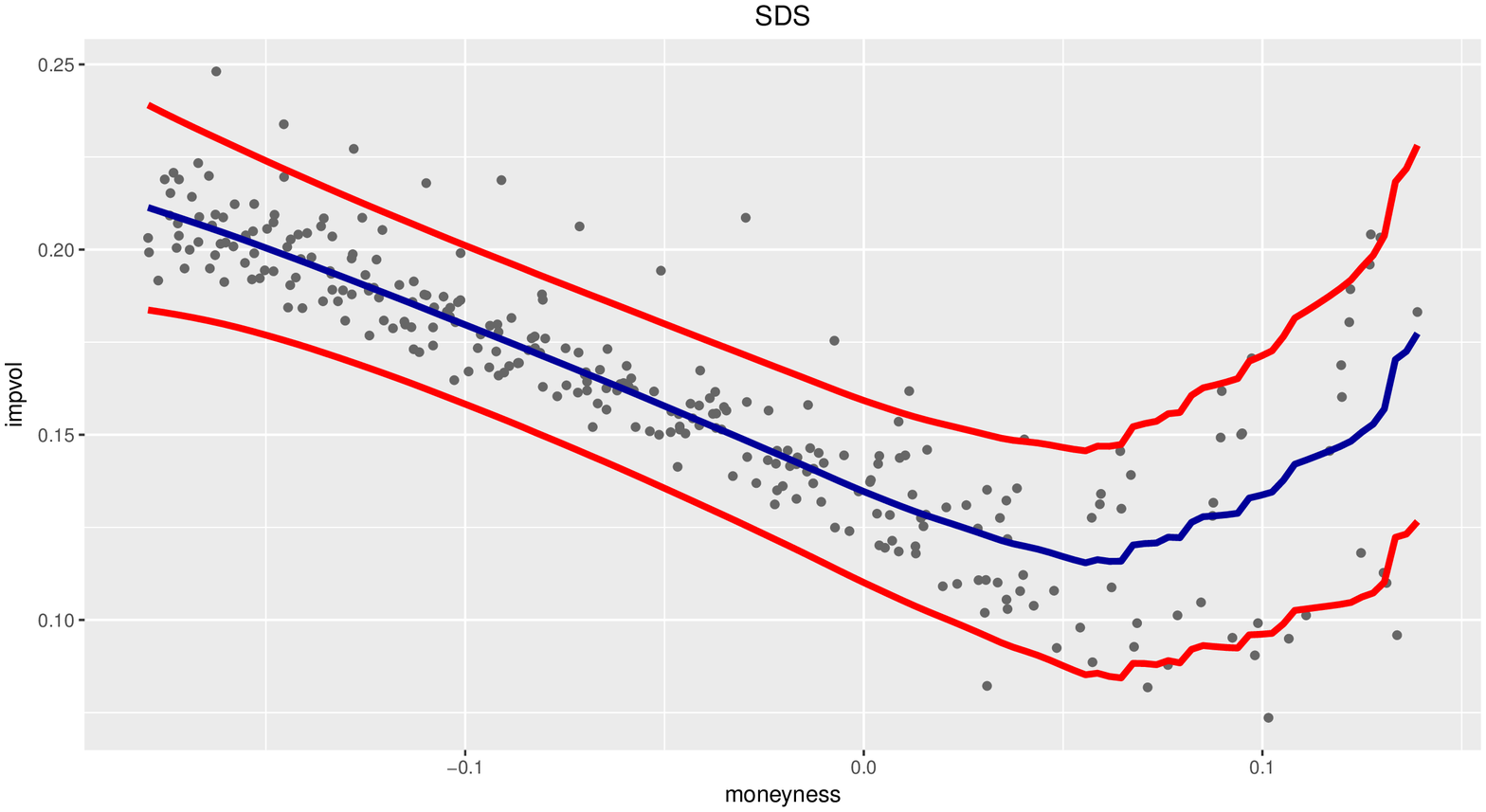}
				\endminipage
			\end{center}
			\caption{\textcolor{dblue}{Fitted implied volatility} and \textcolor{red}{bootstrap uniform confidence bands} for 4 (L)ETFs on S\&P500; $\tau$: 0.5 years} 
			\label{confbs05}
		\end{figure}
	\end{landscape}

\begin{landscape}
		\begin{figure}[H]
			\begin{center}
				\minipage{0.8\textwidth}
				\includegraphics[width=\linewidth]{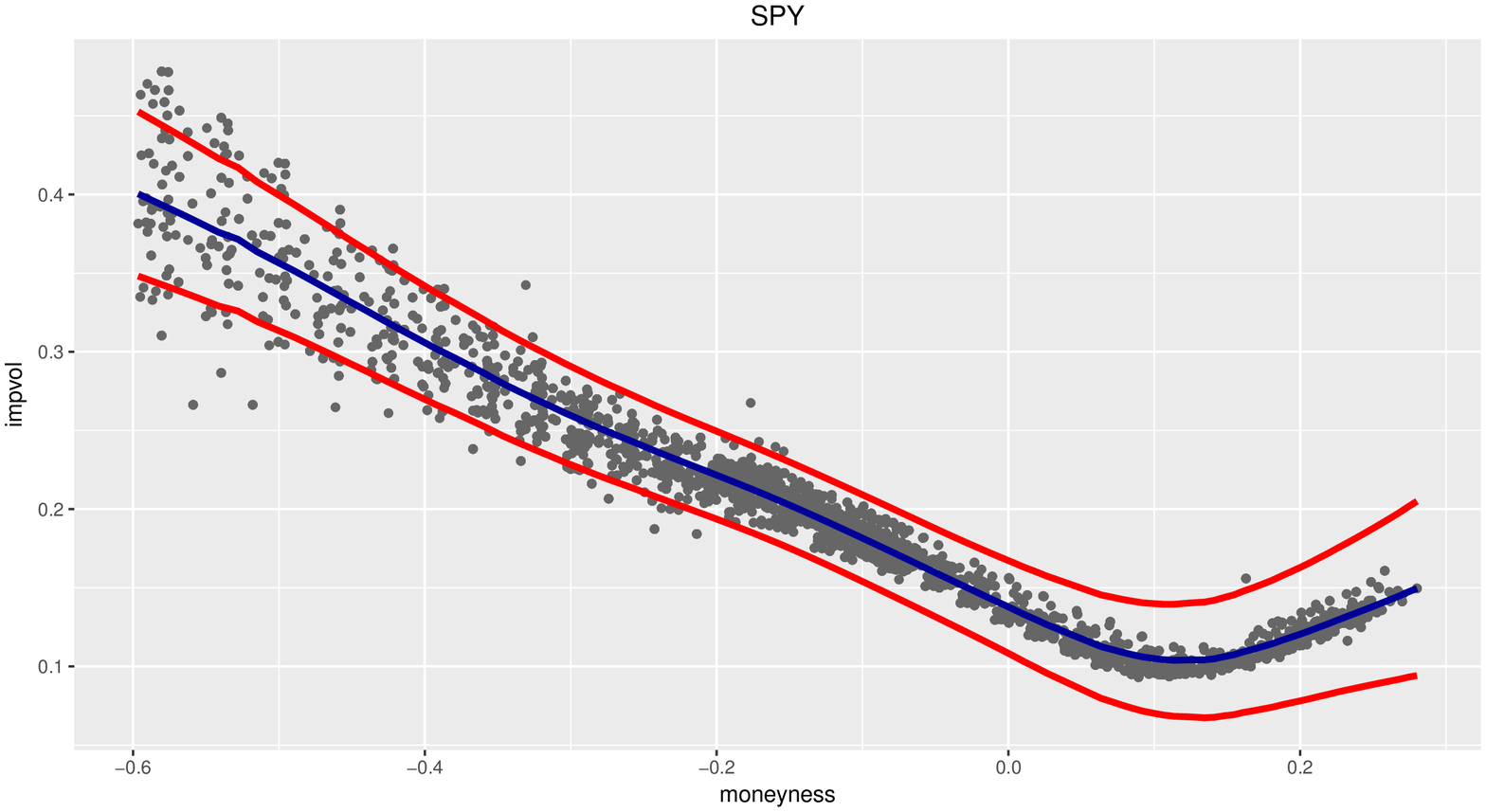}
				\endminipage
				\minipage{0.8\textwidth}
				\includegraphics[width=\linewidth]{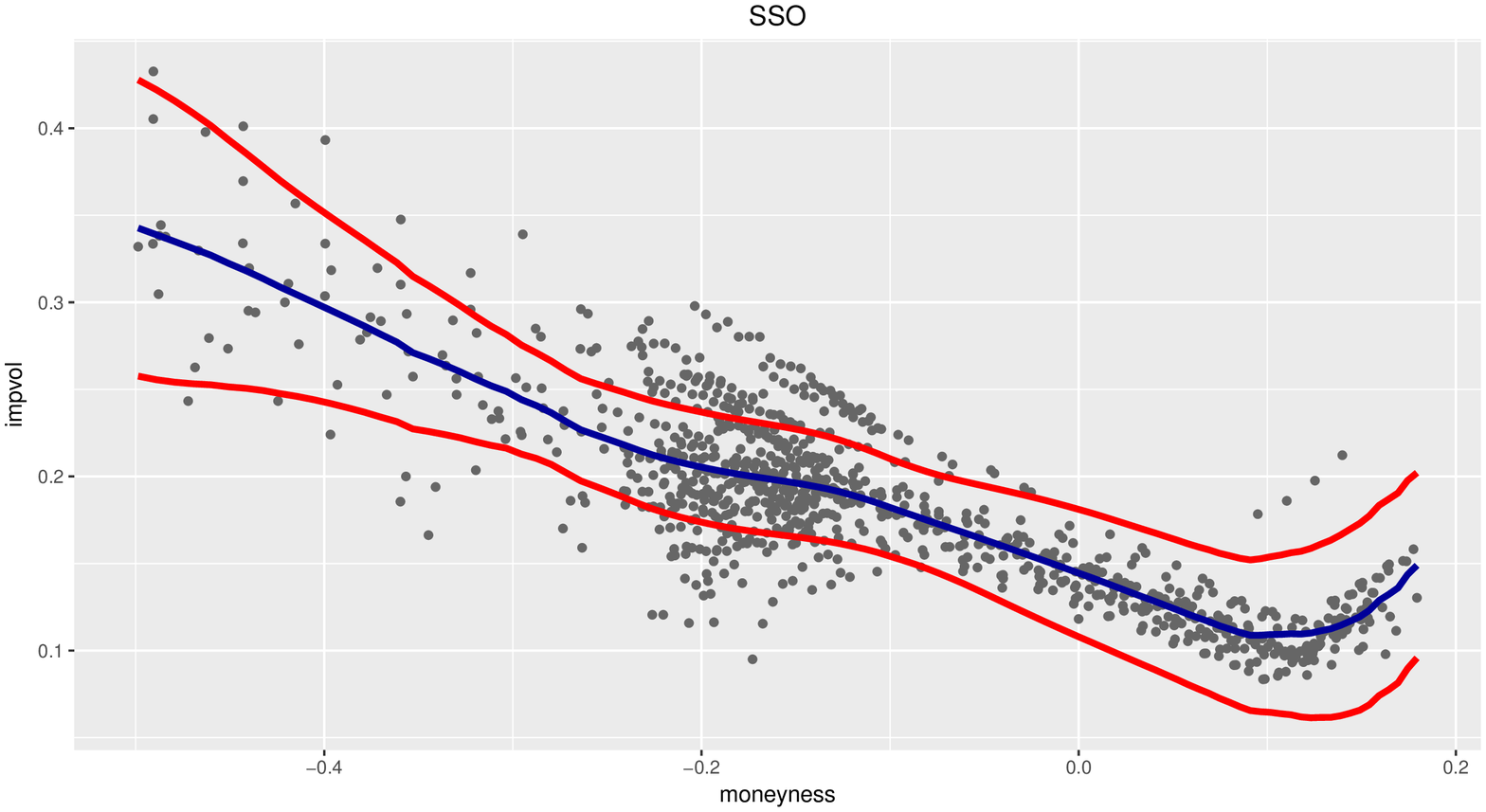}
				\endminipage\\
				\minipage{0.8\textwidth}
				\includegraphics[width=\linewidth]{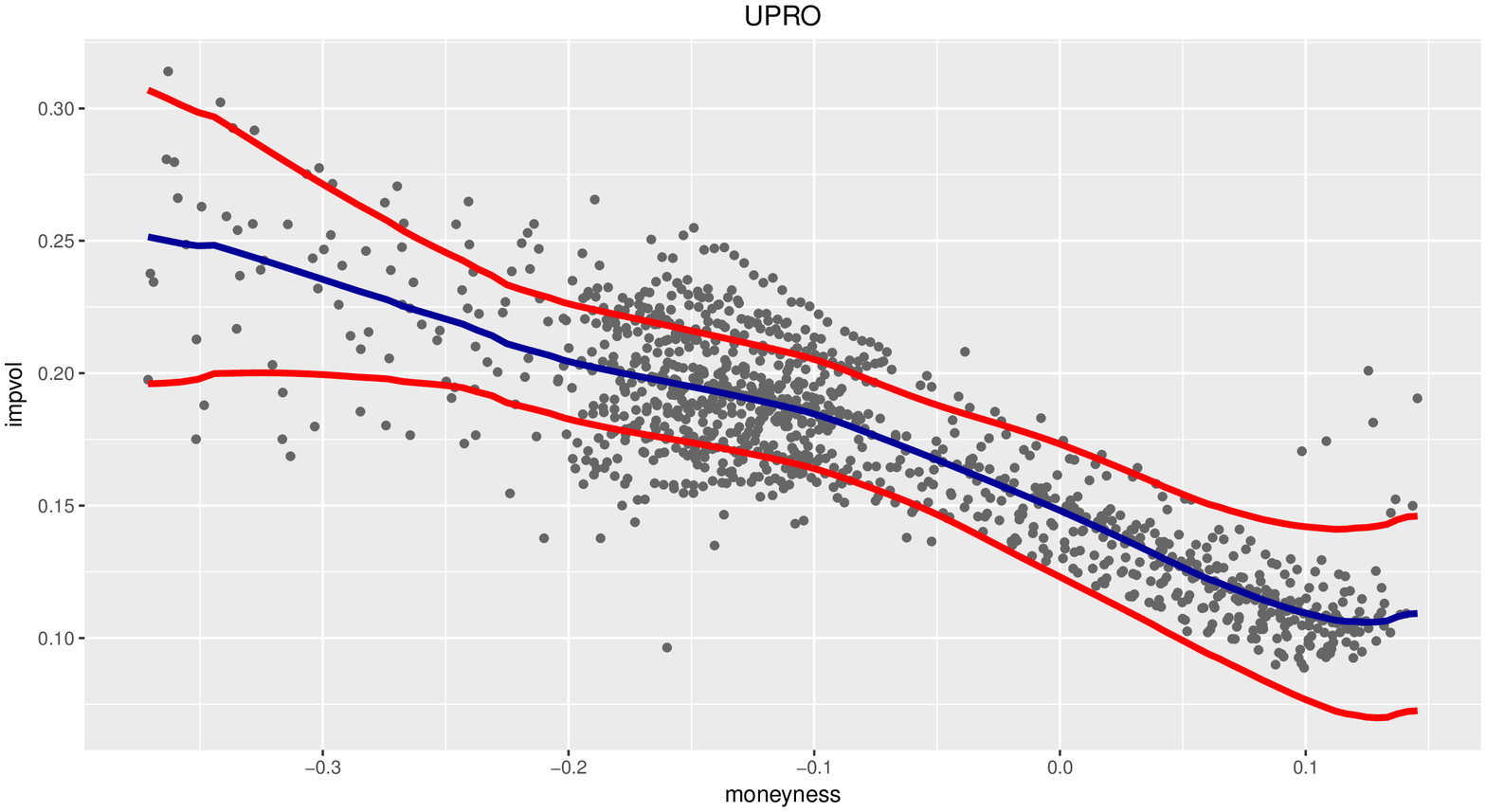}
				\endminipage
				\minipage{0.8\textwidth}
				\includegraphics[width=\linewidth]{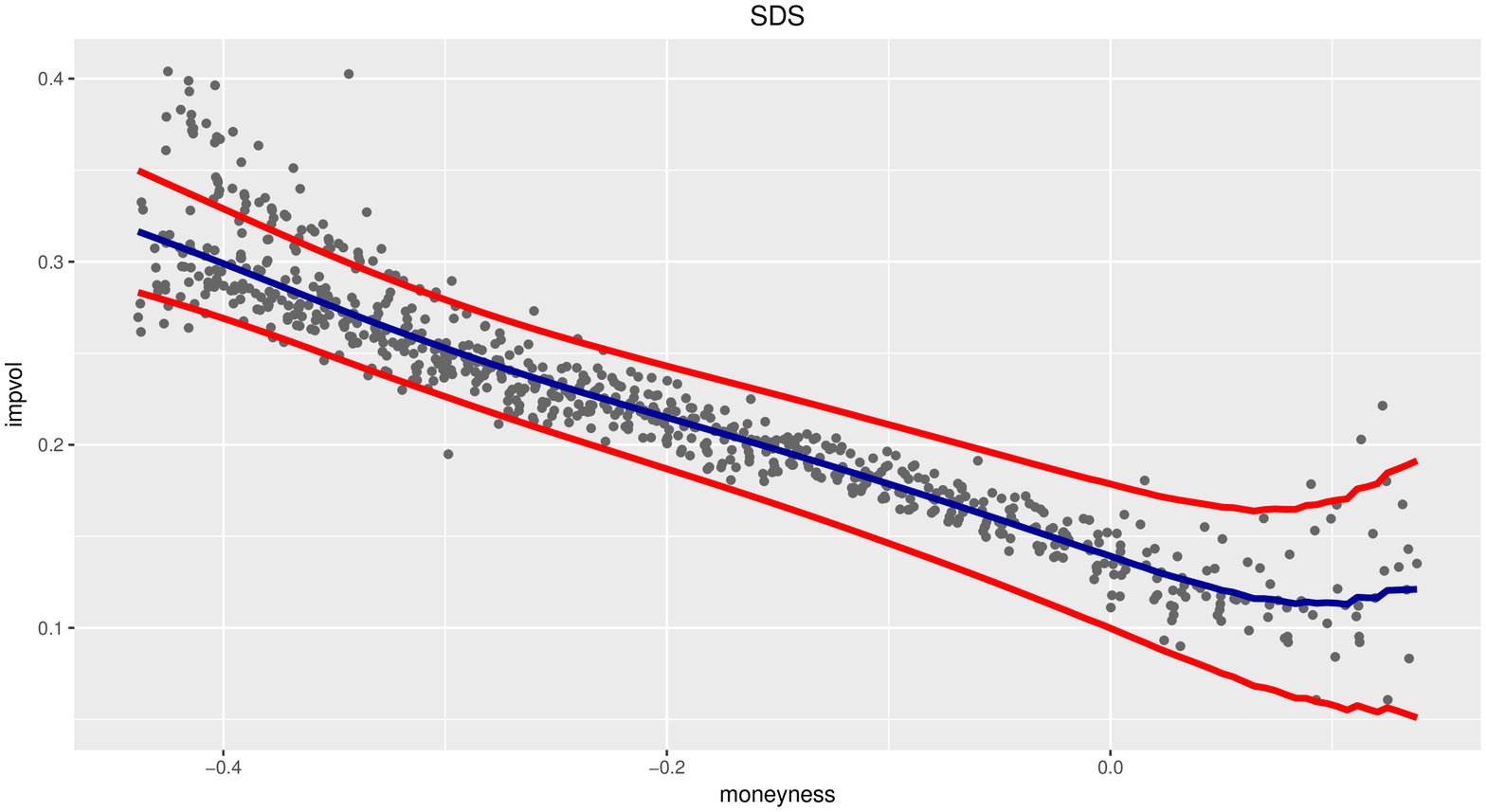}
				\endminipage
			\end{center}
			\caption{\textcolor{dblue}{Fitted implied volatility} and \textcolor{red}{bootstrap uniform confidence bands} for 4 (L)ETFs on S\&P500; $\tau$: 0.6 years}  
			\label{confbs06}
		\end{figure} 
	\end{landscape}

\begin{figure}[H]
		\begin{center}
			\minipage{0.82\textwidth}
			\includegraphics[width=\linewidth]{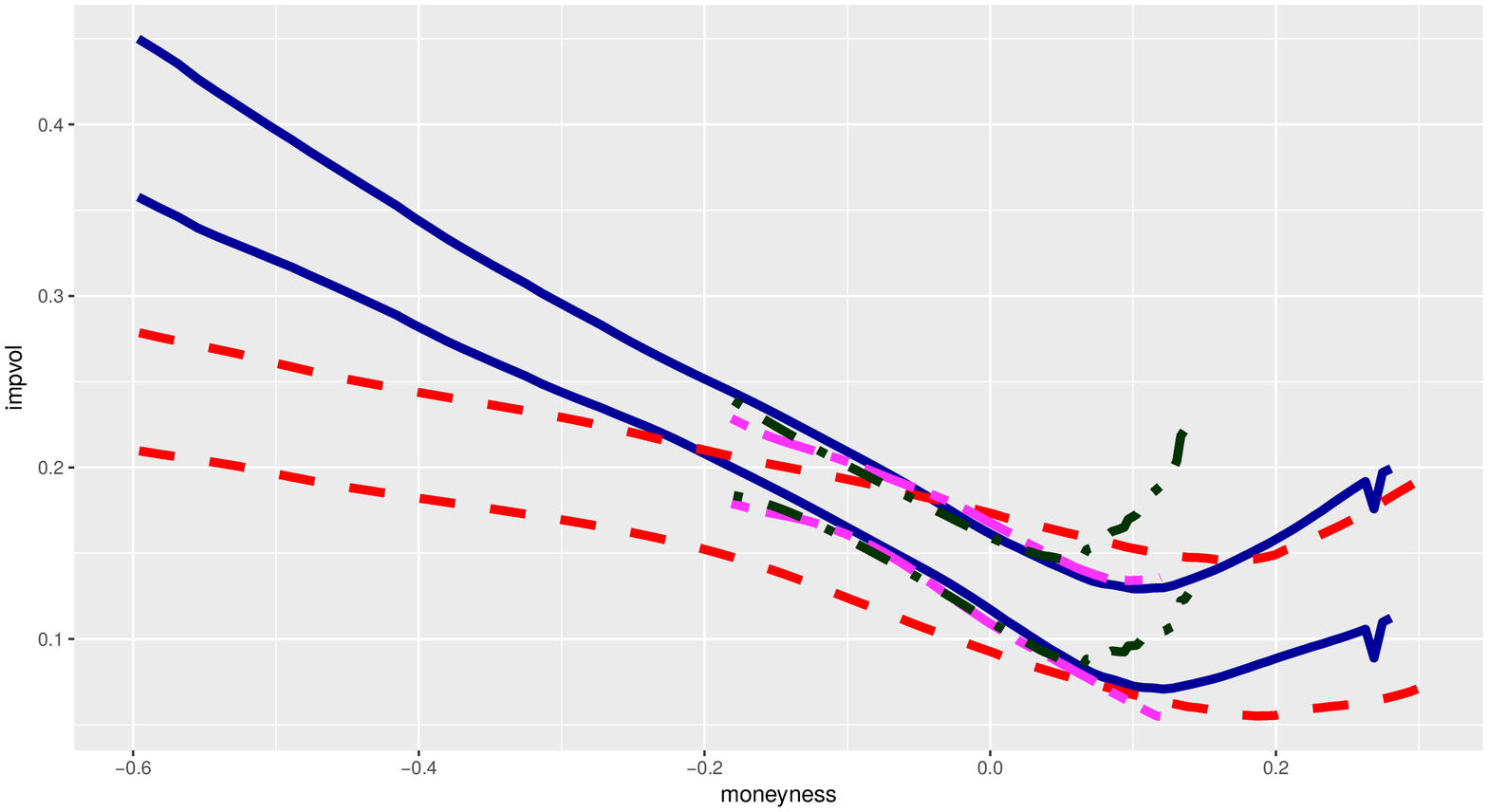}
			\endminipage\\
			\minipage{0.82\textwidth}
			\includegraphics[width=\linewidth]{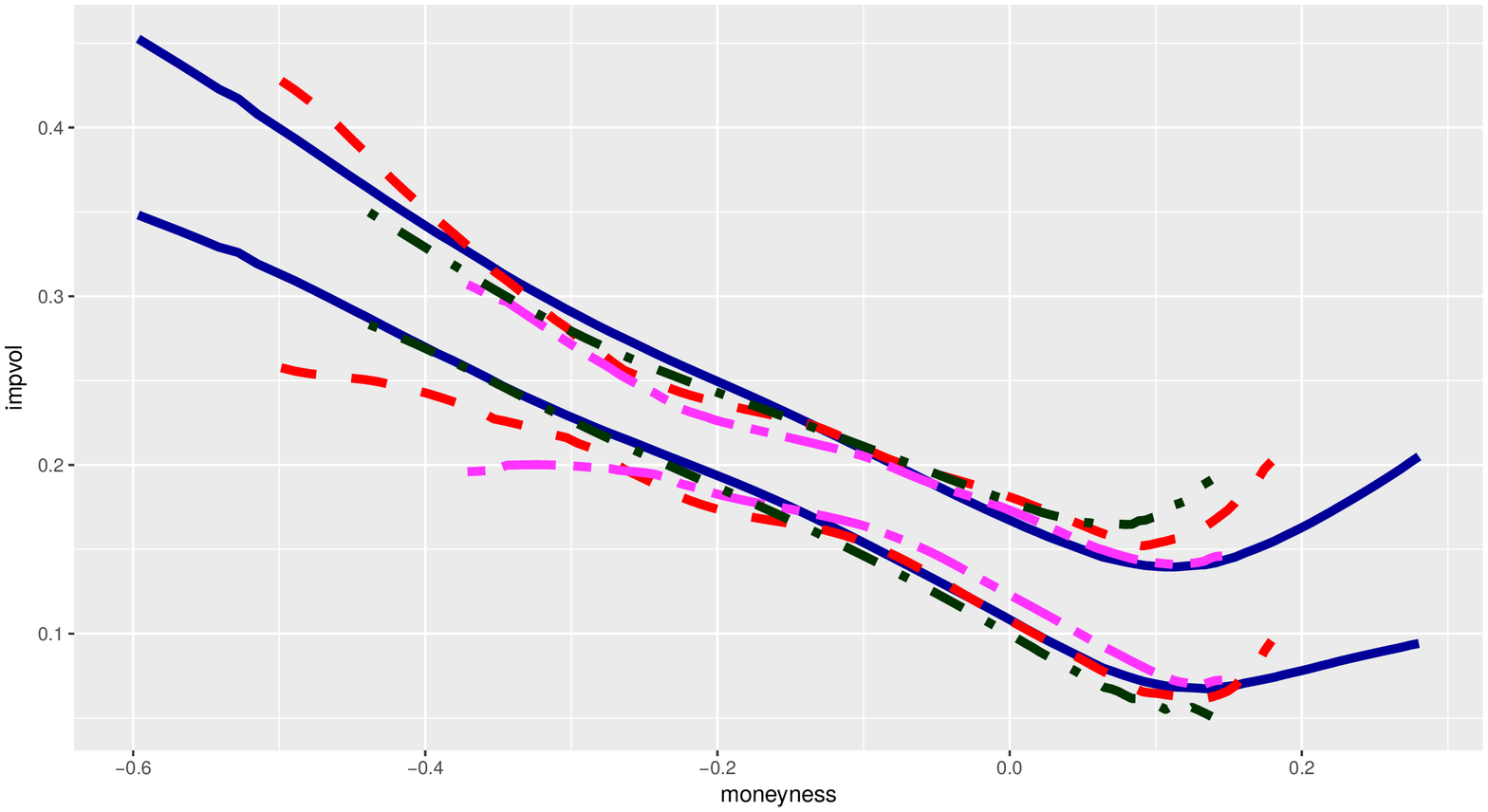}
			\endminipage
		\end{center}
		\caption{Combined uniform bootstrap confidence bands for \textcolor{spycol}{SPY}, \textcolor{ssocol}{SSO}, \textcolor{uprocol}{UPRO} and \textcolor{sdscol}{SDS} after moneyness scaling ($\tau = 0.5$ (top) and $\tau = 0.6$ years (bottom), respectively)} 
		\label{bndscomb}
	\end{figure}

	We use the data described above to construct bootstrap confidence bands for the $M$-smoother of implied volatility $Y$ given log-moneyness $X$, according to methodology described in Section \ref{confb}. Accordingly, $X$ is transformed using (\ref{monsc}). The results are shown in Figures \ref{confbs05} and \ref{confbs06} for time-to-maturity 0.5 and 0.6 years, respectively. In Figure \ref{bndscomb} combined bands are provided.
	
	We can observe clear discrepancy between the implied volatilities of leveraged ETFs and their unleveraged counterpart SPY. For all LETFs, non-overlapping confidence bands imply that there is a statistically significant difference between IV functions at the significance level $\alpha = 0.05$. It is more pronounced for in- and out-of-the-money options. This phenomenon may occur due to lower liquidity of in- and out-of-the-money options compared to at-the-money options. On the other hand, as shown, e.g. in \citet{etling_miller}, the relationship between option moneyness and liquidity is more complex than quadratic, maximized for at-the-money options. Therefore, liquidity need not be the only reason for this fact.
	
	We can see from Figure \ref{bndscomb} that the bands for SSO demonstrate particularly strong deviation from those of SPY. This implies that discrepancies not removed by the moneyness scaling procedure are the largest for this LETF. Therefore we conclude this section with a trading strategy which is meant to exploit such statistical discrepancies on the market of SPY and SSO options.

	\subsection{DSFM estimation and forecasting}
	\label{emptest}
	
	The $EV$, $RMSE$ and $RMSPE$ criteria are displayed in Table \ref{evrmse}. The model order $L = 3$ is chosen for estimation. The data for the SPY ETF option are used with parameters $n_{\kappa}, n_{\tau} = 3$; $M = 9$, $N = 7$, so that $U = 6$, $V = 4$, $K = 6 \times 4 = 24$.  
	
\begin{landscape}
		\begin{figure}[H]
			\begin{center}
				\includegraphics[scale=0.7]{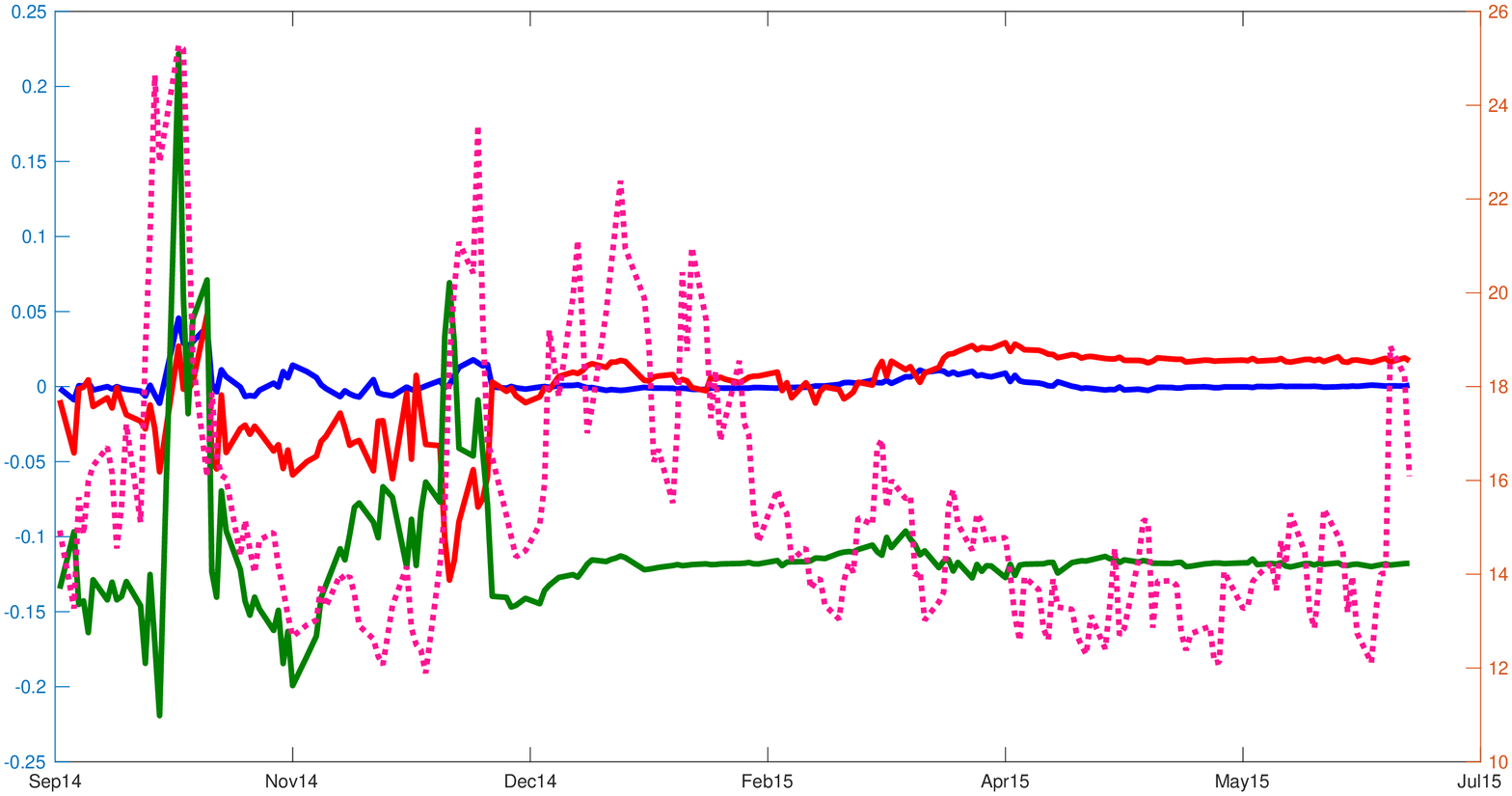}
			\end{center}
			\caption{Time dynamics of \textcolor{blue}{$\widehat{\mathcal{Z}}_{t,1}$}, \textcolor{red}{$\widehat{\mathcal{Z}}_{t,2}$}, \textcolor{darkgreen}{$\widehat{\mathcal{Z}}_{t,3}$}, \textcolor{vixcol}{VIX index} }
			\label{zfacs}
		\end{figure}
	\end{landscape}

	Figure \ref{zfacs} shows the dynamics of $\widehat{\mathcal{Z}}_{t}$ in time. Two largest "spikes" in the value of the third stochastic loading in the beginning of the period correspond to the period of relatively large values of the CBOE volatility index (VIX). The second of the "spikes" precedes in time an increase in the VIX value implying that the model has predictive value with respect to market instability dynamics. This shows that DSFM captures leading dynamic effects as well as can explain effects like skew or term structure changes.
	
	Theoretical and simulation results in \citet{park_et_al} justify using vector autoregression (VAR) analysis to model $\widehat{\mathcal{Z}}_{t}$. To select a VAR model, we computed the Schwarz (SC), the Hannan-Quinn (HQ) and the Akaike (AIC) criteria, as shown in Table \ref{zcrit}. All three criteria select the VAR(1) model. Furthermore, the roots of the characteristic polynomial all lie inside the unit circle, which shows that the specified model is stationary. Portmanteau and Breusch-Godfrey LM test results with 12 lags for the autocorrelations of the error term fail to reject residual autocorrelation at 10\% significance level.   
	
	The degenerate nature of implied volatility data is reflected by the fact that empirical observations do not cover estimation grids at given time points. This is due to the fact that contracts at certain maturities or strikes are not always traded. The DSFM fitting procedure introduces basis functions which approximate a high-dimensional space and depend on time. This allows to account for all information in the dataset simultaneously in one minimization procedure which runs over all $\widehat{m}_l$ and $\widehat{\mathcal{Z}}_{t}$ and avoid bias problems which would inevitably occur if some kernel smoothing procedure such as Nadaraya-Watson were applied for this type of degenerate data.

	\subsection{Option trading strategy}
	
	\subsubsection{Description} 
	\label{stratdescr}
	
	Ability to forecast the whole surface of implied volatility can be used in combination with the moneyness scaling technique to exploit potential discrepancies in ETF and LETF option prices or implied volatilities to build a trading strategy. A suitable strategy would be the so-called "trade-with-the-smile/skew" strategy adapted for the special case of ETF-LETF option IV discrepancy. It would use the ETF option data to estimate the model (theoretical) smile of the leveraged counterpart and the information from the IV surface forecast to recognize the future (one-period-ahead) possible IV discrepancy. 
	
	Going back to the results in Section \ref{confb_appl}, we see that the largest statistical discrepancy between leveraged and unleveraged ETF implied volatilities is the one between SPY and SSO, so we consider these two options in the strategy setup. The strategy can be outlined as follows: choose a moving window width $w$; then for each $t = w,\ldots,T$ ($T$ is the final time point in the sample) do the following: 
	
	\begin{enumerate}
		\item
		given two leverage ratios $\beta_{SPY} = 1$, $\beta_{SSO} = 2$, re-scale the log-moneyness coordinate $LM^{(\beta_{SPY})}$ according to the moneyness scaling formula (\ref{monsc}) to obtain $\widehat{LM}^{(\beta_{SSO})}$. This will be the "model" moneyness coordinate for DSFM estimation,
		\item
		map the space $[\widehat{LM}_{min}^{(\beta_{SSO})}, \widehat{LM}_{max}^{(\beta_{SSO})}] \times [\tau_{min}^{SPY}, \tau_{max}^{SPY}]$ to $[0,1] \times [0,1]$ using marginal transformation,
		\item
		estimate the DSFM model (\ref{dsfm}) on $[0,1] \times [0,1]$. This will yield the IV surface estimates $\widehat{IV}_1^{SSO},\ldots,\widehat{IV}_t^{SSO}$,
		\item
		forecast the IV surface estimate $\widehat{IV}_{t+1}^{SSO}$ using the VAR structure of the estimated stochastic loadings $\widehat{\mathcal{Z}}_{t}$ and the factor functions $\widehat{m}_{l}$,
		\item
		choose a time-to-maturity $\tau^*$ at time point $t$, take the corresponding real-world values of SSO log-moneyness $LM^{(\beta_{SSO})}$ and map them to $[0,1]$ using the marginal distribution of $\widehat{LM}^{(\beta_{SSO})}$; denote the output as $LM^{(\beta_{SSO})}_{ \tau^*; M }$,
		\item
		using the marginally re-scaled grid $[\widehat{LM}_{min}^{(\beta_{SSO})}, \widehat{LM}_{max}^{(\beta_{SSO})}] \times [\tau^*, \tau^*]$ and $\widehat{IV}_{t+1}^{SSO}$, obtain interpolated values $\widehat{IV}^{SSO}_{t+1; LM^{(\beta_{SSO})}_{\tau^*; M}, \tau^*}$ corresponding to $LM^{(\beta_{SSO})}_{\tau^*; M}$, $\tau^{*}$,
		\item
		compare the "theoretical" values $\widehat{IV}^{SSO}_{t+1; LM^{(\beta_{SSO})}_{\tau^*; M}, \tau^*}$ with known real-world implied volatilities $IV^{SSO}_{t; LM^{(\beta_{SSO})}_{\tau^*; M}, \tau^*}$ corresponding to $LM^{(\beta_{SSO})}_{\tau^*; M}$, and construct a delta-hedged option portfolio: 
		\begin{itemize}
			\item
			if $\widehat{IV}^{SSO}_{t+1; LM^{(\beta_{SSO})}_{\tau^*; M}, \tau^*} > IV^{SSO}_{t; LM^{(\beta_{SSO})}_{\tau^*; M}, \tau^*}$ for all $LM^{(\beta_{SSO})}_{\tau^*; M}$, then buy (long) options corresponding to the largest difference $D_{long} \stackrel{\operatorname{def}}{=} \widehat{IV}^{SSO}_{t+1; LM^{(\beta_{SSO})}_{\tau^*; M}, \tau^*} - IV^{SSO}_{t; LM^{(\beta_{SSO})}_{\tau^*; M}, \tau^*}$,
			\item
			if $\widehat{IV}^{SSO}_{t+1; LM^{(\beta_{SSO})}_{\tau^*; M}, \tau^*} < IV^{SSO}_{t; LM^{(\beta_{SSO})}_{\tau^*; M}, \tau^*}$ for all $LM^{(\beta_{SSO})}_{\tau^*; M}$, then sell (short) options corresponding to the largest difference $D_{short} \stackrel{\operatorname{def}}{=} IV^{SSO}_{t; LM^{(\beta_{SSO})}_{\tau^*; M}, \tau^*} - \widehat{IV}^{SSO}_{t+1; LM^{(\beta_{SSO})}_{\tau^*; M}, \tau^*}$,
			\item
			if it holds that both $\widehat{IV}^{SSO}_{t+1; LM^{(\beta_{SSO})}_{\tau^*; M}, \tau^*} > IV^{SSO}_{t; LM^{(\beta_{SSO})}_{\tau^*; M}, \tau^*}$ and $\widehat{IV}^{SSO}_{t+1; LM^{(\beta_{SSO})}_{\tau^*; M}, \tau^*} < IV^{SSO}_{t; LM^{(\beta_{SSO})}_{\tau^*; M}, \tau^*}$ for different $LM^{(\beta_{SSO})}_{\tau^*; M}$, then buy (long) options with the largest $D_{long}$  and sell (short) options with the largest $D_{short}$. In all three cases use the underlying SSO LETF asset to make the portfolio delta-neutral,
		\end{itemize}
		\item
		at time point $t+1$, terminate the portfolio via an offsetting sale/purchase, calculate profit/loss and repeat until time $T$.
	\end{enumerate} 
	
	The strategy described above aims to exploit the information from the statistical discrepancies between the forecast "theoretical" (model) SSO LETF implied volatilities and the historical ("true") ones. It protects the portfolio against unfavorable moves in the underlying asset $L_t$ through delta-hedging and aims to gain from forecast moves in another option risk factor, the implied volatility via its explicit estimation and forecasting. 
	
	The key transformation $\widehat{LM}^{(\beta_{SSO})}$ can be also perceived to drive a statistical equilibrium for $IV^{SSO}_{t+1}$ through $\widehat{IV}^{SSO}_{t+1}$, deviations from which induce entry and exit points for trading. This reasoning is in line with that of \citet{avellaneda_lee} who introduce a model-driven pairs-trading strategy in US equities.
	
	The real-world SSO implied volatility $IV^{SSO}_{t; LM^{(\beta_{SSO})}_{\tau^*; M}, \tau^*}$ at time step $t$ is expected to converge to the forecast implied volatility $\widehat{IV}^{SSO}_{t+1; LM^{(\beta_{SSO})}_{\tau^*; M}, \tau^*}$, which has been constructed using scaled moneyness and implied volatility input from the unleveraged LETF, i.e., SPY.
	
	Such a strategy would fall into the class of model-driven statistical arbitrage in equities and equity options. It has the three characteristic features of statistical arbitrage described by \citet{avellaneda_lee}: (i) trading signals are systematic or rules-based, as opposed to driven by fundamentals, (ii) the trading portfolio is market-neutral, i.e. has zero sensitivity to the market, and (iii) the algorithm for generating excess returns is statistical. The market here is defined by the SSO LETF and delta neutrality implies market neutrality.
	
	In the following section, we present the strategy's performance as well as a numerical example. It occurs that the existing statistical discrepancies between implied volatilities of leveraged and unlevered ETF options together with predictive capacity of the DSFM model can provide non-negative trading gains on the option market.

	\subsubsection{Numerical example}
	\label{strat_ex}
	
	For the purpose of the estimation of the strategy from the previous section, the DSFM model parameters are taken to be the same as in Section \ref{emptest}. The rolling window width is assumed to be $w = 100$ and the forecasting horizon is 1 day ahead.
	
\begin{figure}[H]
		\begin{center}
			\includegraphics[scale=0.5]{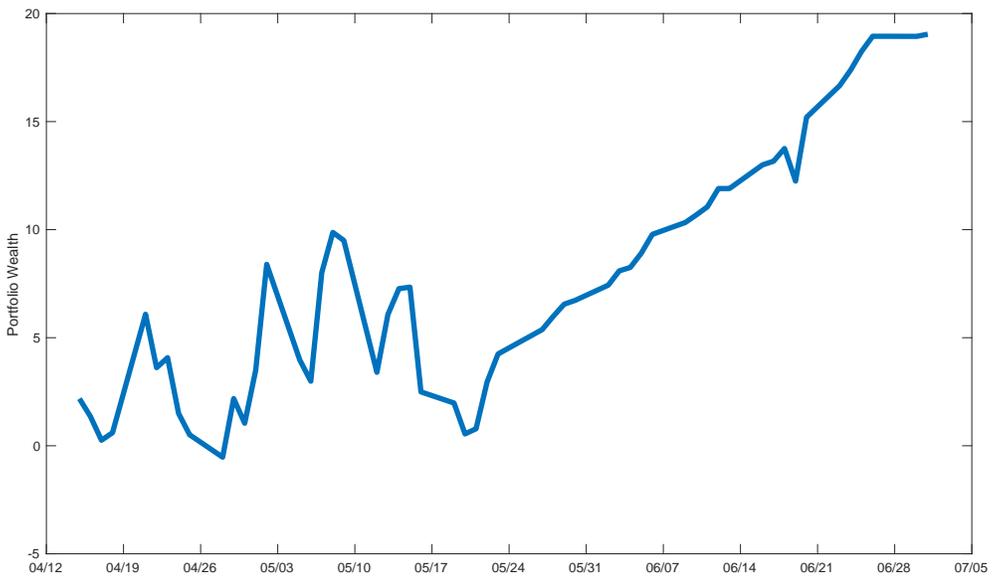}
		\end{center}
		\caption{Cumulative performance of the trading strategy}
		\label{strperf}
	\end{figure}

	The dynamic strategy performance in the period April 2015 - June 2015 is displayed in Figure \ref{strperf}. Out of 55 investment periods, in 30 cases long-only portfolios were constructed, the remaining 25 cases short and long positions were taken; net portfolios were short portfolios in 42 cases, long portfolios in the remaining 13 cases. 
	
\begin{figure}[H]
		\begin{center}
			\minipage{0.80\textwidth}
			\includegraphics[width=\linewidth]{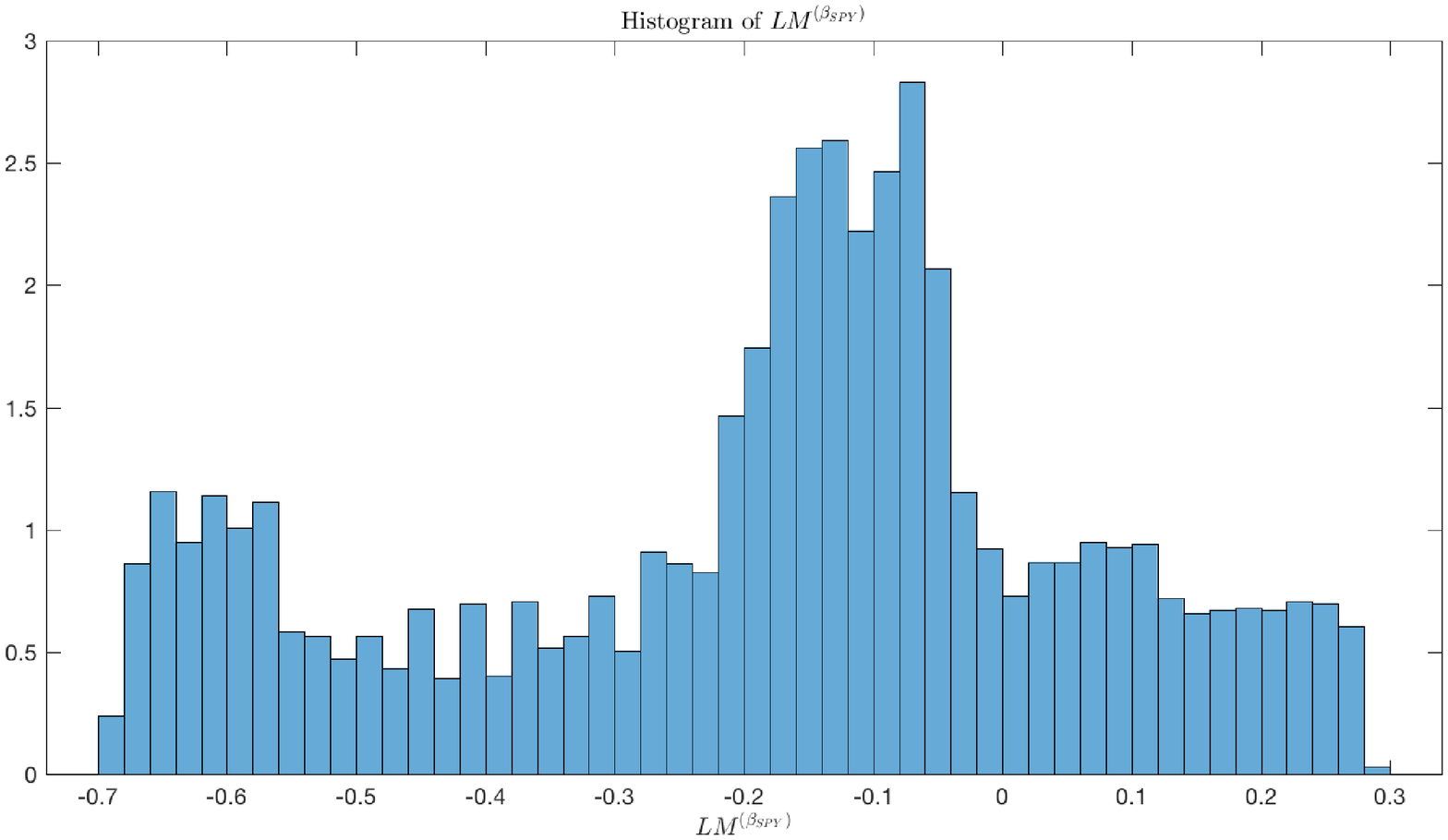}
			\endminipage\\
			\minipage{0.80\textwidth}
			\includegraphics[width=\linewidth]{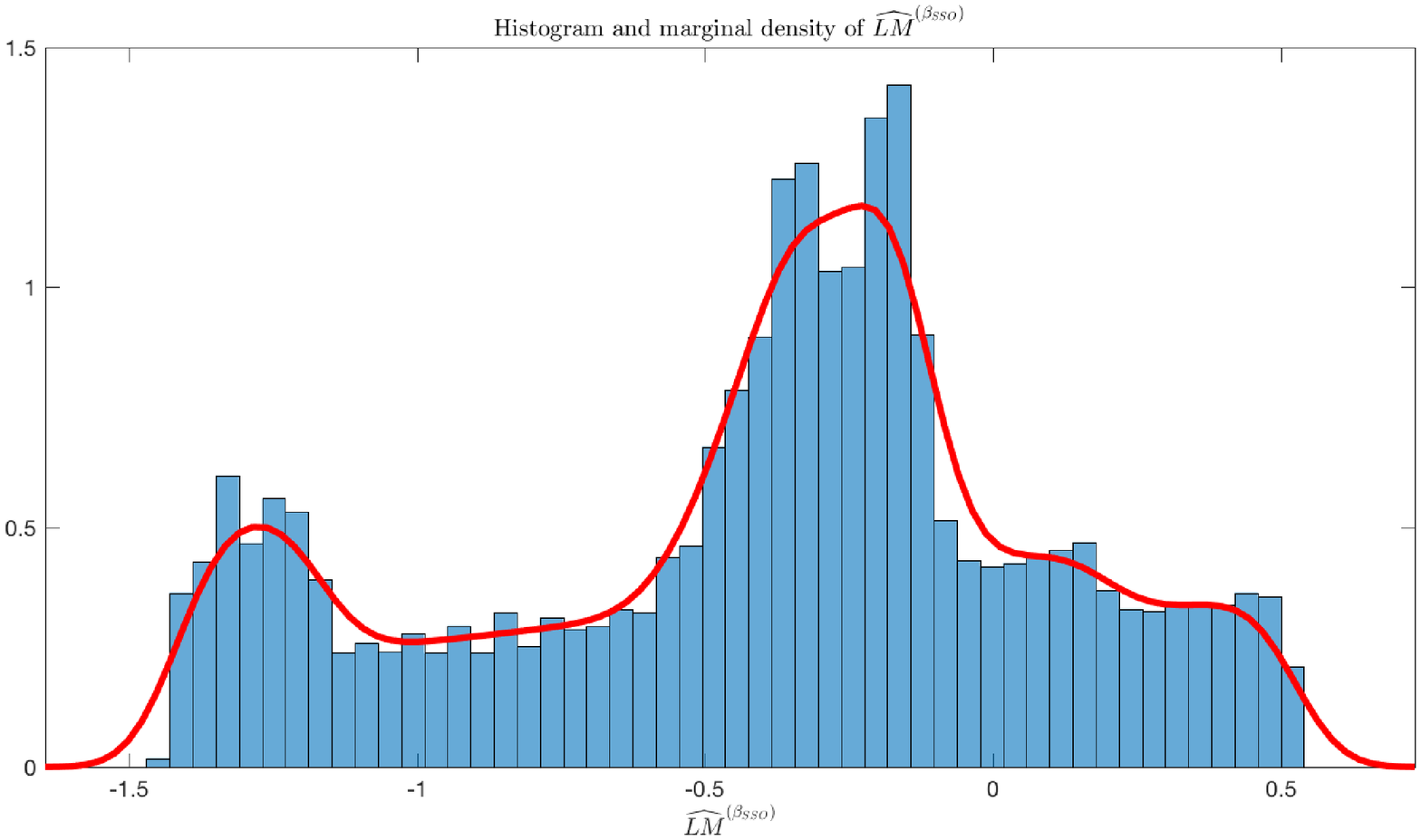}
			\endminipage\\
			\minipage{0.80\textwidth}
			\includegraphics[width=\linewidth]{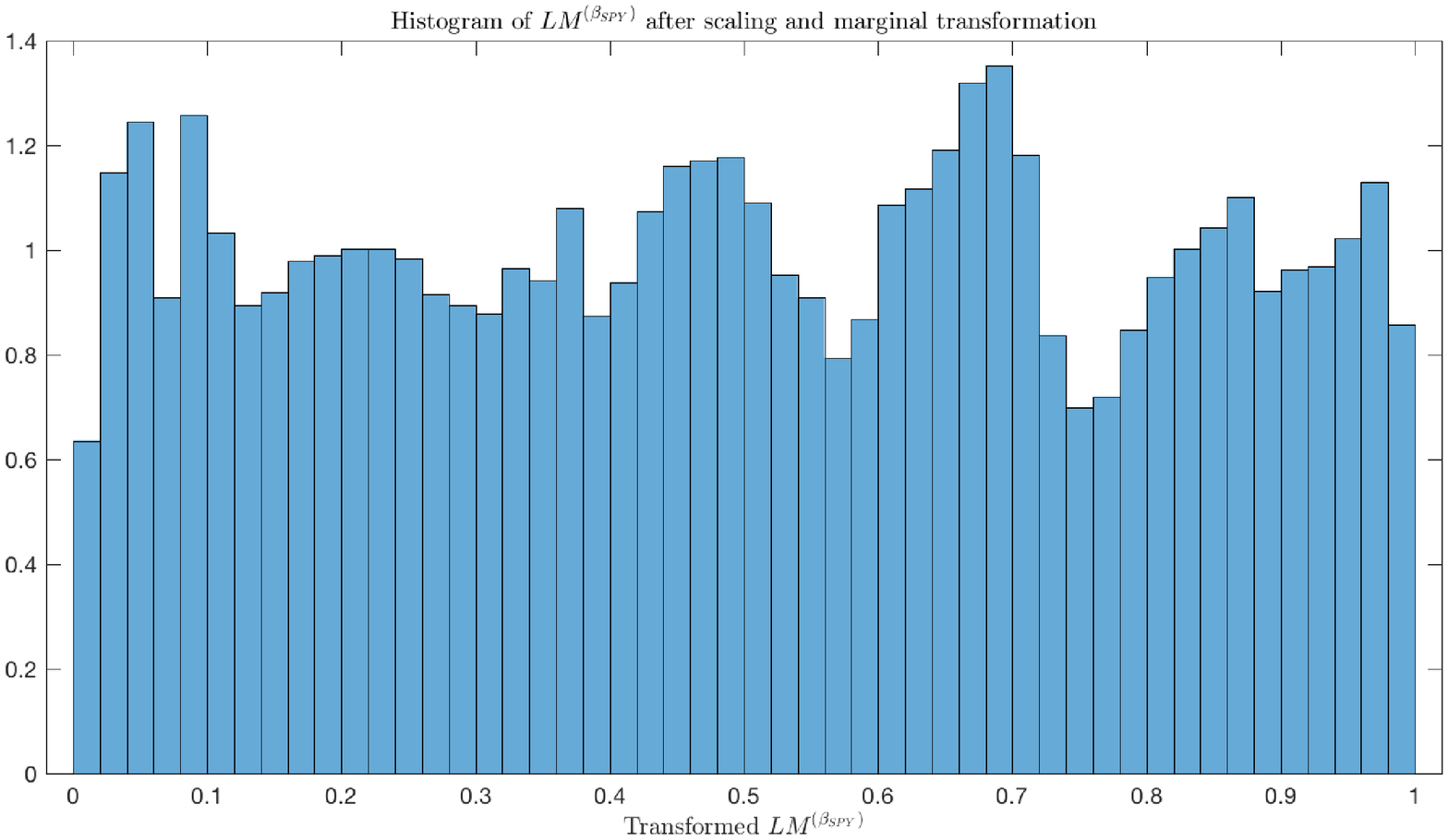}
			\endminipage
		\end{center}
		\caption{$LM^{(\beta_{SPY})}$ (top panel), $\widehat{LM}^{(\beta_{SSO})}$ (middle panel), $\widehat{LM}^{(\beta_{SSO})}$ after marginal transformation (bottom panel) } 
		\label{spy_logmon}
	\end{figure}

	For the sake of illustration, let us go through one step from the outlined strategy in a numerical example. Assume that we are at the step $t = 147$ of the sample, which corresponds to June 18, 2015. At this point, we have a training sample of 100 days for DSFM estimation, encompassing 14,859 observations of the option data for contracts with various strike prices and time-to-maturity. The histogram and density estimates for SPY log-moneyness $LM^{(\beta_{SPY})}$, "theoretical" SSO log-moneyness $\widehat{LM}^{(\beta_{SSO})}$ (that is, rescaled $LM^{(\beta_{SPY})}$) and its marginally transformed version are given in Figure \ref{spy_logmon}. Further we proceed as proposed in the strategy above:
	
	\begin{enumerate}
		\item
		estimate (\ref{dsfm}) and perform a forecast to obtain $\widehat{IV}_{148}^{SSO}$ on June 19, 2015 (day 148 in the sample),
		\item
		choose $\tau^{*} = 0.6$; we have 37 values of $LM^{(\beta_{SSO})}$ for $\tau^{*} = 0.6$. We use the marginal distribution of $\widehat{LM}^{(\beta_{SSO})}$ shown in Figure \ref{spy_logmon} to calculate the corresponding  "theoretical" values $LM^{(\beta_{SSO})}_{ \tau^{*}; M }$ implied by SPY data and the moneyness scaling procedure, both shown in Figure \ref{logmon_transf_result},
		\item
		using the forecast IVS $\widehat{IV}_{148}^{SSO}$, we can determine "theoretical" IV values corresponding to $LM^{(\beta_{SSO})}_{ \tau^{*}; M } \in [0, 1]$, $\widehat{IV}^{SSO}_{148; LM^{(\beta_{SSO})}_{\tau^*; M}, \tau^*}$ and the real-world IV values $IV^{SSO}_{147; LM^{(\beta_{SSO})}_{\tau^*; M}, \tau^*}$ corresponding to the same $LM^{(\beta_{SSO})}_{ \tau^{*}; M }$ through the mapping of $LM^{(\beta_{SSO})}$, described above,
		\item
		the resulting $\widehat{IV}^{SSO}_{148; LM^{(\beta_{SSO})}_{\tau^*; M}, \tau^*}$ and $IV^{SSO}_{147; LM^{(\beta_{SSO})}_{\tau^*; M}, \tau^*}$ are demonstrated in Figure \ref{ivs_real_theor}. We buy an option corresponding to the largest $D_{long} = 0.052$ at the closing price $C_{long; 147} = \$27.375$ and sell an option corresponding to the largest $D_{short} = 0.163$ at the closing price $C_{short; 147} = \$2.740$,
		\item
		additionally, we delta-hedge the resulting option portfolio using portfolio delta $\Delta_{Portf; 147} = \Delta_{long; 147} - \Delta_{short; 147} = 0.859 - 0.461 = 0.398$ and the underlying SSO which has a closing price of $L = \$66.960$; the total portfolio value on June 18, 2015 is: $C_{long; 147} - C_{short; 147} - \Delta_{Portf; 147} \times L_{147} = \$27.375 - \$2.740 - 0.398 \times \$66.960 = -2.015$,
		\item
		at the start of the next day, i.e., June 19, 2015, we check the prices of the options $C_{long; 148}$ and $C_{short; 148}$ as well as the price of the underlying $L_{148}$. We find that on June 19, 2015, $C_{long; 148} = \$28.450$ and $C_{short; 148} = \$0.320$, $L_{148} = \$68.300$. With the share of $L_{148}$ still equal to the previous-day portfolio delta $\Delta_{Portf; 147}$, the portfolio is now worth $C_{long; 148} - C_{short; 148} - \Delta_{Portf; 147} \times L_{148} = \$28.450 - \$0.320 - 0.398 \times \$68.300 = 0.947$. We sell that portfolio in an offsetting trade and have secured a gain of $\$2.962$. As expected, $C_{long}$ gained in value while $C_{short}$ went down in value. The coupled gain was larger than an offsetting loss of $\$0.533$ from the delta hedge which resulted in the total gain. 
	\end{enumerate}
	
	In the end, the cumulative gain of the strategy applied daily as shown in the illustrating example above, is 19.043 after 55 investment periods. It occurs that 39 out of 55 investment decisions correctly determined the direction of one-step-ahead implied volatility smile change. In the strategy, this smile change is anticipated according the relation between the "model" IVS computed using the moneyness scaling approach and the real-world IVS of a LETF option. There is a high positive chance of generating positive cumulative returns exploiting statistical deviations of leveraged and unlevered implied volatility smiles in the ETF option market.
	
	It should be mentioned that the sample period includes the day of an underlying SSO stock 2-for-1 split which took place on May 20, 2015. The split was implemented to attract a wider range of buyers at the resulting lower price per share. It has been shown by many researchers, such as \citet{ohlson_penman}, \citet{aamir_sheikh}, \citet{desai_nimalendran} that stock splits result in post-split increases of implied stock volatilities. For instance, \citet{ohlson_penman} show that stock splits cause short-term increases in volatility upon announcement and long-term increases in volatility after the date the split is effective.
	
\begin{figure}[H]
		\begin{center}
			\minipage{1.00\textwidth}
			\includegraphics[width=\linewidth]{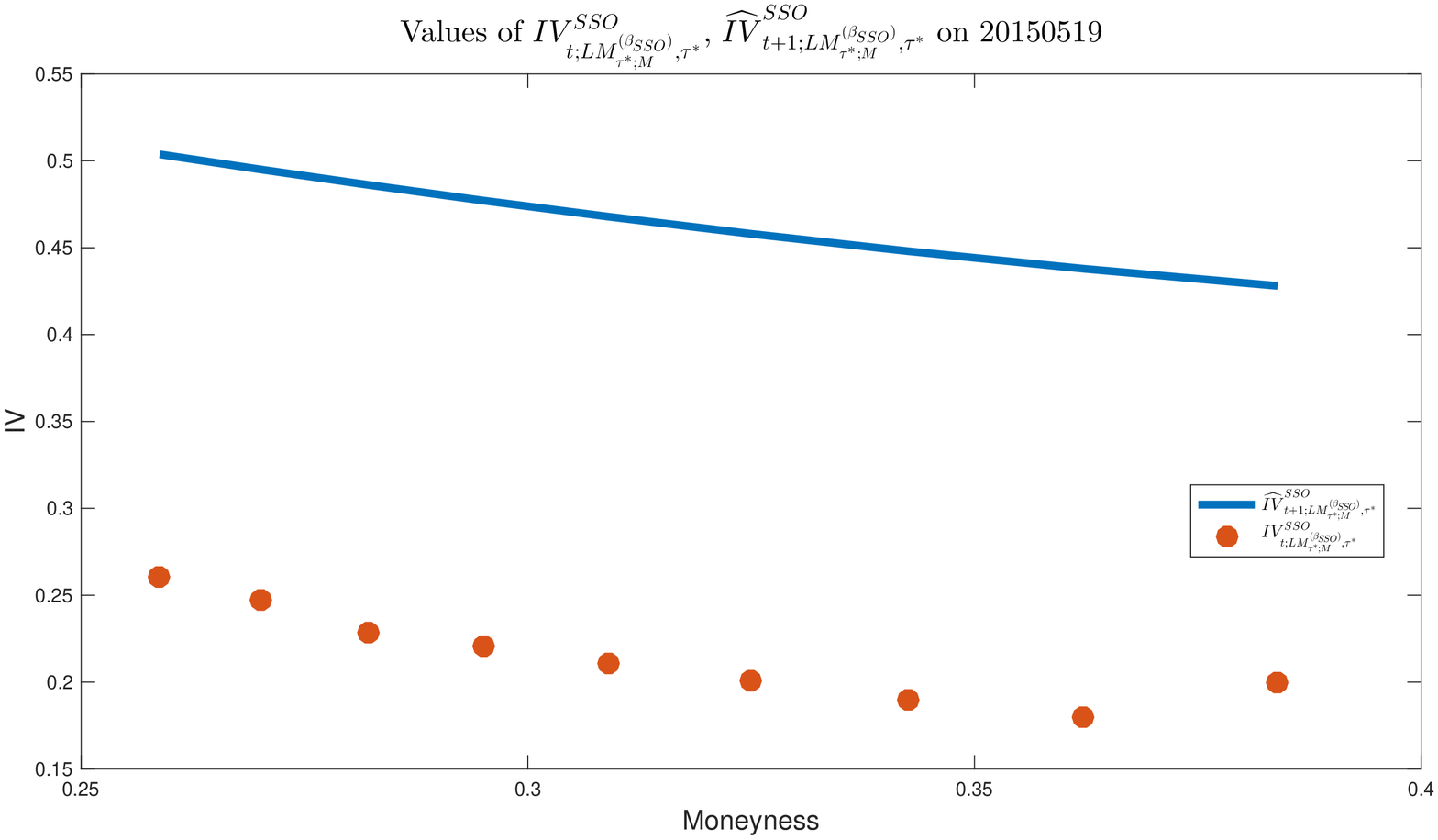}
			\endminipage\\
			\vspace{3mm}
			\minipage{1.00\textwidth}
			\includegraphics[width=\linewidth]{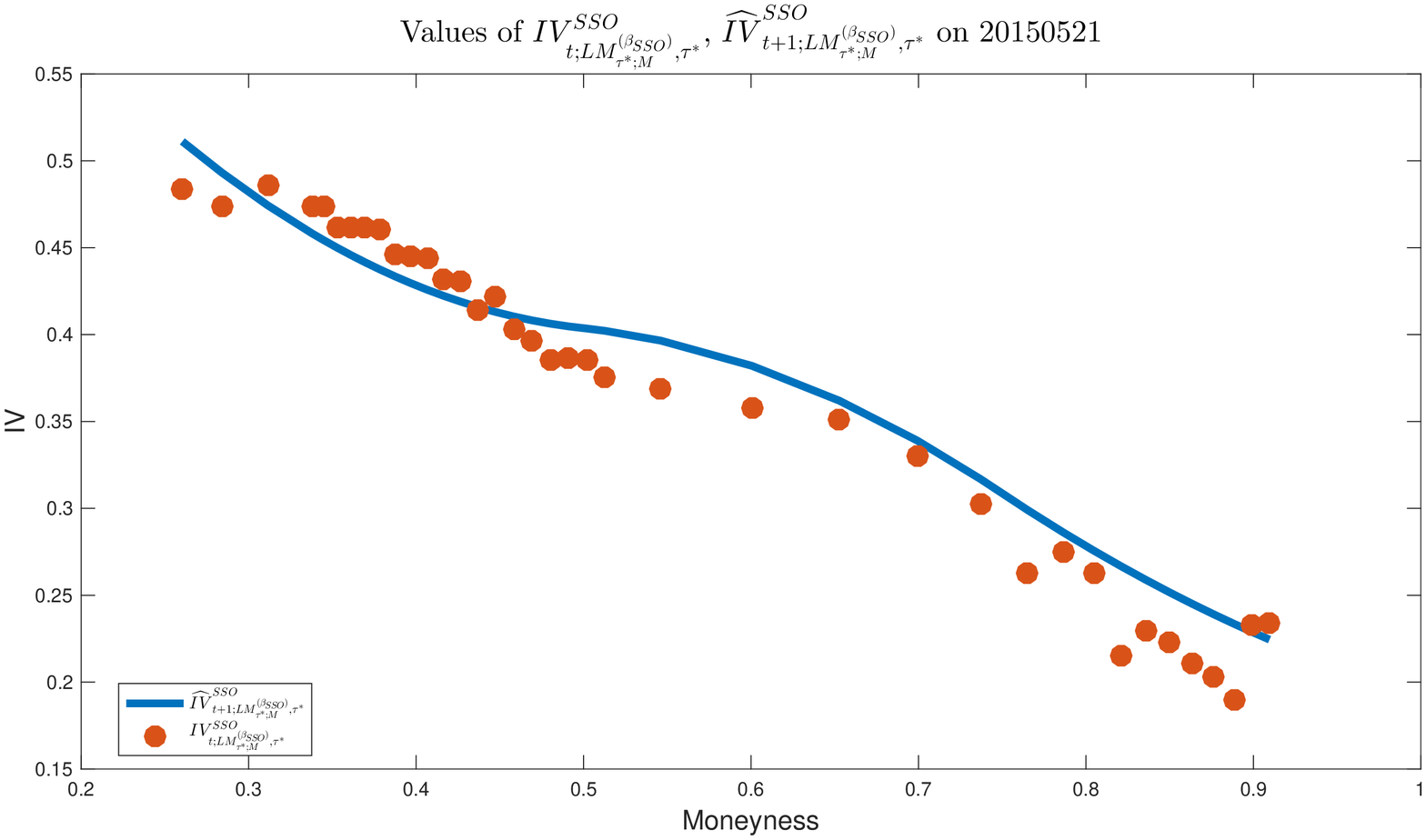}
			\endminipage
		\end{center}
		\caption{Model and real-world IVs before (top panel) and after (bottom panel) the split on May 20, 2015 } 
		\label{split_increase}
	\end{figure}

	In Figure \ref{split_increase}, we show real-world and DSFM-forecast IVS on two different dates: before (19 June, 2015) and after the split (21 June, 2015). It can be seen that the model anticipates a significant increase in implied volatilities after the split which indeed takes place.

	\subsubsection{Robustness check}
	The performance of the option trading strategy obtained in Section \ref{strat_ex} above may seem to have occurred purely by chance. Therefore some sort of a robustness check is necessary.
	
	We perform a bootstrap resampling exercise on the time series of the underlying prices of SPY and SSO (L)ETFs and re-run the strategy on the resampled data. Overlapping block bootstrap approach proposed by \citet{kuensch} is applied. It works as follows: given the data observations $\{X_i: i = 1,\ldots,T\}$, a block size $b$ is specified. With overlapping blocks of length $b$, block 1 is then observations $\{X_j: j = 1,\ldots,b\}$, block 2 is $\{X_{j+1}: j = 1,\ldots,b\}$ and so on. Random sampling is then performed on the level of blocks.
	
	Overlapping block bootstrap assumes the data to be stationary. However, in the current case we do not do inference on the resampled data, so this stringent assumption is less relevant here. Nevertheless, we run standard stationarity tests on SPY and SSO price series in the period from November 2014 to June 30, 2015 such as Phillips-Perron, augmented Dickey-Fuller and KPSS tests. The first two tests have presence of unit root in the series as a null hypothesis, while the KPSS test tests trend stationarity as the null. 
	
	We perform a series of tests for each approach using the number of lags from 1 to 10 in the Newey-West estimator of the long-run variance. In Table \ref{stat_tests} the results of stationarity tests are demonstrated. Most of this evidence does not reject the hypothesis of possible trend stationarity of price series of the (L)ETFs. Therefore we proceed with the bootstrap.
	
	We run 500 bootstrap iterations on 2-dimensional series of SPY and SSO prices in the period from November 2014 to June 30, 2015 and take the block size equal to 5. In Figure \ref{bootstrap_perf} cumulative performance of the strategy on bootstrapped time series is shown. At each of the 155 time steps the values of 2.5\% and 97.5\% empirical percentiles are found. We can see that at the end of the period, positive performance occurs with more than a 95\% probability. Positive performance with this probability occurs from period 23 onwards until the end of of the test sample, which yields 32 periods out of 55 in total.

	\section{Conclusion}
	
	In this paper, we provide statistical and econometric analysis of the moneyness scaling transformation for leveraged and unlevered exchange-traded funds' options' implied volatility smiles. This transformation adjusts the moneyness coordinate of the smile in an attempt to remove the discrepancy between the levered and unlevered counterparts. 
	
	We incorporate stochastic volatility into the moneyness scaling method by explicit estimation of the conditional expectation of the realized variance. We present two approaches to implement this estimate: via an analytical approach and using a Monte-Carlo method.
	
	We construct bootstrap uniform confidence bands which reveal a statistically significant discrepancy between the implied volatility smiles, even after moneyness scaling has been performed. We find that this discrepancy is stronger for in- and out-of-the-money options which, however, is unlikely to be explained by liquidity issues alone.  
	
	This discrepancy allows to define a theoretical statistical equilibrium value of LETF moneyness. Based on deviations from this equilibrium, possible trading gain opportunities on the (L)ETF market which can be exploited. We construct a trading strategy based on a dynamic semiparametric factor model. This model-based statistical arbitrage strategy utilizes the dynamic structure of implied volatility surface allowing out-of-sample forecasting and information on unleveraged ETF options to construct theoretical one-step-ahead implied volatility surfaces. 
	
	The proposed strategy has the potential to generate trading gains due to simultaneous use of the information from the discrepancies between the forecast "theoretical" (model) SSO LETF implied volatilities and the historical ("true") ones. It protects the portfolio against unfavorable moves in the underlying asset through delta-hedging and aims to gain from forecast moves in volatility. The strategy is shown via bootstrap technique to generate positive returns with a high probability.

	\section*{Funding}
	This research was supported by the Deutsche Forschungsgemeinschaft under Grant "IRTG 1792".
	
	\newpage
	
	\bibliographystyle{plainnat}
	\bibliography{manuscript}
	
	\newpage

	\section{Appendix}
	
	\subsection{The local linear M-smoothing estimator} 
	\label{app2}
	$M$-type smoothers apply a non-quadratic loss function $\rho(\cdot)$ to make estimation more robust. Given the model 
	\begin{equation}
	Y_i = m(X_i) + \varepsilon_i,
	\label{varestr}
	\end{equation}
	where $Y_i \in \mathbb{R}$, $X_i \in \mathbb{R}^d$, $\varepsilon_i \stackrel{\operatorname{def}}{=} \sigma (X_i) u_i$, $u_i \sim (0,1)$, i.i.d, $\mathcal{X} \stackrel{\operatorname{def}}{=} \{(X_i,Y_i); 1 \leq i \leq n\}$, the local linear $M$-smoothing estimator is obtained from:
	\begin{equation}
	\underset{\alpha \in \mathbb{R}, \beta \in \mathbb{R}^p}{\min} \sum_{i=1}^n \rho\left\{ Y_i - \alpha - \beta^{\top}(X_i - x) \right\} W_{ih}(x),
	\end{equation}
	where 
	\begin{equation}
	W_{hi}(x) \stackrel{\operatorname{def}}{=} \frac{h^{-2} K^{'} \{(x-X_i)/h\}}{\widehat{f}_h(x)} - \frac{K_h (x-X_i) \widehat{f}^{'}_h(x)}{\widehat{f}^{2}_h(x)}
	\end{equation}
	is a kernel weight sequence with $\widehat{f}^{'}_h(x) \stackrel{\operatorname{def}}{=} n^{-1} \sum_{i=1}^n K_h^{'}(x-X_i)$, $h$ is the bandwidth, $K$ is a kernel function; $\int K(u)du = 1$, $K_h(\cdot) \stackrel{\operatorname{def}}{=} h^{-1}K(\cdot/h)$. The function $\rho(\cdot)$ is designed to provide more robustness than the quadratic loss. An example of such a function is given by \citet{huber}, see also \citet{haerdle_paper_asy}:
	
	\begin{equation}
	\rho(u) = \left\{ \begin{array}{ll}
	0.5 u^2, & \mbox{if $|u| \leq c$};\\
	c|u|-0.5 c^2 & \mbox{if $|u| > c$}.\end{array} \right. ,
	\end{equation}
	with the constant $c$ regulating the degree of resistance.

	\subsection{Derivation of the conditional density of Heston integrated variance given terminal log-price }
	\label{hestproof}
	As pointed out in (\ref{expconddens}), we require the conditional density $f_{\tilde{V}|X_T}(\tilde{v}|x_T)$, under the measure $\QQ$, to analytically determine the expression (\ref{monscex}). As noted by \citet{broadie_kaya} and \citet{glasserman_kim}, the conditional distribution of $X_T$ is conditionally normal given $\widetilde{V}$ and $\int_{0}^{T}\sqrt{V_t} dW_{V,t}^{\QQ}$:
	\begin{align}
	X_T &\sim \mathbb{N} \left( (r-c-0.5)T - 0.5 \widetilde{V} + \rho \int_{0}^{T}\sqrt{V_t} dW_{V,t}^{\QQ}, \sqrt{1-\rho^2} \widetilde{V}  \right), \nonumber \\
	&\sim \mathbb{N} \left( (r-c-0.5)T - 0.5 \widetilde{V} + \frac{\rho}{\sigma} (V_T - V_0 - \kappa \theta T + \kappa \widetilde{V}), \sqrt{1-\rho^2} \widetilde{V}  \right) \label{terminalpricedist},
	\end{align}
	where (\ref{terminalpricedist}) follows from
	\begin{equation*}
	\int_{0}^{T}\sqrt{V_t} dW_{V,t}^{\QQ} = \sigma^{-1}(V_T - V_0 - \kappa \theta T + \kappa \widetilde{V}),
	\end{equation*}
	which, in turn, follows from (\ref{sz2}).
	
	Expression (\ref{terminalpricedist}) yields the conditional density $f_{X_T|\widetilde{V}, V_T}$ given Heston parameters $\kappa$, $\theta$, $V_0$, $\sigma$ and $\rho$. The joint density $f_{\widetilde{V}, X_T, V_T}$ can then be obtained by simply multiplying $f_{X_T|\widetilde{V}, V_T}$ by $f_{\widetilde{V}, V_T}$, the joint density of $\widetilde{V}$ and $V_T$ given some starting Heston variance level $V_0$. 
	
	\citet{broadie_kaya} have derived the characteristic function of the distribution of $\widetilde{V}$ given variance endpoint values $V_T$ and $V_0$. This function is quite complex and involves modified Bessel functions of a complex variable of the first kind:
	\begin{multline}
	\varphi_{\widetilde{V}|V_T}(\omega) = \frac{\gamma(\omega)e^{-0.5(\gamma(\omega) - \kappa)T}(1 - e^{-\kappa T})}{\kappa (1 - e^{-\gamma(\omega)T})} \exp \left\{ \frac{V_T + V_0}{\sigma^2} \frac{\kappa (1 + e^{-\kappa T})}{1 - e^{-\kappa T}} - \frac{\gamma(\omega) (1 + e^{-\gamma(\omega) T})}{1 - e^{-\gamma(\omega) T}} \right\} \times \\
	\times \frac{B_{2 \kappa \theta \sigma^{-2} - 1} \left( \sqrt{V_T V_0}
		 \frac{4 \gamma(\omega) e^{-0.5 \gamma(\omega) T}}{\sigma^2 (1 - e^{-\gamma(\omega) T})} \right)}{B_{2 \kappa \theta \sigma^{-2} - 1} \left( \sqrt{V_T V_0} \frac{4 \kappa e^{-0.5 \kappa T}}{\sigma^2 (1 - e^{-\kappa T})} \right)},
	\end{multline}
	where $\gamma(\omega) = \sqrt{\kappa^2 - 2\sigma^2 i \omega}$, $i = \sqrt{-1}$ and $B_{\nu}(z)$ is the modified Bessel function of the first kind given by
	\begin{equation*}
	B_{\nu}(z) \stackrel{\operatorname{def}}{=} (z/2)^{\nu} \sum_{j=0}^{\infty} \frac{(z^2/4)^j}{j! \Gamma(\nu + j + 1)},
	\end{equation*}
	where $\Gamma(x) \stackrel{\operatorname{def}}{=} \int_0^{\infty} t^{x-1} e^{-t} dt$ is the gamma function.
	
	Using the inversion formula for characteristic functions, we can compute the density $f_{\widetilde{V}|V_T, V_0}$ as follows:
	\begin{equation*}
	f_{\widetilde{V}|V_T}(\widetilde{v}|v_T) = \frac{1}{2 \pi} \int_{-\infty}^{+\infty} e^{-i\widetilde{v} \omega} \varphi_{\widetilde{V}|V_T} (\omega) d\omega.
	\end{equation*}
	
	To find the joint density $f_{\widetilde{V}, X_T, V_T}$, the transitional density $f_{V_T|V_0}$ is required. As noted by \citet{cox_et_al} in the context of short interest rate process, $V_T$ given $V_0$ follows a scaled non-central chi-squared distribution:
	\begin{equation}
	V_T = \frac{\sigma^2 (1 - e^{-\kappa T})}{4 \kappa} \widetilde{\chi}_d^2 \left( \frac{4 \kappa e^{- \kappa T V_0}}{\sigma^2 (1 - e^{-\kappa T})} \right),
	\end{equation}
	where $\widetilde{\chi}_d^2 (\lambda)$ stands for the non-central chi-squared random variable with $d$ degrees of freedom and non-centrality parameter $\lambda$. The probability density function of $\widetilde{\chi}_d^2 (\lambda)$ is defined using $B_{\nu}(z)$:
	\begin{equation*}
	f_{\widetilde{\chi}_d^2 (\lambda)}(x) = 0.5 e^{-0.5(x+\lambda)} (x \lambda^{-1})^{0.25d-0.5} B_{0.5 d-1}(\sqrt{\lambda x})
	\end{equation*}
	
	Using a change-of-variables technique, it is straightforward to show that the density $f_{V_T|V_0}$ takes the form:
	\begin{multline*}
	f_{V_T|V_0} (v_T|V_0)= \frac{2 \kappa}{\sigma^2 (1 - e^{-\kappa T})} \exp \left\{ \frac{\kappa^2 \theta T}{\sigma^2} - 0.5 \kappa T - \frac{2 \kappa (v_T + e^{-\kappa T} V_0)}{\sigma^2 (1 - e^{-\kappa T})} \right\} \times \\ \times \left( \frac{v_T}{V_0} \right)^{\kappa \theta \sigma^{-2} - 0.5} B_{2 \kappa \theta \sigma^{-2} - 1} \left( \frac{4 \kappa e^{-0.5 \kappa T}}{\sigma^2 (1 - e^{-\kappa T})} \sqrt{V_0 v_T} \right).
	\end{multline*}
	
	Using known rules for computing joint densities via conditional and marginal densities, it follows that
	\begin{align*}
	f_{\widetilde{V}, X_T, V_T} (\widetilde{v}, x_T, v_T) &= f_{X_T|\widetilde{V}, V_T} (x_T|\widetilde{v}, v_T) f_{\widetilde{V}|V_T}(\widetilde{v}|v_T) f_{V_T|V_0} (v_T|V_0) \\
	f_{\widetilde{V}, X_T} (\widetilde{v}, x_T) &= \int_{0}^{\infty} f_{\widetilde{V}, X_T, V_T} (\widetilde{v}, x_T, v_T) d v_T
	\end{align*}
	
	Therefore we have for $f_{\widetilde{V}, X_T}$ estimated at $\widetilde{v}$, $x_T$, given the Heston parameters:
	\begin{multline*}
	f_{\widetilde{V}, X_T} (\widetilde{v}, x_T) = \frac{2 \kappa}{(2 \pi)^{3/2} \sqrt{(1 - \rho^2) \widetilde{v}} \sigma^2 (1 - e^{-\kappa T})} \int_{0}^{\infty} \exp \left\{ \frac{\kappa^2 \theta T}{\sigma^2} - 0.5 \kappa T - \frac{2 \kappa (v_T + e^{-\kappa T} V_0)}{\sigma^2 (1 - e^{-\kappa T})}  \right. \\ \left. + \frac{(v_T + V_0)\kappa (1 + e^{-\kappa T})}{\sigma^2 (1 - e^{-\kappa T})} - \frac{ \left( x_T - \log(S_0) - (r-c-0.5)T + 0.5 \widetilde{v} - \frac{\rho}{\sigma} (v_T - V_0 - \kappa \theta T + \kappa \widetilde{v}) \right)^2 }{2 (1 - \rho^2) \widetilde{v}} \right\} \\ \times \left( \frac{v_T}{V_0} \right)^{\kappa \theta \sigma^{-2} - 0.5} \int_{-\infty}^{+\infty} e^{-i \widetilde{v} \omega} \left[ \frac{\gamma(\omega)e^{-0.5(\gamma(\omega) - \kappa)T}(1 - e^{-\kappa T})}{\kappa (1 - e^{-\gamma(\omega)T})} \exp \left\{ - \frac{\gamma(\omega) (1 + e^{-\gamma(\omega) T}) (v_T + V_0)}{\sigma^2 (1 - e^{-\gamma(\omega) T})} \right\} \right. \\
	\left. \times B_{2 \kappa \theta \sigma^{-2} - 1} \left( \sqrt{v_T V_0} \frac{4 \gamma(\omega) e^{-0.5 \gamma(\omega) T}}{\sigma^2 (1 - e^{-\gamma(\omega) T})} \right) \right] d \omega d v_T
	\end{multline*}
	
	Finally, the density $f_{\widetilde{V}|X_T} (\widetilde{v}|x_T)$ is found as
	\begin{equation*}
	f_{\widetilde{V}|X_T} (\widetilde{v}|x_T) = \frac{f_{\widetilde{V}, X_T} (\widetilde{v}, x_T)}{f_{X_T}(x_T)},
	\end{equation*}
	where $f_{X_T}(x_T)$ is the probability density of $X_T$ estimated at $x_T$. This marginal density under the risk-neutral measure $\QQ$ is again found via inversion of the characteristic function, see \citet{rouah}:
	\begin{equation*}
	\varphi(\omega)_{X_T} = \exp\left\{ C(\omega) + D(\omega)V_0 + i\omega \log(S_0) \right\},
	\end{equation*}
	where
	\begin{align*}
	C(\omega) &= r i \omega T + \frac{\kappa \theta}{\sigma^2} \left\{(\kappa - \rho \sigma i \omega + d(\omega)) T - 2 \log \left(\frac{1-g(\omega) e^{-d(\omega) T}}{1-g(\omega)}\right)\right\},\\
	D(\omega) &= \frac{(\kappa - \rho \sigma i \omega + d(\omega))}{\sigma^2} \left(\frac{1-e^{-d(\omega)T}}{1-g(\omega)e^{-d(\omega)T}}\right), \\
	g(\omega) &= \frac{\kappa - \rho \sigma i \omega + d(\omega)}{\kappa - \rho \sigma i \omega - d(\omega)},\\
	d(\omega) &= \sqrt{(\rho \sigma i \omega - \kappa)^2 + \sigma^2 (i \omega - \omega^2)}.
	\end{align*}

	\newpage
	\section{Tables}
	
	\vspace*{\fill}
	\begin{table}[ht!]
		\centering
		\begin{tabular}{lccccc}
			\hline \hline
			(L)ETF & \multicolumn{1}{c}{Ticker} & \multicolumn{1}{c}{Lev. ratio} & \multicolumn{1}{c}{Exp. ratio (\%)} & \multicolumn{1}{c}{Div. yield (\%)} \\ 
			\hline
			SPDR S\&P 500                                & SPY    & $+1$  & 0.090  & 1.867   \\
			ProShares Ultra S\&P500                 & SSO    & $+2$  & 0.900  & 0.440   \\ 
			ProShares UltraPro S\&P500            & UPRO & $+3$  & 0.950  & 0.263   \\ 
			ProShares UltraShort S\&P500         & SDS    & $-2$  & 0.890   & 0.000   \\ 
			ProShares UltraPro Short S\&P500  & SPXU  & $-3$  & 0.900  & 0.000   \\ 
			\hline \hline
		\end{tabular}
		\caption{Summary financial information on (leveraged) ETFs on S\&P 500 underlying index}
		\label{letfs_desc_info}
	\end{table}
	\vspace*{\fill}
	
	\newpage
	
	\vspace*{\fill}
	\begin{table}[ht!]
		\centering
		\begin{tabular}{l|l|......}
			\hline \hline
			\multicolumn{2}{c}{} & \multicolumn{1}{c}{Min.} & \multicolumn{1}{c}{Max.} & \multicolumn{1}{c}{Mean} & \multicolumn{1}{c}{Stdd.} & \multicolumn{1}{c}{Skewn.} & \multicolumn{1}{c}{Kurt.}  \\
			\hline
			\multirow{4}[1]{*}{\textbf{SPY}}  & $\tau$       & 0.258  & 2.364 & 1.202  & 0.515 & 0.421  & 2.316  \\
			& $LM$         & -3.061 & 0.477 & -0.381 & 0.574 & -1.513 & 5.871  \\
			& $\sigma_{I}$ & 0.086  & 2.677 & 0.271  & 0.195 & 3.228  & 18.522 \\
			\hline
			\multirow{4}[1]{*}{\textbf{SSO}}  & $\tau$       & 0.208  & 2.236 & 1.239  & 0.585 & -0.044 & 1.795  \\
			& $LM$         & -1.704 & 0.558 & -0.484 & 0.461 & -0.089 & 2.264  \\
			& $\sigma_{I}$ & 0.154  & 1.340 & 0.363  & 0.091 & 1.774  & 12.224 \\
			\hline
			\multirow{4}[1]{*}{\textbf{UPRO}} & $\tau$       & 0.208  & 2.236 & 1.205  & 0.585 & 0.043  & 1.795  \\
			& $LM$         & -1.182 & 0.665 & -0.168 & 0.331 & -0.360 & 2.719  \\
			& $\sigma_{I}$ & 0.250  & 1.669 & 0.503  & 0.099 & 1.335  & 9.080  \\
			\hline
			\multirow{4}[1]{*}{\textbf{SDS}}  & $\tau$       & 0.208  & 2.236 & 1.146  & 0.581 & 0.196  & 1.852  \\
			& $LM$         & -0.738 & 0.858 & 0.187  & 0.344 & -0.276 & 2.226  \\
			& $\sigma_{I}$ & 0.107  & 1.262 & 0.424  & 0.129 & 0.792  & 4.830 \\
			\hline \hline
		\end{tabular}
		\caption{ Summary statistics on (L)ETF options data ($\tau$ is time to maturity, $LM$ log-moneyness, $\sigma_I$ implied volatility)}
		\label{sumstat}
	\end{table} 
	\vspace*{\fill}
	
	\newpage

	\vspace*{\fill}
	\begin{table}[ht!]
		\centering
		\begin{tabular}{l....}
			\hline \hline
			Criterion & \multicolumn{1}{c}{$L=2$} & \multicolumn{1}{c}{$L=3$} & \multicolumn{1}{c}{$L=4$} & \multicolumn{1}{c}{$L=5$} \\ 
			\hline
			$EV$     & 0.915 & 0.921 & 0.925 & 0.930 \\ 
			$RMSE$    & 0.090 & 0.088 & 0.087 & 0.082 \\ 
			$RMSPE$  & 0.095 & 0.096 & 0.099 & 0.102 \\
			\hline \hline
		\end{tabular}
		\caption{ $EV$, $RMSE$ and $RMSPE$ criteria for different model order sizes}
		\label{evrmse}
	\end{table}
	\vspace*{\fill}
	
	\newpage

	\vspace*{\fill}
	\begin{table}[ht!]
		\centering
		\begin{tabular}{l...}
			\hline \hline
			Model order $n$ & \multicolumn{1}{c}{AIC($n$)} & \multicolumn{1}{c}{HQ($n$)} & \multicolumn{1}{c}{SC($n$)} \\ 
			\hline
			1 & -4.20^* & -4.10^* & -3.96^*  \\ 
			2 & -4.13 & -3.96 & -3.72  \\ 
			3 & -4.07 & -3.83 & -3.48  \\ 
			4 & -4.03 & -3.72 & -3.27  \\ 
			5 & -3.97 & -3.59 & -3.03 \\ 
			\hline \hline
		\end{tabular}
		\caption{The VAR model selection criteria. The smallest value is marked by an asterisk}
		\label{zcrit}
	\end{table}
	\vspace*{\fill}
	
	\newpage
	
	\vspace*{\fill}
	\begin{table}[ht!]
		\centering
		\caption{Stationarity tests' statistics for SPY, SSO price series}
		\renewcommand*{\arraystretch}{1.5}
		\begin{tabular}{l......}
			\toprule
			\multicolumn{1}{l}{\multirow{2}{*}{Lags }} & \multicolumn{3}{c}{SPY} & \multicolumn{3}{c}{SSO} \\ 
			\cmidrule(l{20pt}r){2-4} \cmidrule(ll{20pt}r){5-7}
			& \multicolumn{1}{c}{PP} & \multicolumn{1}{c}{ADF} & \multicolumn{1}{c}{KPSS} & \multicolumn{1}{c}{PP} & \multicolumn{1}{c}{ADF} & \multicolumn{1}{c}{KPSS} \\
			\midrule
			1    & -3.621^{**} & -3.505^{**}   & 0.216^{**}      & -3.685^{**}  & -3.578^{**}   & 0.200^{**}      \\
			2    & -3.693^{**}  & -3.593^{**}  & 0.157^{**}      & -3.784^{**}  & -3.779^{**}   & 0.145^{*}     \\
			3    & -3.752^{**}  & -3.725^{**}  & 0.127^{*}      & -3.855^{**}  & -3.940^{**}   & 0.118     \\
			4    & -3.730^{**}  & -3.402^{*}    & 0.110            & -3.837^{**}  & -3.601^{**}   & 0.102     \\
			5    & -3.716^{**}  & -3.400^{*}    & 0.098           & -3.824^{**}  & -3.622^{**}   & 0.092     \\
			6    & -3.664^{**}  & -3.133         & 0.091           & -3.770^{**}  & -3.360^{*}    & 0.085     \\
			7    & -3.610^{**}  & -2.970         & 0.085           & -3.706^{**}  & -3.152^{*}    & 0.081     \\
			8    & -3.575^{**}  & -2.908         & 0.081           & -3.660^{**}  & -3.094   & 0.077     \\
			9    & -3.537^{**}  & -2.897         & 0.078           & -3.613^{**}  & -3.103   & 0.074     \\
			10   & -3.513^{**}  & -3.003        & 0.075           & -3.572^{**}  & -3.125   & 0.072 \\
			\bottomrule
		\end{tabular}
		\begin{tablenotes}
			\item[*] PP: Phillips-Perron test; ADF: augmented Dickey-Fuller test; KPSS: KPSS test for trend stationarity
			\item[**] $***, **, *$: significant on 1\%, 5\%, 10\% level, respectively
		\end{tablenotes}
		\label{stat_tests}
	\end{table}
	\vspace*{\fill}

	\newpage
	\section{Figures}
	
	\vfill
	\vfill

	\newpage
	\pagebreak

	\pagebreak

	\vspace*{\fill}
	\vspace*{\fill}
	
	\newpage
	
	\vspace*{\fill}
	\vspace*{\fill}
	
	\newpage
	
	\newpage
	
	\vspace*{\fill}
	\vspace*{\fill}
	
	\newpage
	
	\vspace*{\fill}
	\vspace*{\fill}
	
	\newpage
	
	\begin{landscape}
		
		\begin{figure}[H]
			\begin{center}
				\includegraphics[scale=0.9]{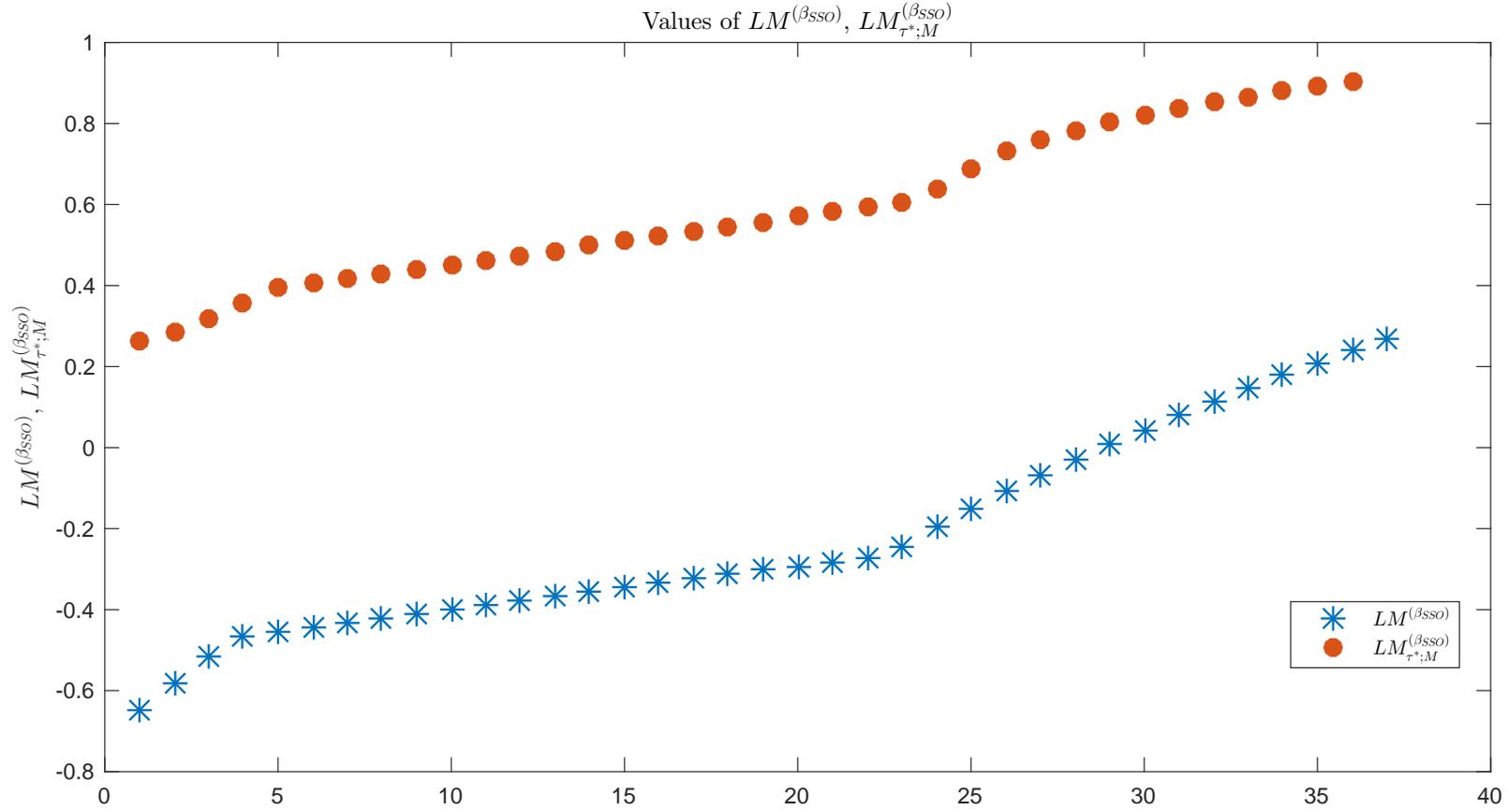}
			\end{center}
			\caption{Values of $LM^{(\beta_{SSO})}$ and $LM^{(\beta_{SSO})}_{ \tau^{*}; M }$}
			\label{logmon_transf_result}
		\end{figure}
		\vspace*{\fill}
		
		\begin{figure}[H]
			\begin{center}
				\includegraphics[scale=0.7]{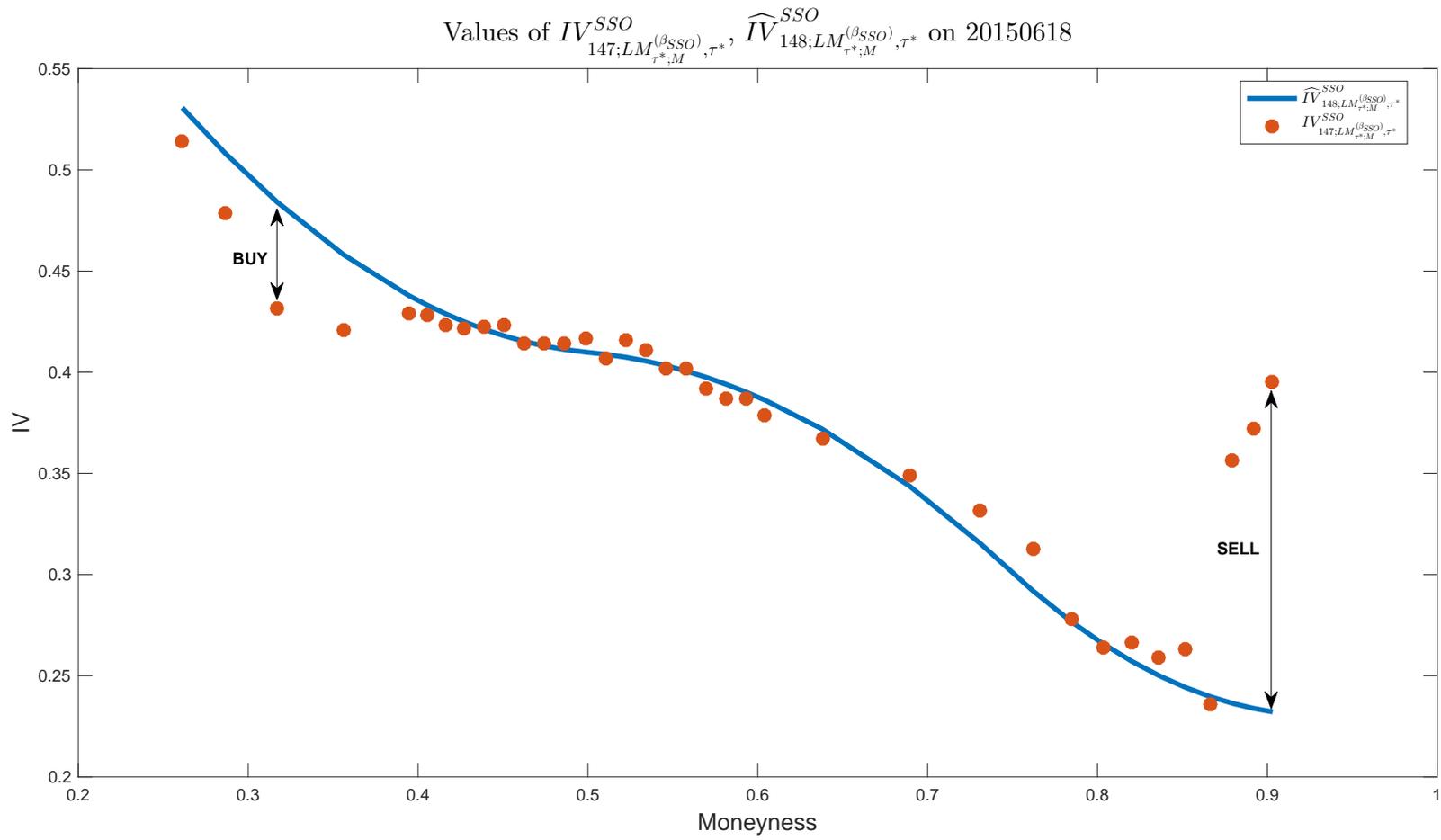}
			\end{center}
			\caption{Values of real-world $IV^{SSO}_{147; LM^{(\beta_{SSO})}_{\tau^*; M}, \tau^*}$ and predicted $\widehat{IV}^{SSO}_{148; LM^{(\beta_{SSO})}_{\tau^*; M}, \tau^*}$; the IV points at which long and short trades are done, are indicated by arrows }
			\label{ivs_real_theor}
		\end{figure}
		
	\end{landscape}
	
	\newpage
	
	\vspace*{\fill}
	\vspace*{\fill}
	
	\newpage
	
	\begin{landscape}
		\vspace*{\fill}
		\begin{figure}[H]
			\begin{center}
				\includegraphics[scale=0.6]{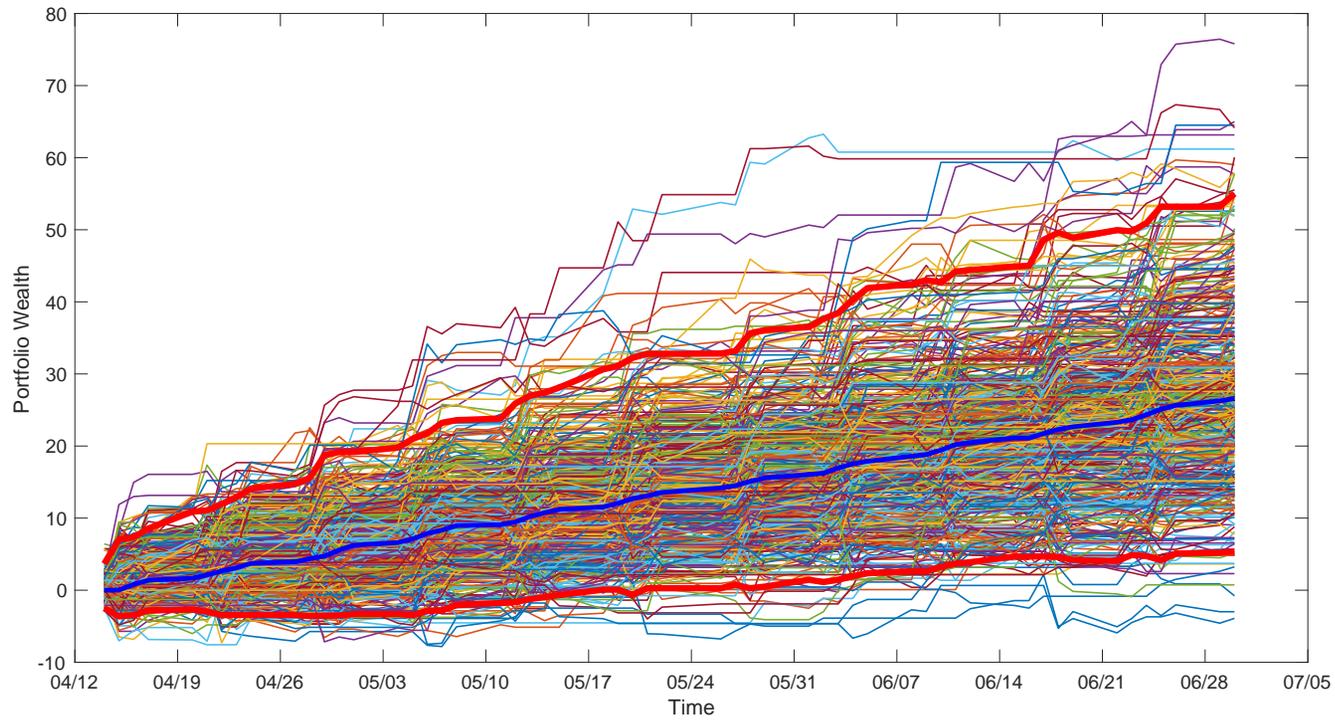}
			\end{center}
			\caption{Bootstrapped performance of the strategy; in red: 2.5\% and 97.5\% empirical percentiles of the cumulative performance; in blue: median of the cumulative performance}
			\label{bootstrap_perf}
		\end{figure}
		\vspace*{\fill}
	\end{landscape}
	
	\newpage
	
	\vspace*{\fill}
	\vspace*{\fill}
	
	\newpage
	
	\vspace*{\fill}
	\vspace*{\fill}
	
	\newpage
	
	\vspace*{\fill}
	\vspace*{\fill}

\end{document}